\documentclass[%
 reprint,
 amsmath,amssymb,
 aps,
prb,
]{revtex4-2}

\usepackage{graphicx}
\usepackage{dcolumn}
\usepackage{bm}

\usepackage[colorlinks=true, allcolors=blue]{hyperref}
\usepackage{mathtools}
\usepackage{tensor}
\usepackage{multirow}
\usepackage{amsmath}
\usepackage{amssymb}
\usepackage{accents}
\usepackage{caption}
\usepackage{braket}
\usepackage{lipsum}
\usepackage{mwe}
\usepackage{comment}
\usepackage{bigdelim}
\DeclareSymbolFont{extraup}{U}{zavm}{m}{n}
\DeclareMathSymbol{\varheart}{\mathalpha}{extraup}{86}
\DeclareMathSymbol{\vardiamond}{\mathalpha}{extraup}{87}

\begin{document}

\preprint{APS/123-QED}

\title{Classifying Order-Two Spatial Symmetries in Non-Hermitian Hamiltonians: Point-gapped AZ and AZ$^\dag$ Classes}
\author{Yifan Wang}
\email{Ivan.wang@emory.edu}
\affiliation{Department  of  Physics,  Emory  University, Atlanta, GA 30322, USA}

\begin{abstract}
    Crystalline topological insulators and superconductors have been a prominent topic in the field of condensed matter physics. These systems obey certain crystalline (spatial) symmetries that depend on the geometry of the lattice. The presence of spatial symmetries can lead to shift in the classification of ten-fold Altland–Zirnbauer class, given rise to new symmetry-protected topological phases. If the constraint of Hermiticity is broken, the classification expand into 38-fold. In this paper, following procedures in Hermitian systems, we classify all possible types of order-two spatial symmetries for point-gapped non-Hermitian systems within $16$ out of $38$ non-Hermitian topological classes. These $16$ classes are denoted by AZ and AZ$^\dag$ classes. We show that, similar to the Hermitian case, spatial symmetries will also lead to a shift in the classification of AZ and AZ$^\dag$ classes. There also exist novel symmetry-protected topological phases exclusive to point-gapped non-Hermitian Hamiltonians. Toy models are also given based on our classifications. 
\end{abstract}
\maketitle

\tableofcontents
\section{Introduction}

Topological insulators (TI) and superconductors (TSC) are some of the most intriguing discovery in condensed matter physics~\cite{Qi_2011,Hasan_2010,Moore_2010,Sato_2017}.  Ref.~\cite{Kitaev_Lebedev_Feigel’man_2009} proposed a systematic way of classifying TI and TSC based on three internal symmetries: time-reversal symmetry (TRS), particle-hole symmetry (PHS), and chiral symmetry (CS). Together, they composed 10 different classes which are called periodic table of TI and TSC or Altland–Zirnbauer (AZ) symmetry class~\cite{Altland_1997}. AZ classes give a complete classification of gapped phases of non-interacting fermionic systems with respect to internal symmetries. Different AZ classes are distinct from each other by some topological invariants that could exist in the presence or absence of certain internal symmetries. A phase transition between systems that have different topological invariants is achieved through the gap-closing of energy bands. For example, Chern insulator, which breaks TRS, can exist in 2D of class A~\cite{Kitaev_Lebedev_Feigel’man_2009,Chiu_2016_classification}. It has a $\mathbb{Z}$ invariant called the Chern number. When a system with a non-trivial Chern number is in contact with the vacuum (which has Chern number $0$), a gap closing of energy bands at the interface (or boundary) must happen. The states that localized at the boundary that closed the energy gap are the chiral edge states of Chern insulators. Similarly, quantum spin hall effect, which requires the existence of spinful TRS, can exist in class AII. It has a $\mathbb{Z}_2$ invariant called Kane-Mele $\mathbb{Z}_2$ invariant~\cite{Kane_1,Kane_2}. When a non-trivial quantum spin hall system is in contact with the vacuum, helical edge states appear at the interface. These edge modes show that topological numbers can predict the phenomena of quantum systems.

The boundary of a system can be more generally understand in the context of topological defects~\cite{Teo_2010,Teo_Hughes_2017,Shiozaki_2014}. Topological defects are defined to be lattice dislocation and disinclination that cannot be gapped out by continuous deformation. Under the protection of symmetries, defects may also host gapless topological modes. Therefore, under this view, boundary of a system is just another form of defects. 

Meanwhile, Ref.~\cite{Fu_2007,Fu_2010} discovered TIs that are based on inversion and crystalline symmetries, which are non-local spatial symmetries that depends on the geometry of the lattice. And, as shown in Ref.~\cite{Fu_2007,Fu_2010}, non-trivial TIs exist under the protection of spatial symmetries which are outside the classification of AZ symmetry classes. Therefore, a complete classification that includes spatial symmetries is called for. 
Works that classifies AZ symmetry class in addition to spatial symmetries are Ref.~\cite{Chiu_2013,Shiozaki_2014,Morimoto_2013,Shiozaki_2016,Shiozaki_2017,Cano_2017,Bradlyn_Elcoro_Cano_Vergniory_Wang_Felser_Aroyo_Bernevig_2017}. 


Another intriguing topics discovered in recent years is non-Hermitian (NH) TIs and TSCs. These systems break the conventional condition that the Hamiltonian describing the system must be Hermitian. A Hermitian Hamiltonian means the energy of the system is conserved. NH Hamiltonians counts in the exchange of energy between the original system and the environment, resulting in an exotic phenomenon called non-Hermitian skin effect (NHSE)~\cite{Yao_2018,Song_2019,Okuma_2019,Kohei_classificaion_2019}. NHSE is protected by a $\mathbb{Z}$ invariant called winding number. When a system with non-trivial winding number is in contact with the vacuum, \emph{all} states will localize at the boundary. This phenomenon is unique to NH systems which received vast attention in the recent years~\cite{Xiujuan_2022,Zhang_Yang_Fang_2022,Lin_2023,Okugawa_2020,Kawabata_higer_order_2020,Liu_second_order_2019,Zhang_2020,Yang_2020,Schindler_2021,Bhargava_2021,Panigrahi_2022,Ammari_Barandun_Cao_Davies_Hiltunen_2024}. Another interesting aspect of non-Hermiticity is that every internal symmetries would split into two (See Sec.~\ref{Sec:struture of classification}): TRS splits into TRS and TRS$^\dag$; PHS splits into PHS and PHS$^\dag$; CS splits into CS and sub-lattice symmetry (SLS). A complete classification of NH Hamiltonian is made in Ref.~\cite{Kohei_classificaion_2019,Zhou_2019} shortly after the $\mathbb{Z}$ topological invariant of NHSE is discovered. The bifurcation of internal symmetries results in 38 different classes, which greatly enriched the phenomenon of TIs and TSCs. Although much attention has been put into NH systems, little progress has been made to include spatial symmetries in the 38-fold classification. In this paper, following Ref.\cite{Shiozaki_2014} which classify all order-two spatial symmetries for Hermitian Hamiltonians, we aim to incorporate order-two spatial symmetries into the classification of $16$ out of $38$ classes of NH Hamiltonians.

This paper is organized as the following: in Sec,~\ref{sec:K group hermitian}, we review the topological classification of Hermitian Hamiltonian without and with order-two spatial symmetries. In Sec.~\ref{Sec:struture of classification}, we review the classification of point-gapped NH Hamiltonians without order-two spatial symmetries. In Sec.~\ref{Sec:order-two spatial symmetry AZ and AZ dag}, we introduce the classification scheme for point-gapped Hamiltonians of $16$ out of $38$ classes with spatial symmetries. Periodic tables for spatial symmetries are introduced in Sec.~\ref{sec:periodic table}. Finally, in Sec.~\ref{Sec:examples}, we introduce several examples to illustrate our classification. We conclude the paper with several remarks in Sec.~\ref{sec:discussion}. Mathematical details are provided in Appendix. 
\section{Classification of order-two spatial symmetry for Hermitian Hamiltonians}\label{sec:K group hermitian}
To understand the process of classification of spatial symmetries in NH systems, we first review the classification of stable equivalent Hermitian Hamiltonian and spatial symmetries of Hermitian systems. 
\subsection{Review of topological defects and $K$ group classification of stable equivalent Hermitian Hamiltonians}
First, we briefly review the process of including topological defects into the classification. In the absence of defects, we consider Hamiltonians that are defined on $d$-dimensional Brillouin zone (BZ) which is a $d$-Torus $T^d$. In general, we may consider BZ as a simpler space $S^d$ which is a $d$-sphere. This substitution will not affect the strong topological invariants~\cite{Shiozaki_2014}. 

Topological defects are defined to be discontinuity or distortion in lattices that cannot be removed. This discontinuity may be a point, a line, or a surface. In real space, we let a $D$-dimensional sphere $S^D$ surrounds the defects. Combining with the $d$-dimensional BZ, the total space of classification is given by $S^{d+D}$. $K$ group classification is a common method for classifying topological invariants for topological insulators and superconductors in different dimensions~\cite{Teo_2010,Shiozaki_2014,Chiu_2013,Chiu_2016_classification}. 
The classification of stable equivalent Hamiltonian is then equivalent to the classification of maps from base space $(\textbf{k},\textbf{r})\in S^{d+D}$ of a $d+D$ dimensional sphere to the classifying space of Hamiltonian $H(\textbf{k,\textbf{r}})$. This belongs to the problem of homotopy classification. As we will see, this classification can be simplified by considering the $K$ group $K=\pi_0$, which is the zeroth homotopy group of classifying space. As we will see, the $K$ group in fact has an Abelian group structure, which allows us to classify Hamiltonians that are topologically distinct from each other. 

We consider Hamiltonians that are gapped in the bulk i.e. far away from defects. This gap is often chosen to be the Fermi level. Two Hamiltonians are considered topologically equivalent if they can be continuously deformed into each other without closing the band gap. This means the two Hamiltonians have the same topological invariants. This allows us to deform a given Hamiltonian into two flat bands without closing the band gap: all the $n$ ($m$) bands below (above) the Fermi level are deformed into the same energy $-1$ ($1$). Then, according to the symmetries that the Hamiltonian has, we can find the classifying space of this Hamiltonian. Then by computing $\pi_0$ of the classifying space, we can derive the topological invariant of the corresponding Hamiltonian in $0$ dimension. Topological invariants in higher dimension can be found by using the \emph{dimensional hierarchy}, which we will introduce later in Eqs.~\eqref{eq:K group Hermitian complex} and~\eqref{eq:K group Hermitian real}.


Throughout the paper, we consider Hamiltonians that are stable equivalent: Hamiltonian $H_1$ and $H_2$ are stable equivalent $H_1 \sim H_2$ if they can be continuously deformed into each other by adding trivial bands. This ensures that the dimension of the Hamiltonian will not affect the classification. In other words, the addition of trivial bands will not affect the topology of the Hamiltonian (Hamiltonians that violate this condition are called Fragile insulators or Hopf insulators which we will not discuss in this paper~\cite{Ahn_2019,Hwang_2019,Zhida_2020,Po_2018,Brouwer_2023,Moore_2008}).  This allows us to identify a set of Hamiltonian $[H]$ that includes all Hamiltonians that are stable equivalent to $H$. We define the addition of two sets to be $[H_1]\oplus [H_2]=[H_1\oplus H_2]$, where $\oplus$ is the direct sum. Then the inverse of $[H]$ is given by $[H]\oplus[-H]=[0]$, where $[0]$ is the set of trivial Hamiltonians. Since we are discussing stable equivalent Hamilonians, the addition of $[0]$ to any Hamiltonians will not change the set it belongs to i.e. $[H]\oplus [0]=[H]$. We see that $[0]$ acts as the identity in this operation. The direct sum $\oplus$ is associative and commutative. Therefore, different classes of Hamiltonians $[H]$ form an Abelian group, which is called $K$ group. 

Now we consider symmetries that separate these classes. The internal symmetries that govern Hermitian Hamiltonians are
\begin{align}
   \text{TRS}: \quad &\mathcal{T} H(\textbf{k},\textbf{r})\mathcal{T}^{-1}=H(-\textbf{k},\textbf{r}), \mathcal{T}^2=\pm 1\nonumber\\
  \text{PHS}: \quad  &\mathcal{C} H(\textbf{k},\textbf{r})\mathcal{C}^{-1}=-H(-\textbf{k},\textbf{r}),\mathcal{C}^2=\pm 1\nonumber\\
    \text{CS}: \quad &\Gamma H(\textbf{k},\textbf{r}) \Gamma^{-1}=-H(\textbf{k},\textbf{r}), \Gamma^2=1,
\end{align}
which are called Time-Reversal Symmetry (TRS), Particle-hole symmetry (PHS), and Chiral Symmetry (CS), respectively. Specifically, $\mathcal{T}$ and $\mathcal{C}$ are antiunitary operators, meaning they can be written in the form of a unitary matrix times complex conjugate operator $\mathcal{K}$. And $\Gamma$ is a unitary operator. The combination of these symmetries will result in $10$ classes. If no antiunitary symmetries (TRS and PHS) are present, the symmetry classes are called the complex AZ class. Since in the absence of complex conjugate operator $\mathcal{K}$, the Clifford generators that generate the classifying space of each class is defined up to a phase. So these generators are in general complex. Hence the name complex AZ class is given to classes without antiunitary symmetries. If one or both antiunitary symmetries are present, the symmetry classes are called real AZ class. In this case, generators of each class can only be real or imaginary. Hence the name real AZ class. 

Having discussed $K$ group and symmetries that govern each class, we now show the $K$ group relationship of each class. The AZ symmetry classes may summarized as~\cite{Teo_2010}
\begin{align}
    K_{\mathbb{C}}(s;d,D)&=K_{\mathbb{C}}(s-d+D;0,0)=\pi_0 (\mathcal{C}_{s-d+D})\nonumber\\
    &s=0,1 \mod{2}
    \label{eq:K group Hermitian complex}
\end{align}
for complex AZ classes (A and AIII), and
\begin{align}
    K_{\mathbb{R}}(s;d,D)&=K_{\mathbb{R}}(s-d+D;0,0)=\pi_0 (\mathcal{R}_{s-d+D})\nonumber\\
    &s=0,\cdots,7 \mod{8},
    \label{eq:K group Hermitian real}
\end{align}
for real AZ classes (the rest of the eight classes). The $\mathcal{C}_{s-d+D}$ and $\mathcal{R}_{s-d+D}$ are classifying space of the Hamiltonians. Their exact expressions can be found in Ref.~\cite{Kitaev_Lebedev_Feigel’man_2009}. The number $s$ labels the symmetry classes. Notice that complex AZ class has periodicity $2$ i.e. the classifying space obeys $\mathcal{C}_s \simeq \mathcal{C}_{s-2}$. Hence $s$ is defined mod 2 in Eq.~\eqref{eq:K group Hermitian complex}. This periodicity is the origin of two complex AZ classes. Similarly, for real AZ class, the classifying space obeys $\mathcal{R}_s\simeq \mathcal{R}_{s-8}$. Hence $s$ is defined mod 8 in Eq.~\eqref{eq:K group Hermitian real}. This periodicity gives eight real AZ classes. Periodicity of Eqs.~\eqref{eq:K group Hermitian complex} and~\eqref{eq:K group Hermitian real} are called Bott periodicity, which is an essential mathematical property of $K$ group in the classification of TIs and TSCs. The resulting table from Eqs.~\eqref{eq:K group Hermitian complex} and~\eqref{eq:K group Hermitian real} is the AZ symmetry class, which we do not repeat it here. 

\begin{figure}
    \centering
    \includegraphics[width=\linewidth]{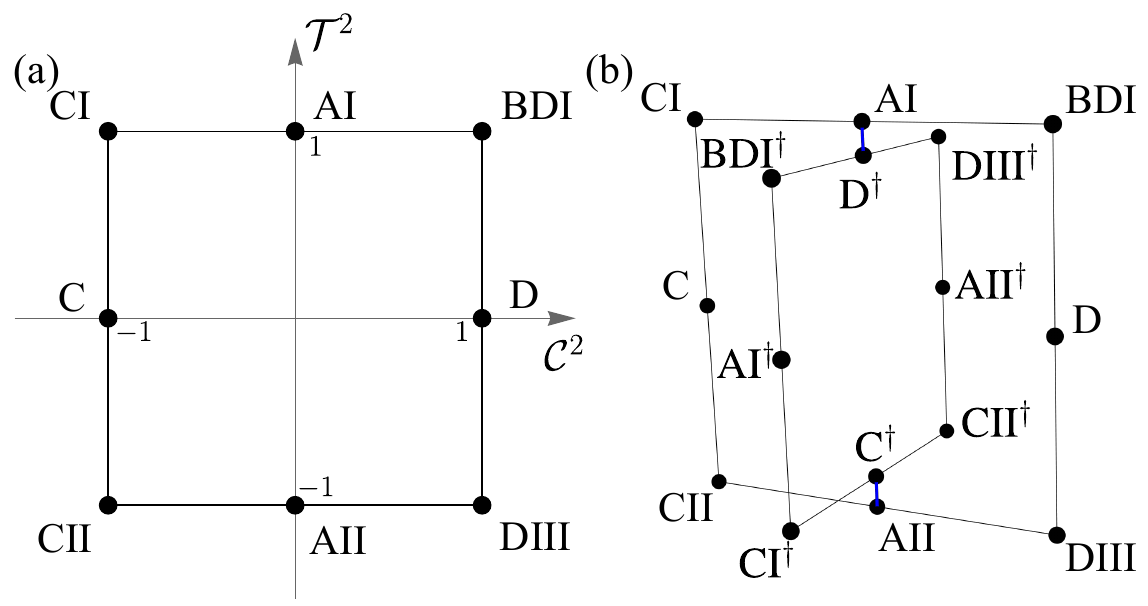}
    \caption{Visualization of dimensional hierarchy of real Hermitian AZ class and point-gapped NH AZ/AZ$^\dag$. (a) real Hermitian AZ class can be considered as a "clock" where eight classes are "hours" on the "clock". Two axes indicate the squared value of $\mathcal{T}^2$ and $\mathcal{C}^2$. (b) For NH class, we may consider two "clocks" such that one of them indicates AZ the other indicates AZ$^\dag$. Two "clocks" are connected by the mapping Eq.~\eqref{eq:PHS dag to TRS and vice versa}, indicated by two blue lines. }
    \label{fig:Dimensional Hierarchy clocks}
\end{figure}

Finally, we remark that Eqs.~\eqref{eq:K group Hermitian complex} and~\eqref{eq:K group Hermitian real} are referred to as dimensional hierarchy of AZ symmetry class, which is first proved in Ref.~\cite{Teo_2010}. A neat way to visualize Eq.~\eqref{eq:K group Hermitian real} is by putting real AZ classes on a coordinate where $x$ ($y$) axis indicates the value of $\mathcal{C}^2$ ($\mathcal{T}^2$). We plot this visualization in Fig.~\ref{fig:Dimensional Hierarchy clocks} (a). In this way, eight real AZ classes would form a "clock". In Sec.~\ref{Sec:struture of classification}, we also provide a similar relation and visualization for point-gapped Hamilonians. 
\subsection{Classification under order-two spatial symmetry}

The spatial symmetries (or crystalline symmetries) are symmetries that are dependent on the geometry of the lattice~\cite{Chiu_2013,Shiozaki_2014,Shiozaki_2016,Shiozaki_2017}. In the paper, we consider spatial symmetries that are of order two, meaning apply them twice on the Hamiltonian will keep the Hamiltonian unchanged. These spatial symmetries can be unitary ($U$) or antiunitary ($A$). Furthermore, they can commute or anti-commute with the Hamiltonians. For the later, a bar is placed on top of the spatial operator. To sum up, possible order-two spatial symmetries for Hermitian systems are 
\begin{align}
    &U H(\textbf{k},\textbf{r})U^{-1}=H(-\textbf{k}_{\parallel},
\textbf{k}_{\perp},-\textbf{r}_{\parallel},\textbf{r}_{\perp})\nonumber \\
&\overline{U} H(\textbf{k})\overline{U}^{-1}=-H(-\textbf{k}_{\parallel},
\textbf{k}_{\perp},-\textbf{r}_{\parallel},\textbf{r}_{\perp})\nonumber \\
&A H(\textbf{k})A^{-1}=H(\textbf{k}_{\parallel},
-\textbf{k}_{\perp},-\textbf{r}_{\parallel},\textbf{r}_{\perp})\nonumber \\ 
&\overline{A} H(\textbf{k})\overline{A}^{-1}=-H(\textbf{k}_{\parallel},
-\textbf{k}_{\perp},-\textbf{r}_{\parallel},\textbf{r}_{\perp}),
\label{eq:Hermitian spatial symmetry}
\end{align}
where $\textbf{k}_{\parallel}$ ($\textbf{k}_{\perp}$) denotes $\textbf{k}$ that are flipped (unchanged) under unitary symmetries. Similar notation also applies to $\textbf{r}$. Pioneer works that classify order-two spatial symmetries as an \emph{addition} to internal symmetries are Ref.~\cite{Shiozaki_2014,Chiu_2013}. These additional spatial symmetries will add a finer structure to the dimensional hierarchy Eqs.~\eqref{eq:K group Hermitian complex},~\eqref{eq:K group Hermitian real} that may add or remove certain topological invariants.

In the following sections, we will repeat this process for point-gapped NH Hamiltonians. Before that, we first introduce the 38-fold classification of NH Hamiltonians in the next sections. 

\begin{table*}[t]
	\centering
	\caption{AZ and $\text{AZ}^{\dag}$ symmetry classes for non-Hermitian Hamiltonians from Ref.~\cite{Kohei_classificaion_2019}. Notice that class D$^\dag$ and C$^\dag$ are also in AZ class, so they do not enter the $38$-fold classification classes~\cite{Kohei_classificaion_2019}. }
		\label{tab: AZ}
     \begin{tabular}{cccccccc|ccccccccc} \hline \hline
    \multicolumn{2}{c}{~Symmetry class~} & $s/s^\dag$ & ~$\mathcal{T}_+$~ & ~$\mathcal{C}_{-}$~ & ~$\mathcal{C}_+$~ & ~$\mathcal{T}_{-}$~ & ~$\Gamma$~ & C.L.& $\delta=0$ & $1$ & $2$ & $3$ & $4$ & $5$ & $6$ & $7$ \\ \hline
    \multirow{2}{*}{~Complex AZ~} 
    & A & 1 &$0$ & $0$ & $0$ & $0$ & $0$& $\mathcal{C}_1$& $0$ & $\mathbb{Z}$ & $0$ & $\mathbb{Z}$ & $0$ & 
$\mathbb{Z}$ & $0$ & $\mathbb{Z}$ \\
    & AIII & 0 & $0$ & $0$ & $0$ & $0$ & $1$ &  $\mathcal{C}_0$ & $\mathbb{Z}$ & $0$ & $\mathbb{Z}$ & $0$ & 
$\mathbb{Z}$ & $0$ & $\mathbb{Z}$ & $0$ \\ \hline
    \multirow{9}{*}{Real AZ} 
    & AI & 1 & $+1$ & $0$ & $0$ & $0$ & $0$ & $\mathcal{R}_1$ & $\mathbb{Z}_2$ & $\mathbb{Z}$ & $0$ & $0$ & $0$ & $2\mathbb{Z}$ & $0$ & $\mathbb{Z}_2$\\
    & BDI & 2 & $+1$ & $+1$ & $0$ & $0$ & $1$ & $\mathcal{R}_2$ & $\mathbb{Z}_2$ & $\mathbb{Z}_2$ & $\mathbb{Z}$ & $0$ & $0$ & $0$ & $2\mathbb{Z}$& $0$ \\
    & D & 3 & $0$ & $+1$ & $0$ & $0$ & $0$ & $\mathcal{R}_3$ & $0$ & $\mathbb{Z}_2$ & $\mathbb{Z}_2$ & $\mathbb{Z}$ & $0$ & $0$ & $0$ & $2\mathbb{Z}$ \\
    & DIII & 4 & $-1$ & $+1$ & $0$ & $0$ & $1$ & $\mathcal{R}_4$ & $2\mathbb{Z}$ & $0$ & $\mathbb{Z}_2$ & $\mathbb{Z}_2$ & $\mathbb{Z}$ & $0$ & $0$ & $0$ \\
    & AII & 5 & $-1$ & $0$ & $0$ & $0$ & $0$ & $\mathcal{R}_5$ & $0$ & $2\mathbb{Z}$ & $0$ & $\mathbb{Z}_2$ & $\mathbb{Z}_2$ & $\mathbb{Z}$ & $0$ & $0$ \\
    & CII & 6 & $-1$ & $-1$ & $0$ & $0$ & $1$ & $\mathcal{R}_6$ & $0$ & $0$ & $2\mathbb{Z}$ & $0$ & $\mathbb{Z}_2$ & $\mathbb{Z}_2$ & $\mathbb{Z}$ & $0$ \\
    & C & 7 & $0$ & $-1$ & $0$ & $0$ & $0$ & $\mathcal{R}_7$ & $0$ & $0$ & $0$ & $2\mathbb{Z}$ & $0$ & $\mathbb{Z}_2$ & $\mathbb{Z}_2$ & $\mathbb{Z}$ \\
    & CI & 0 & $+1$ & $-1$ & $0$ & $0$ & $1$ & $\mathcal{R}_0$ & $\mathbb{Z}$ & $0$ & $0$ & $0$ & $2\mathbb{Z}$ & $0$ & $\mathbb{Z}_2$ & $\mathbb{Z}_2$\\ \hline
    \multirow{9}{*}{Real $\text{AZ}^{\dag}$} 
    & $\text{AI}^{\dag}$ & 7 & $0$ & $0$ & $+1$ & $0$ & $0$ & $\mathcal{R}_7$ & $0$ & $0$ & $0$ & $2\mathbb{Z}$ & $0$ & $\mathbb{Z}_2$ & $\mathbb{Z}_2$ & $\mathbb{Z}$\\
    & $\text{BDI}^{\dag}$ & 0 & $0$ & $0$ & $+1$ & $+1$ & $1$ & $\mathcal{R}_0$ & $\mathbb{Z}$ & $0$ & $0$ & $0$ & $2\mathbb{Z}$ & $0$ & $\mathbb{Z}_2$ & $\mathbb{Z}_2$\\
    & $\text{D}^{\dag}$ & 1 & $0$ & $0$ & $0$ & $+1$ & $0$ & $\mathcal{R}_1$ & $\mathbb{Z}_2$ & $\mathbb{Z}$ & $0$ & $0$ & $0$ & $2\mathbb{Z}$ & $0$ & $\mathbb{Z}_2$\\
    & $\text{DIII}^{\dag}$ & 2 & $0$ & $0$ & $-1$ & $+1$ & $1$ & $\mathcal{R}_2$ & $\mathbb{Z}_2$ & $\mathbb{Z}_2$ & $\mathbb{Z}$ & $0$ & $0$ & $0$ & $2\mathbb{Z}$& $0$\\
    & $\text{AII}^{\dag}$ & 3 & $0$ & $0$ & $-1$ & $0$ & $0$ & $\mathcal{R}_3$ & $0$ & $\mathbb{Z}_2$ & $\mathbb{Z}_2$ & $\mathbb{Z}$ & $0$ & $0$ & $0$ & $2\mathbb{Z}$ \\
    & $\text{CII}^{\dag}$ & 4 & $0$ & $0$ & $-1$ & $-1$ & $1$ & $\mathcal{R}_4$ & $2\mathbb{Z}$ & $0$ & $\mathbb{Z}_2$ & $\mathbb{Z}_2$ & $\mathbb{Z}$ & $0$ & $0$ & $0$\\
    & $\text{C}^{\dag}$ & 5 & $0$ & $0$ & $0$ & $-1$ & $0$ & $\mathcal{R}_5$ & $0$ & $2\mathbb{Z}$ & $0$ & $\mathbb{Z}_2$ & $\mathbb{Z}_2$ & $\mathbb{Z}$ & $0$ & $0$ \\
    & $\text{CI}^{\dag}$ & 6 & $0$ & $0$ & $+1$ & $-1$ & $1$ & $\mathcal{R}_6$ & $0$ & $0$ & $2\mathbb{Z}$ & $0$ & $\mathbb{Z}_2$ & $\mathbb{Z}_2$ & $\mathbb{Z}$ & $0$\\ \hline \hline
    \end{tabular}
\end{table*}

\section{Structure of $38$-fold classification of NH Hamiltonian}\label{Sec:struture of classification}
We consider internal symmetries of NH Hamiltonians. Due to non-Hermiticity, $H$ will bifurcate into $H$ and $H^\dag$. Following notations in Ref.~\cite{Kohei_classificaion_2019}, internal symmetries for NH Hamiltonians bifurcate into
\begin{align}
    \text{TRS}:\quad &\mathcal{T}_+ H(\textbf{k},\textbf{r})\mathcal{T}_+^{-1}=H(-\textbf{k},\textbf{r})\nonumber\\
    \text{PHS}:\quad &\mathcal{C}_{-} H(\textbf{k},\textbf{r})  \mathcal{C}_{-}^{-1}=-H^\dag(-\textbf{k},\textbf{r})\nonumber\\
    \text{TRS}^\dag :\quad &\mathcal{C}_+ H(\textbf{k},\textbf{r})\mathcal{C}_+^{-1}=H^\dag(-\textbf{k},\textbf{r}) \nonumber\\
    \text{PHS}^\dag: \quad  &\mathcal{T}_{-}H(\textbf{k},\textbf{r})\mathcal{T}_{-}=-H(-\textbf{k},\textbf{r})\nonumber\\
    \text{CS}:\quad&\Gamma H(\textbf{k},\textbf{r})\Gamma^{-1}=-H(\textbf{k},\textbf{r})^\dag \nonumber\\
    \text{SLS}:\quad &\mathcal{S} H(\textbf{k},\textbf{r})\mathcal{S}^{-1}=-H(\textbf{k},\textbf{r}),
    \label{eq:nH internal symmetries}
\end{align}
where SLS denotes sub-lattice symmetry. The combination of these symmetries will result in $38$ independent classes. In this paper, we consider classes without SLS. As we will see, combinations of other symmetries will results in two sets of classes that are similar to Hermitian AZ symmetry classes. 

We now comment on the structure of classes without SLS. The combination of TRS, PHS, and CS will result in $10$ symmetry classes that are similar to AZ symmetry class for Hermitian Hamiltonians (See Table~\ref{tab: AZ}). These $10$ classes are called as AZ class for NH Hamiltonians. From now on, by AZ class, we mean AZ class for NH Hamiltonians from Table~\ref{tab: AZ}. Next, consider the combination of TRS$^\dag$ and PHS$^\dag$. Symmetry classes formed by these symmetries are called AZ$^\dag$ classes. They constitute $8$ classes in total.
Furthermore, if \emph{only} PHS$^\dag$ is present (class D$^\dag$ and C$^\dag$ in Table~\ref{tab: AZ}), PHS$^\dag$ can be transformed into to TRS (class AI and AII) and vice versa~\cite{Kawabata_Higashikawa_Gong_Ashida_Ueda_2019}. To see this, consider a Hamiltonian $H(\textbf{k},\textbf{r})$ that obeys PHS$^\dag$ ($\mathcal{T}_-$). Then $\textrm{i}H(\textbf{k},\textbf{r})$ obeys TRS with operator $\mathcal{T}_-$:
\begin{equation}
    \mathcal{T}_- \textrm{i}H(\textbf{k},\textbf{r})\mathcal{T}_-^{-1}=\textrm{i}H(-\textbf{k},\textbf{r}).
    \label{eq:PHS dag to TRS and vice versa}
\end{equation}
The mapping $H(\textbf{k},\textbf{r})\rightarrow \textrm{i}H(\textbf{k},\textbf{r})$ can be understood as a 90 degree rotation of energy spectrum on the complex plane. This operation does not change the topology of Hamiltonian $H$. 
Therefore, we see that class D$^\dag$ (C$^\dag$) is unified with AI (AII). As we will see in Appendix.~\ref{Sec:dimensional Hierarchy AZ and AZ dag}, this unification will have some consequences on the dimensional hierarchy of AZ and AZ$^\dag$ classes. Therefore, there are $6$ independent symmetry classes in AZ$^\dag$. Similar to Hermitian classification, class A and AIII are complex classes, meaning they can be represented by complex Clifford algebra, while the rest of the classes are real (due to the existence of complex conjugate operator $\mathcal{K}$ in $\mathcal{T}_\pm$ and $\mathcal{C}_\pm$), meaning they can be represented by real Clifford algebra. In Table~\ref{tab: AZ}, we label AZ (AZ$^\dag$) class by $s$ ($s^\dag$). 
 
The addition of SLS to AZ classes will result in additional $22$ symmetry classes. So there are $10+6+22=38$ symmetry classes in total. The $22$-fold SLS classes can be further divided into five sub-classes (See Ref.~\cite{Kohei_classificaion_2019} for details). However, since the SLS classes are rarely explored in current literature, we do not consider them in this paper. 


\begin{figure}
    \centering
    \includegraphics[width=\linewidth]{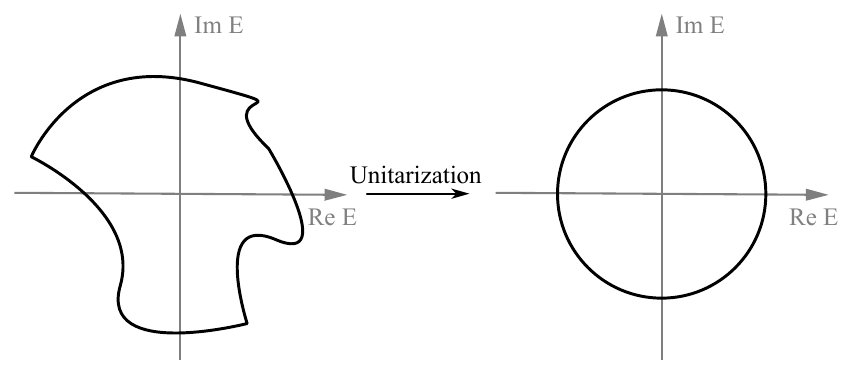}
    \caption{Effect of unitarization of a point-gapped Hamiltonian on energy spectrum. }
    \label{fig:Unitarization}
\end{figure}
As we are interested in point gap in this paper, we also briefly review the classification schemes for point gapped Hamiltonians. For a given point-gapped Hamiltonian $H(\textbf{k},\textbf{r})$, we can continuously deform it into a unitary matrix $\mathcal{U}(\textbf{k},\textbf{r})$ without closing the point gap. This process is called unitary flattening~\cite{Kohei_classificaion_2019,Roy_Floquet_2017}, which we show in Fig.~\ref{fig:Unitarization}. We may then consider the extended Hermitian Hamiltonian
\begin{equation}
    \Tilde{H}(\textbf{k},\textbf{r})=\begin{pmatrix}
        0 & \mathcal{U}(\textbf{k},\textbf{r})\\
         \mathcal{U}^\dag(\textbf{k},\textbf{r}) & 0
    \end{pmatrix}.
    \label{eq:extended Hermitian Hamiltonian}
\end{equation}
If original Hamiltonian $H(\textbf{k},\textbf{r})$ obeys symmetries in Eq.~\eqref{eq:nH internal symmetries}, the extended Hamiltonian $\Tilde{H}(\textbf{k},\textbf{r})$ then obeys
\begin{align}
    &\Tilde{\mathcal{T}}_\pm \Tilde{H}(\textbf{k},\textbf{r}) \Tilde{\mathcal{T}}^{-1}_\pm=\pm \Tilde{H}(-\textbf{k},\textbf{r}), \Tilde{\mathcal{T}}_\pm=\begin{pmatrix}
        \mathcal{T}_\pm & 0\\
        0 & \mathcal{T}_\pm
    \end{pmatrix}\nonumber\\
    &\Tilde{\mathcal{C}}_\pm \Tilde{H}(\textbf{k},\textbf{r}) \Tilde{\mathcal{C}}^{-1}_\pm=\pm \Tilde{H}(-\textbf{k},\textbf{r}), \Tilde{\mathcal{C}}_\pm=\begin{pmatrix}
        0 & \mathcal{C}_\pm\\
        \mathcal{C}_\pm & 0
    \end{pmatrix}\nonumber\\
    &\Tilde{\Gamma}  \Tilde{H}(\textbf{k},\textbf{r}) \Tilde{\Gamma}^{-1}=-\Tilde{H}(\textbf{k},\textbf{r}),\Tilde{\Gamma}=
    \begin{pmatrix}
        0 & \Gamma\\
        \Gamma & 0
    \end{pmatrix}\nonumber\\
    &\Tilde{\mathcal{S}}\Tilde{H}(\textbf{k},\textbf{r})\Tilde{\mathcal{S}}^{-1}=-\Tilde{H}(\textbf{k},\textbf{r}),\Tilde{\mathcal{S}}=
    \begin{pmatrix}
       \mathcal{S}& 0 \\
       0 & \mathcal{S}
    \end{pmatrix}.
    \label{eq:extended internal symmetries}
\end{align}
Furthermore, Eq.~\eqref{eq:extended Hermitian Hamiltonian} also obeys an additional CS
\begin{equation}
    \Sigma \Tilde{H}(\textbf{k},\textbf{r})\Sigma^{-1}=-\Tilde{H}(\textbf{k},\textbf{r}), \Sigma=
    \begin{pmatrix}
        1 &0\\
        0&-1
    \end{pmatrix}.
    \label{eq:additional CS Sigma}
\end{equation}
For clearance, Eq.~\eqref{eq:additional CS Sigma} will be called $\Sigma$ symmetry in the rest of the paper. Notably, $\Sigma$ symmetry arise since we are using extended Hamiltonian for the classification. In fact, the existence of $\Sigma$ is the biggest difference between classifying a Hermitian Hamiltonian and a point-gapped NH Hamiltonian. Now, the classification of point-gapped Hamiltonian $H(\textbf{k},\textbf{r})$ is mapped to the classification of extended Hermitian Hamiltonian~\eqref{eq:extended Hermitian Hamiltonian}
 with symmetries~\eqref{eq:extended internal symmetries} and~\eqref{eq:additional CS Sigma}. This allows us to use the classification process on $\Tilde{H}$ for Hermitian Hamiltonians discussed in Sec.~\ref{sec:K group hermitian}. 
 The $K$ group relationships are given by
 \begin{align}
    &K_\mathbb{C}(s;d,D)=K_\mathbb{C}(s-d+D;0,0)=\pi_0 (\mathcal{C}_{s-d+D}), \nonumber\\
    &s=0,1 \mod{2}
    \label{eq:Hierarchy complex AZ miantext}
\end{align}
for complex AZ classes and
\begin{align}
    &K_{\mathbb{R}}(s;d,D)=K_{\mathbb{R}}(s-d+D;0,0)=\pi_0 (\mathcal{R}_{s-d+D}),\nonumber\\
    &s=0,\cdots,7 \mod{8}
    \label{eq:K_R AZ class}
\end{align}
for real AZ classes and finally
\begin{align}
    &K_{\mathbb{R}}^\dag(s^\dag;d,D)=K_{\mathbb{R}}^\dag(s^\dag-d+D;0,0)=\pi_0 (\mathcal{R}_{s^\dag-d+D}),\nonumber\\
    &s^\dag=0,\cdots,7 \mod{8}
    \label{eq:Hierarchy AZ dag miantext}
\end{align}
for AZ$^\dag$ classes; The superscript $\dag$ is put to distinguish $K$ group of AZ$^\dag$ class from that of AZ class. 
The unification of AI and D$^\dag$, AII and C$^\dag$ also gives
\begin{align}
    K_{\mathbb{R}}(1;d,D)=K_{\mathbb{R}}^\dag(1;d,D)\nonumber\\
    K_{\mathbb{R}}(5;d,D)=K_{\mathbb{R}}^\dag(5;d,D).
\end{align}
We proof these relationships rigorously in Appendix~\ref{Sec:dimensional Hierarchy AZ and AZ dag}.

The dimensional hierarchy of AZ and AZ$^\dag$ class for point-gapped Hamiltonian can be understood as two "clocks":  one for AZ and the other for AZ$^\dag$. Two clocks are connected by the mapping Eq.~\eqref{eq:PHS dag to TRS and vice versa}. We draw these two clocks in Fig.~\ref{fig:Dimensional Hierarchy clocks}. 
 
In order to include spatial symmetries in the classification, we also need to consider the extension of spatial symmetries. In the following, we focus on order-two spatial symmetries as an additional symmetry to $16$-fold AZ and AZ$^\dag$ classes in Table~\ref{tab: AZ}. 

\section{Order-two spatial symmetries in AZ and AZ$^\dag$}\label{Sec:order-two spatial symmetry AZ and AZ dag}
Similar to the bifurcation of internal symmetries, each type of spatial symmetry Eqs.~\eqref{eq:Hermitian spatial symmetry} will also bifurcate into two. We label them according to  
\begin{align}
    &\prescript{c}{}{U} H(\textbf{k},\textbf{r})\prescript{c}{}{U^{-1}}=H(-\textbf{k}_{\parallel},
\textbf{k}_{\perp},-\textbf{r}_{\parallel},\textbf{r}_{\perp})\nonumber \\
&\prescript{g}{}{U} H(\textbf{k},\textbf{r})\prescript{g}{}{U^{-1}}=H^\dag(-\textbf{k}_{\parallel},
\textbf{k}_{\perp},-\textbf{r}_{\parallel},\textbf{r}_{\perp})\nonumber \\
&\prescript{c}{}{\overline{U}} H(\textbf{k},\textbf{r})\prescript{c}{}{\overline{U}^{-1}}=-H(-\textbf{k}_{\parallel},
\textbf{k}_{\perp},-\textbf{r}_{\parallel},\textbf{r}_{\perp})\nonumber \\
&\prescript{g}{}{\overline{U}} H(\textbf{k},\textbf{r})\prescript{g}{}{\overline{U}^{-1}}=-H^\dag(-\textbf{k}_{\parallel},
\textbf{k}_{\perp},-\textbf{r}_{\parallel},\textbf{r}_{\perp})\nonumber \\
&\prescript{c}{}{A} H(\textbf{k},\textbf{r})\prescript{c}{}{A^{-1}}=H(\textbf{k}_{\parallel},
-\textbf{k}_{\perp},-\textbf{r}_{\parallel},\textbf{r}_{\perp})\nonumber \\ 
&\prescript{g}{}{A} H(\textbf{k},\textbf{r})\prescript{g}{}{A^{-1}}=H^\dag(\textbf{k}_{\parallel},
-\textbf{k}_{\perp},-\textbf{r}_{\parallel},\textbf{r}_{\perp})\nonumber \\ 
&\prescript{c}{}{\overline{A}} H(\textbf{k},\textbf{r})\prescript{c}{}{\overline{A}^{-1}}=-H(\textbf{k}_{\parallel},
-\textbf{k}_{\perp},-\textbf{r}_{\parallel},\textbf{r}_{\perp})\nonumber\\
&\prescript{g}{}{\overline{A}} H(\textbf{k},\textbf{r})\prescript{g}{}{\overline{A}^{-1}}=-H^\dag(\textbf{k}_{\parallel},
-\textbf{k}_{\perp},-\textbf{r}_{\parallel},\textbf{r}_{\perp}),
\label{eq:nH spatial symmetry}
\end{align}
where $c$ and $g$ refer to conventional and generalized spatial symmetries, respectively; $\textbf{k}_{\parallel}$ ($\textbf{k}_{\perp}$) denotes $\textbf{k}$ that are flipped (unchanged) under unitary symmetries. Similar notation also applies to $\textbf{r}$. Similar to the extension of internal symmetries for point-gapped Hamiltonians in Eq.~\eqref{eq:extended internal symmetries}, to include spatial symmetries in the classification, we also consider the extension of Eq.~\eqref{eq:nH spatial symmetry}. For the conventional spatial symmetries (those with $c$ at the front), their extensions are given by
\begin{equation}
    \prescript{c}{}{\Tilde{L}}=\begin{pmatrix}
        \prescript{c}{}{L} & 0\\
        0 & \prescript{c}{}{L}
    \end{pmatrix},
    \label{eq:conventional spatial extension}
\end{equation}
where $L=\prescript{c}{}{U},\prescript{c}{}{\overline{U}},\prescript{c}{}{A}$, or $\prescript{c}{}{\overline{A}}$. For generalized spatial symmetries (those with $g$ at the front), their extensions are given by
\begin{equation}
    \prescript{g}{}{\Tilde{L}}=\begin{pmatrix}
        0 & \prescript{g}{}{L}\\
        \prescript{g}{}{L} & 0
    \end{pmatrix},
    \label{eq:generalized spatial extension}
\end{equation}
where $L=\prescript{g}{}{U},\prescript{g}{}{\overline{U}},\prescript{g}{}{A}$, or $\prescript{g}{}{\overline{A}}$. These extended spatial symmetries would have the same effect on $\Tilde{H}$ in Eq.~\eqref{eq:extended Hermitian Hamiltonian} as original spatial symmetries on original $H$. Therefore, we again mapped the classification problem of point gapped Hamiltonian to that of extended Hamiltonian~\eqref{eq:extended Hermitian Hamiltonian}. 

Before continuing, we would like to comment on an important effect of the $\Sigma$ symmetry [Eq.~\eqref{eq:additional CS Sigma}] that would "unify" certain spatial symmetries. As an example, consider spatial symmetry $\prescript{c}{}{U}$ and its extension $\prescript{c}{}{\Tilde{U}}$, which operate on extended Hermitian Hamiltonian
\begin{equation}
    \prescript{c}{}{\Tilde{U}} \Tilde{H}(\textbf{k},\textbf{r})  \prescript{c}{}{\Tilde{U}}^{-1}=\Tilde{H}(-\textbf{k}_{\parallel},\textbf{k}_{\perp},-\textbf{r}_{\parallel},\textbf{r}_{\perp}). 
\end{equation}
Combining with $\Sigma$ symmetry, which is always present for point gap Hamiltonians, $ \prescript{c}{}{\Tilde{U}}$ will result in the following symmetry:
\begin{equation}
    \Sigma \prescript{c}{}{\Tilde{U}} \Tilde{H}(\textbf{k},\textbf{r})  (\Sigma\prescript{c}{}{\Tilde{U}})^{-1}=-\Tilde{H}(-\textbf{k}_{\parallel},\textbf{k}_{\perp},-\textbf{r}_{\parallel},\textbf{r}_{\perp}).
\end{equation}
Therefore, we see that $\Sigma  \prescript{c}{}{\Tilde{U}}$ has the exact same expression as $\prescript{c}{}{\Tilde{\overline{U}}}$ on Hamiltonian. Hence $\Sigma$ symmetry would map a spatial symmetry that commutes with extended Hamiltonian to its anti-commuting counterpart and vice versa. In this sense, $\Sigma$ symmetry "unifies" commuting and anti-commuting symmetries. However, this unification is not physical. We can understand this fact in two ways: First, $\Sigma$ symmetry is only present for extended Hamiltonians instead of the original Hamiltonians. Therefore, this unification is only a by-product of adopting extended Hamiltonian formalism.  Second, if we write $\Sigma  \prescript{c}{}{\Tilde{U}}$ explicitly, we would find
\begin{equation}
    \Sigma \prescript{c}{}{\Tilde{U}}=\begin{pmatrix}
        \prescript{c}{}{U}&0\\
        0& -\prescript{c}{}{U}
    \end{pmatrix}
\end{equation}
which does not follow the form of Eq.~\eqref{eq:conventional spatial extension}. Hence, even if commuting symmetries will always have the same classification as anti-commuting symmetries, we should bear in mind that such unification is only a mathematical coincidence, which \emph{does not} implies that both symmetries exist in the original system. However, since the classification problem is mapped to $\Tilde{H}$, $\Sigma$ symmetry \emph{will} cause the same classification for $\prescript{c}{}{U}$ and $\prescript{c}{}{\overline{U}}$, as they have the same effect on $\Tilde{H}$ up to a multiplication of $\Sigma$. In the following sections, we will discuss the effects of $\Sigma$ symmetry in more detail.

In the present of internal symmetries Eqs.~\ref{eq:nH internal symmetries}, not all these spatial symmetries are independent. Following the process of Ref.~\cite{Shiozaki_2014}, we first find equivalent spatial symmetries under the presenting internal symmetries for a certain symmetry class, then we give a finer classification of the corresponding symmetry classes due to spatial symmetries.

\subsection{Complex classes A and AIII with order-two unitary symmetry}
We first consider classes A and AIII in the present of order-two unitary symmetry ($\prescript{c/g}{}{U}$ and $\prescript{c/g}{}{\overline{U}}$ in Eqs.~\eqref{eq:nH spatial symmetry}).  Due to the absent of antiunitary symmetries, symmetries $\prescript{c/g}{}{U}$ and $\prescript{c/g}{}{\overline{U}}$ are defined up to a phase that does not affect the classification. Therefore, we may set $(\prescript{c/g}{}{U})^2=(\prescript{c/g}{}{\overline{U}})^2=1$. 

For class A in which no internal symmetries are present, all $\prescript{c/g}{}{U}$ and $\prescript{c/g}{}{\overline{U}}$ are independent of each other. 

Now consider class AIII where only CS in Eq.~\eqref{eq:nH internal symmetries} is present. The CS operator $\Gamma$ might commute or anti-commute with spatial symmetry operators. For example, 
\begin{equation}
    \Gamma \prescript{g}{}{U}=\eta_{\Gamma} \prescript{g}{}{U} \Gamma,
\end{equation}
where $\eta_{\Gamma}\in \{0,1\}$. To indicate this commutation relation between operators, we put a subscript $\eta_\Gamma$ to spatial operator such that  $\Gamma \prescript{g}{}{U_{\eta_\Gamma}}=\eta_{\Gamma} \prescript{g}{}{U_{\eta_\Gamma}} \Gamma$. Next, we investigate the combination of CS operator and spatial symmetry operators. Based on the effect of CS in Eq.~\eqref{eq:nH internal symmetries}, CS would flip spatial symmetries $g/c$ to $c/g$ and add another minus sign. Therefore, we see that $\Gamma \prescript{c/g}{}{U}=\prescript{g/c}{}{\overline{U}}$ and $\Gamma \prescript{c/g}{}{\overline{U}}=\prescript{g/c}{}{U}$. We summarize all possible order-two unitary symmetries for A and AIII in Table~\ref{Tab:Unitary+Complex Classes A and AIII}.

Notice that all the conclusions up until now are applicable to all kinds of energy gaps. Following the discussion at the beginning of Sec.~\ref{Sec:order-two spatial symmetry AZ and AZ dag}, by considering the effect of $\Sigma$ symmetry, we limit our scope to point gap. As an example, consider $\prescript{g}{}{U}$ in class A. Then $\Sigma \prescript{g}{}{\Tilde{U}}$ will have the same effect on extended Hamiltonian as $\prescript{g}{}{\Tilde{\overline{U}}}$. As another example, consider $\prescript{g}{}{U}_+$ in class AIII. Notice that $\Sigma$ anti-commute with extended CS operator $\Tilde{\Gamma}$. Therefore, $\Sigma \prescript{g}{}{\Tilde{U}}_+$ will have the same effect on the extended Hamiltonian as $\prescript{g}{}{\Tilde{\overline{U}}}_-$. By going over all possible order-two unitary symmetries for A and AIII in similar process, we can find all symmetries that are independent or related to each other through $\Sigma$ symmetry. We label those symmetries that are independent by $t=0$ and $t=1$, where $t$ is defined mod$2$. For simplicity, we abbreviate them into $t_0$ and $t_1$ in Table~\ref{Tab:Unitary+Complex Classes A and AIII}. We emphasize again that such connection is a purely mathematical coincidence which does not imply both symmetries exist in the system. 

In Appendix~\ref{Sec:DH AZ and AZ dag with spatial}, we show the following dimensional hierarchy of complex AZ classes with unitary spatial symmetries
\begin{align}
    &K_{\mathbb{C}}^U(s,t;d+D,d_{\parallel}+D_{\parallel},0,0)\nonumber\\
    &=K_{\mathbb{C}}^U(s-d+D,t-d_{\parallel}+D_{\parallel};0,0,0,0),
    \label{eq:K group complex AZ unitary}
\end{align}
where the superscript $U$ indicates this relationship works for unitary spatial symmetries. Recall that $d$ is the dimension of the system and $D$ is the dimension of the sphere surrounds the defect. Here, $d_{\parallel}$($D_{\parallel}$) is the dimension of the system(sphere) that is being flipped under the spatial symmetry.
Furthermore,
\begin{align}
    &K_{\mathbb{C}}^U(s,t=0;0,0,0,0)=\pi_0 (\mathcal{C}_s)\times \pi_0 (\mathcal{C}_s)\nonumber\\
    &K_{\mathbb{C}}^U(s,t=1;0,0,0,0)=\pi_0 (\mathcal{C}_{s-1}),
    \label{eq:homotopy complex AZ unitary}
\end{align}
which we show in Appendix~\ref{Appendix:classifying space} using Clifford Algebra. We record some properties of Clifford Algebra in Appendix~\ref{Appendix:Properties of Clifford Algebra}. 
For both Eqs, $s$ and $t$ are defined mod$2$. 
Eqs.~\ref{eq:K group complex AZ unitary} and~\ref{eq:homotopy complex AZ unitary} allow us to conclude periodic table for complex AZ class with unitary spatial symmetries, which we will introduce in Sec.~\ref{sec:periodic table}.  

\begin{table}[]
\caption{Possible types of order-two unitary symmetry for complex classes A and AIII. The symmetries in brackets indicate that they are related by $\Gamma$ operators. The subscript $\eta_\Gamma$ indicates commutation relation $\Gamma U_{\eta_\Gamma}=\eta_{\Gamma}U_{\eta_\Gamma}\Gamma$. Symmetries that are related by $\Sigma$ symmetry Eq.~\eqref{eq:additional CS Sigma} are labeled by identical $t_n,n\in\{0,1\}$ at the end.}
\begin{tabular}{ccc}\hline \hline
Class &  &  \\ \hline
A     & $\prescript{g}{}{U} t_1$ & $\prescript{g}{}{\overline{U}} t_1$ \\
A     & $\prescript{c}{}{U} t_0$ & $\prescript{c}{}{\overline{U}} t_0$ \\
\hline
AIII  & $(\prescript{g}{}{U}_+,\prescript{c}{}{\overline{U}_{+}})t_1$ & $(\prescript{g}{}{U_-},\prescript{c}{}{\overline{U}_{-}})t_0$ \\
AIII  & $(\prescript{c}{}{U_+},\prescript{g}{}{\overline{U}_{+}})t_0$ &  $(\prescript{c}{}{U_-},\prescript{g}{}{\overline{U}_{-}})t_1$  \\ \hline\hline
\end{tabular}
\label{Tab:Unitary+Complex Classes A and AIII}
\end{table}


\subsection{Complex classes A and AIII with order-two antiunitary symmetry}\label{sec:A and AIII+antyiunitary}
Now we consider the presence of antiunitary spatial symmetries [$\prescript{c/g}{}{A}$ and $\prescript{c/g}{}{\overline{A}}$ in Eqs.~\eqref{eq:nH spatial symmetry}].

For both $g$ and $c$ types of antiunitary symmetry, we use superscript $\epsilon_A$ to denote their squared value i.e. $(A^{\epsilon_A})^2=\epsilon_A$, $(\overline{A}^{\epsilon_A})^2=\epsilon_A$; The subscript $\eta_\Gamma$ (if present) denote their commutation relationship between CS operator $\Gamma$ i.e.  $A^{\epsilon_A}_{\eta_\Gamma} \Gamma=\eta_\Gamma \Gamma A^{\epsilon_A}_{\eta_\Gamma}$. Furthermore, in the present of CS, CS would flip $g/c$ to $c/g$ and add another minus sign. Therefore, the following relationship hold $\prescript{g/c}{}{A}^{\epsilon_A}_{\eta_\Gamma}=\Gamma \prescript{c/g}{}{\overline{A}}^{\epsilon_A \eta_\Gamma}_{\eta_\Gamma}$. 

Similar to the antiunitary spatial symmetry with Hermitian Hamiltonians (See Ref.~\cite{Shiozaki_2014}), the presence of antiunitary spatial symmetries will map complex classes A and AIII into one of real AZ and AZ$^\dag$ classes. To see this, one may consider antiunitary symmetries as effective TRS$^{(\dag)}$ or PHS$^{(\dag)}$ symmetries.

As an example, consider the addition of $\prescript{c}{}{A}^+$ to class A (First row of Table~\ref{Tab:antiunitary+Complex Classes A and AIII}), which add the symmetry
\begin{equation}
    \prescript{c}{}{A}^+ H(\textbf{k},\textbf{r}) [\prescript{c}{}{A}^+]^{-1}=H(\textbf{k}_{\parallel},
-\textbf{k}_{\perp},-\textbf{r}_{\parallel},\textbf{r}_{\perp}),(\prescript{c}{}{A}^+)^2=1 
\label{eq:conventional A class A}
\end{equation}
to class A. If we treat $(\textbf{k}_{\perp},\textbf{r}_{\parallel})$ as "momentum" and $(\textbf{k}_{\parallel},\textbf{r}_{\perp})$ as "positions", we see that Eq.~\eqref{eq:conventional A class A} has exactly the same form as TRS $\mathcal{T}_+$ with $\mathcal{T}_+^2=1$. Now, the dimension of the BZ is given by the combined dimension of $\textbf{k}_{\perp}$ and $\textbf{r}_{\parallel}$, which is $d-d_{\parallel}+D_{\parallel}$. The dimension of the sphere that surrounds the defect is given by the combined dimension of $\textbf{k}_{\parallel}$ and $\textbf{r}_{\perp}$, which is $D-D_{\parallel}+d_{\parallel}$. Therefore, with the addition of $\prescript{c}{}{A}^+$, class A is mapped to class AI.

As another example, consider the addition of $\prescript{g}{}{\overline{A}}^+_+$ to class AIII (Second row of Table~\ref{Tab:antiunitary+Complex Classes A and AIII}). By considering $(\textbf{k}_{\perp},\textbf{r}_{\parallel})$ as "momentum" , the antiunitary symmetry $\prescript{g}{}{\overline{A}}^+_+$ has the same form as PHS with $(\prescript{g}{}{\overline{A}}^+_+)^2=1$. Next, consider $\Gamma \prescript{g}{}{\overline{A}}^+_+$.  The CS symmetry $\Gamma$ takes Hermitian conjugate of the Hamiltonian again and give another minus sign i.e. $\Gamma  \prescript{g}{}{\overline{A}}^+_+  H(\textbf{k},\textbf{r}) [\Gamma  \prescript{g}{}{\overline{A}}^+_+]^{-1}=H(\textbf{k}_{\parallel},
-\textbf{k}_{\perp},-\textbf{r}_{\parallel},\textbf{r}_{\perp})$. Finally, since $\Gamma$ and $\prescript{g}{}{\overline{A}}^+_+$ commute, we have $(\Gamma \prescript{g}{}{\overline{A}}^+_+)^2=1$. This suggest we can define $\prescript{c}{}{A}^+_+=\Gamma \prescript{g}{}{\overline{A}}^+_+$ which has the same form as TRS. We see that the addition of $\prescript{g}{}{\overline{A}}^+_+$ will map AIII to BDI. 

Following similar process, we enumerate all possible types of order-two antiunitary symmetries in class A and AIII and their mapped classes in Table~\ref{Tab:antiunitary+Complex Classes A and AIII}.

Finally, we comment on the effect of $\Sigma$ symmetry. As an example, we consider again $\prescript{c}{}{A}^+$ in class AI. Due to the $\Sigma$ symmetry, the operator $\Sigma \prescript{c}{}{A}^+$ has the same effect on extended Hamiltonian as $\prescript{c}{}{\overline{A}}^+$. According to Table~\ref{Tab:antiunitary+Complex Classes A and AIII}, $\Sigma$ symmetry would unify class AI and D$^\dag$ i.e. these two classes would have the same topological invariants and classification. Indeed, in the 38-fold classification, AI and D$^\dag$ have the same classification under point gap. We mark symmetry classes that are equivalent under $\Sigma$ symmetry by identical symbols at the end of the symmetry operator in Table~\ref{Tab:antiunitary+Complex Classes A and AIII}. Notice that our results are consistent with 38-fold classification in Ref.~\cite{Kohei_classificaion_2019} i.e. the mapped classes marked with same symbol in Table~\ref{Tab:antiunitary+Complex Classes A and AIII} have the same topological classification under point gap. 

The dimensional hierarchy of complex AZ class with antiunitary spatial symmetry is thus given by
\begin{align}
    K_{\mathbb{C}}^A(s;d,d_{\parallel},D,D_{\parallel})&=K_{\mathbb{R}}(s;d-d_{\parallel}+D_{\parallel},D+d_{\parallel}-D_{\parallel})\nonumber\\
    &=K_{\mathbb{R}}(s-d+D+2(d_{\parallel}-D_{\parallel});0,0)\nonumber\\
    &=\pi_0(\mathcal{R}_{s-d+D+2(d_{\parallel}-D_{\parallel})})
    \label{eq:dimensional hierarchy complex AZ antiunitary}
\end{align}
where $K_{\mathbb{R}}$ is given in Eq.~\eqref{eq:K_R AZ class}. 

\begin{table}[]
\caption{Possible types of order-two antiunitary symmetry for complex classes A and AIII. The superscript $A^{\epsilon_A}$ denote the sign of  $(A^{\epsilon_A})^2=\epsilon_A$, and subscript $A^{\epsilon_A}_{\eta_\Gamma}$ (if present) denotes commutation relationship between $A$ and CS $\Gamma$: $A^{\epsilon_A}_{\eta_\Gamma} \Gamma=\eta_\Gamma \Gamma A^{\epsilon_A}_{\eta_\Gamma}$. Symmetries in the brackets are equivalent upon being multiplied by CS operator $\Gamma$. Symmetries that are related by $\Sigma$ symmetry Eq.~\eqref{eq:additional CS Sigma} are labeled by identical symbols at the end.}
\begin{tabular}{ccc} \hline \hline
Class & Symmetry & Map to class \\ \hline 
      A&  $\prescript{c}{}{A}^+ \square$        & AI           \\
      AIII&    $(\prescript{g}{}{\overline{A}^{+}_{+}},\prescript{c}{}{A^{+}_{+}})\heartsuit$      & BDI          \\
      A&  $\prescript{g}{}{\overline{A}^{+}}\triangle$        & D            \\
      AIII&          $(\prescript{g}{}{\overline{A}^{+}_{-}},\prescript{c}{}{A^{-}_{-}})\varheart$& DIII         \\
      A&          $\prescript{c}{}{A}^-\diamondsuit$& AII          \\
      AIII&          $(\prescript{g}{}{\overline{A}^{-}_{+}},\prescript{c}{}{A^{-}_{+}})\vardiamond$& CII          \\
      A&     $\prescript{g}{}{\overline{A}^-}\circ$     & C            \\
      AIII&          $(\prescript{g}{}{\overline{A}^{-}_{-}},\prescript{c}{}{A^{+}_{-}})\bullet$& CI           \\
      A&          $\prescript{g}{}{A}^+ \circ$& AI$^\dag$\\
      AIII&    $(\prescript{g}{}{A}^{+}_{+},\prescript{c}{}{\overline{A}}^{+}_{+})\bullet$      & BDI$^\dag$\\
      A& $\prescript{c}{}{\overline{A}}^+ \square$& D$^\dag$\\
      AIII&    $(\prescript{g}{}{A}^{-}_{-},\prescript{c}{}{\overline{A}}^{+}_{-}) \heartsuit$      & DIII$^\dag$\\
      A&   $\prescript{g}{}{A}^{-}\triangle$       & AII$^\dag$\\
      AIII&   $(\prescript{g}{}{A}^{-}_{+},\prescript{c}{}{\overline{A}}^{-}_{+})\varheart$       & CII$^\dag$\\
      A&   $\prescript{c}{}{\overline{A}}^{-}\diamondsuit$       & C$^\dag$\\
      AIII&   $(\prescript{g}{}{A}^{+}_{-},\prescript{c}{}{\overline{A}}^{-}_{-})\vardiamond$       & CI$^\dag$\\ \hline \hline
      \end{tabular}
      \label{Tab:antiunitary+Complex Classes A and AIII}
\end{table}

\subsection{Real AZ classes with order-two symmetry}
\begin{table*}[!]
\begin{center}
\caption{Possible order-two spatial symmetries for real AZ classes. The superscript of the operators indicates sign the squared value of the operator ; The subscript indicates the commutation relationship between the spatial symmetry operators and TRS $\mathcal{T}_+$ or PHS $\mathcal{C}_-$: $+$ for commutation and $-$ for anti-commutation. If both TRS and PHS are present, the first (second) subscript indicates the commutation relationship between the spatial symmetry operator and TRS (PHS). Symmetries inside the same circular bracket are equivalent. For each AZ class, spatial symmetries that are related by $\Sigma$ symmetry~\eqref{eq:additional CS Sigma} are labeled by the same $t_n, n\in\{0,1,2,3\}$ at the end of the circular bracket.}
\resizebox{\textwidth}{!}{
\begin{tabular}[t]{cclllllllll}
\hline \hline
$s$ & AZ class &  &  &  &  \\
\hline
\multirow{4}{*}{1} 
& \multirow{4}{*}{AI} & ($\prescript{g}{}{U}^+_+$, $\prescript{g}{}{U}^-_-$)$t_1$ 
& ($\prescript{g}{}{\overline{U}}^+_-$, $\prescript{g}{}{\overline{U}}^-_+$)$t_1$
& ($\prescript{g}{}{U}^+_-$, $\prescript{g}{}{U}^-_+$)$t_3$
& ($\prescript{g}{}{\overline{U}}^+_+$, $\prescript{g}{}{\overline{U}}^-_-$)$t_3$
\\ 
& & ($\prescript{g}{}{A}^+_+$, $\prescript{g}{}{A}^+_-$) 
& ($\prescript{g}{}{\overline{A}}^-_+$, $\prescript{g}{}{\overline{A}}^-_-$) 
& ($\prescript{g}{}{A}^-_-$, $\prescript{g}{}{A}^-_+$) 
& ($\prescript{g}{}{\overline{A}}^+_+$, $\prescript{g}{}{\overline{A}}^+_-$)  
\\ [3pt]
\cline{3-6}
& & ($\prescript{c}{}{U}^+_+$, $\prescript{c}{}{U}^-_-$)$t_0$ 
& ($\prescript{c}{}{\overline{U}}^+_-$, $\prescript{c}{}{\overline{U}}^-_+$)$t_2$
& ($\prescript{c}{}{U}^+_-$, $\prescript{c}{}{U}^-_+$)$t_2$
& ($\prescript{c}{}{\overline{U}}^+_+$, $\prescript{c}{}{\overline{U}}^-_-$)$t_0$
\\ 
& & ($\prescript{c}{}{A}^+_+$, $\prescript{c}{}{A}^+_-$) 
& ($\prescript{c}{}{\overline{A}}^-_+$, $\prescript{c}{}{\overline{A}}^-_-$) 
& ($\prescript{c}{}{A}^-_-$, $\prescript{c}{}{A}^-_+$) 
& ($\prescript{c}{}{\overline{A}}^+_+$, $\prescript{c}{}{\overline{A}}^+_-$)  
\\ [3pt]
\hline 
\multirow{4}{*}{2} 
& \multirow{4}{*}{BDI} 
& ($\prescript{g}{}{U}_{++}^+$, $\prescript{g}{}{U}_{--}^-$, $\prescript{c}{}{\overline{U}}_{++}^+$, $\prescript{c}{}{\overline{U}}_{--}^-$)$t_1$
& ($\prescript{g}{}{U}^+_{+-}$, $\prescript{g}{}{U}^-_{-+}$, $\prescript{c}{}{\overline{U}}^+_{-+}$, $\prescript{c}{}{\overline{U}}^-_{+-}$)$t_2$
& ($\prescript{g}{}{U}^+_{--}$, $\prescript{g}{}{U}^-_{++}$, $\prescript{c}{}{\overline{U}}^+_{--}$, $\prescript{c}{}{\overline{U}}^-_{++}$)$t_3$
& ($\prescript{g}{}{U}^+_{-+}$, $\prescript{g}{}{U}^-_{+-}$, $\prescript{c}{}{\overline{U}}^+_{+-}$, $\prescript{c}{}{\overline{U}}^-_{-+}$)$t_0$
\\
& & ($\prescript{g}{}{A}^+_{++}$, $\prescript{g}{}{A}^+_{--}$, $\prescript{c}{}{\overline{A}}^+_{++}$, $\prescript{c}{}{\overline{A}}^+_{--}$)
& ($\prescript{g}{}{A}^+_{+-}$, $\prescript{g}{}{A}^+_{-+}$, $\prescript{c}{}{\overline{A}}^-_{-+}$, $\prescript{c}{}{\overline{A}}^-_{+-}$)
& ($\prescript{g}{}{A}^-_{--}$, $\prescript{g}{}{A}^-_{++}$, $\prescript{c}{}{\overline{A}}^-_{--}$, $\prescript{c}{}{\overline{A}}^-_{++}$)
& ($\prescript{g}{}{A}^-_{-+}$, $\prescript{g}{}{A}^-_{+-}$, $\prescript{c}{}{\overline{A}}^+_{+-}$, $\prescript{c}{}{\overline{A}}^+_{-+}$)
\\[3pt] 
\cline{3-6}
& & ($\prescript{c}{}{U}_{++}^+$, $\prescript{c}{}{U}_{--}^-$, $\prescript{g}{}{\overline{U}}_{++}^+$, $\prescript{g}{}{\overline{U}}_{--}^-$)$t_0$
& ($\prescript{c}{}{U}^+_{+-}$, $\prescript{c}{}{U}^-_{-+}$, $\prescript{g}{}{\overline{U}}^+_{-+}$, $\prescript{g}{}{\overline{U}}^-_{+-}$)$t_1$
& ($\prescript{c}{}{U}^+_{--}$, $\prescript{c}{}{U}^-_{++}$, $\prescript{g}{}{\overline{U}}^+_{--}$, $\prescript{g}{}{\overline{U}}^-_{++}$)$t_2$
& ($\prescript{c}{}{U}^+_{-+}$, $\prescript{c}{}{U}^-_{+-}$, $\prescript{g}{}{\overline{U}}^+_{+-}$, $\prescript{g}{}{\overline{U}}^-_{-+}$)$t_3$
\\
& & ($\prescript{c}{}{A}^+_{++}$, $\prescript{c}{}{A}^+_{--}$, $\prescript{g}{}{\overline{A}}^+_{++}$, $\prescript{g}{}{\overline{A}}^+_{--}$) 
& ($\prescript{c}{}{A}^+_{+-}$, $\prescript{c}{}{A}^+_{-+}$, $\prescript{g}{}{\overline{A}}^-_{-+}$, $\prescript{g}{}{\overline{A}}^-_{+-}$) 
& ($\prescript{c}{}{A}^-_{--}$, $\prescript{c}{}{A}^-_{++}$, $\prescript{g}{}{\overline{A}}^-_{--}$, $\prescript{g}{}{\overline{A}}^-_{++}$) 
& ($\prescript{c}{}{A}^-_{-+}$, $\prescript{c}{}{A}^-_{+-}$, $\prescript{g}{}{\overline{A}}^+_{+-}$, $\prescript{g}{}{\overline{A}}^+_{-+}$) 
\\[3pt]
\hline
\multirow{4}{*}{3}
& \multirow{4}{*}{D}
& ($\prescript{g}{}{U}^+_+$, $\prescript{g}{}{U}^-_-$)$t_1$
& ($\prescript{g}{}{\overline{U}}^+_+$, $\prescript{g}{}{\overline{U}}^-_-$) $t_1$
& ($\prescript{g}{}{U}^+_-$, $\prescript{g}{}{U}^-_+$)$t_3$ 
& ($\prescript{g}{}{\overline{U}}^+_-$, $\prescript{g}{}{\overline{U}}^-_+$)$t_3$ 
\\
& & ($\prescript{c}{}{\overline{A}}^+_+$, $\prescript{c}{}{\overline{A}}^+_-$) 
& ($\prescript{c}{}{A}^+_+$, $\prescript{c}{}{A}^+_-$) 
& ($\prescript{c}{}{\overline{A}}^-_-$, $\prescript{c}{}{\overline{A}}^-_+$) 
& ($\prescript{c}{}{A}^-_+$, $\prescript{c}{}{A}^-_-$) 
\\[3pt]
\cline{3-6}
& & ($\prescript{c}{}{U}^+_+$, $\prescript{c}{}{U}^-_-$)$t_0$
& ($\prescript{c}{}{\overline{U}}^+_+$, $\prescript{c}{}{\overline{U}}^-_-$)$t_2$ 
& ($\prescript{c}{}{U}^+_-$, $\prescript{c}{}{U}^-_+$)$t_2$ 
& ($\prescript{c}{}{\overline{U}}^+_-$, $\prescript{c}{}{\overline{U}}^-_+$)$t_0$
\\
& & ($\prescript{g}{}{\overline{A}}^+_+$, $\prescript{g}{}{\overline{A}}^+_-$) 
& ($\prescript{g}{}{A}^+_+$, $\prescript{g}{}{A}^+_-$) 
& ($\prescript{g}{}{\overline{A}}^-_-$, $\prescript{g}{}{\overline{A}}^-_+$) 
& ($\prescript{g}{}{A}^-_+$, $\prescript{g}{}{A}^-_-$) 
\\[3pt]
\hline
\multirow{4}{*}{4} 
& \multirow{4}{*}{DIII} 
& ($\prescript{g}{}{U}^+_{++}$, $\prescript{g}{}{U}^-_{--}$, $\prescript{c}{}{\overline{U}}^-_{++}$, $\prescript{c}{}{\overline{U}}^+_{--}$)$t_1$
& ($\prescript{g}{}{U}^+_{-+}$, $\prescript{g}{}{U}^-_{+-}$, $\prescript{c}{}{\overline{U}}^+_{-+}$, $\prescript{c}{}{\overline{U}}^-_{+-}$)$t_2$
& ($\prescript{g}{}{U}^+_{--}$, $\prescript{g}{}{U}^-_{++}$, $\prescript{c}{}{\overline{U}}^-_{--}$, $\prescript{c}{}{\overline{U}}^+_{++}$)$t_3$
& ($\prescript{g}{}{U}^+_{+-}$, $\prescript{g}{}{U}^-_{-+}$, $\prescript{c}{}{\overline{U}}^+_{+-}$, $\prescript{c}{}{\overline{U}}^-_{-+}$)$t_0$
\\
& & ($\prescript{g}{}{A}^-_{++}$, $\prescript{g}{}{A}^-_{--}$, $\prescript{c}{}{\overline{A}}^+_{++}$, $\prescript{c}{}{\overline{A}}^+_{--}$)
& ($\prescript{g}{}{A}^+_{-+}$, $\prescript{g}{}{A}^+_{+-}$, $\prescript{c}{}{\overline{A}}^+_{-+}$, $\prescript{c}{}{\overline{A}}^+_{+-}$)
& ($\prescript{g}{}{A}^+_{--}$, $\prescript{g}{}{A}^+_{++}$, $\prescript{c}{}{\overline{A}}^-_{--}$, $\prescript{c}{}{\overline{A}}^-_{++}$)
& ($\prescript{g}{}{A}^-_{+-}$, $\prescript{g}{}{A}^-_{-+}$, $\prescript{c}{}{\overline{A}}^-_{+-}$, $\prescript{c}{}{\overline{A}}^-_{-+}$)
\\[3pt] 
\cline{3-6}
& & ($\prescript{c}{}{U}^+_{++}$, $\prescript{c}{}{U}^-_{--}$, $\prescript{g}{}{\overline{U}}^-_{++}$, $\prescript{g}{}{\overline{U}}^+_{--}$)$t_0$
& ($\prescript{c}{}{U}^+_{-+}$, $\prescript{c}{}{U}^-_{+-}$, $\prescript{g}{}{\overline{U}}^+_{-+}$, $\prescript{g}{}{\overline{U}}^-_{+-}$)$t_1$
& ($\prescript{c}{}{U}^+_{--}$, $\prescript{c}{}{U}^-_{++}$, $\prescript{g}{}{\overline{U}}^-_{--}$, $\prescript{g}{}{\overline{U}}^+_{++}$)$t_2$
& ($\prescript{c}{}{U}^+_{+-}$, $\prescript{c}{}{U}^-_{-+}$, $\prescript{g}{}{\overline{U}}^+_{+-}$, $\prescript{g}{}{\overline{U}}^-_{-+}$)$t_3$
\\
& & ($\prescript{c}{}{A}^-_{++}$, $\prescript{c}{}{A}^-_{--}$, $\prescript{g}{}{\overline{A}}^+_{++}$, $\prescript{g}{}{\overline{A}}^+_{--}$)
& ($\prescript{c}{}{A}^+_{-+}$, $\prescript{c}{}{A}^+_{+-}$, $\prescript{g}{}{\overline{A}}^+_{-+}$, $\prescript{g}{}{\overline{A}}^+_{+-}$)
& ($\prescript{c}{}{A}^+_{--}$, $\prescript{c}{}{A}^+_{++}$, $\prescript{g}{}{\overline{A}}^-_{--}$, $\prescript{g}{}{\overline{A}}^-_{++}$)
& ($\prescript{c}{}{A}^-_{+-}$, $\prescript{c}{}{A}^-_{-+}$, $\prescript{g}{}{\overline{A}}^-_{+-}$, $\prescript{g}{}{\overline{A}}^-_{-+}$)
\\[3pt]
\hline
\multirow{4}{*}{5}
& \multirow{4}{*}{AII}
& ($\prescript{g}{}{U}^+_+$, $\prescript{g}{}{U}^-_-$)$t_1$
& ($\prescript{g}{}{\overline{U}}^+_-$, $\prescript{g}{}{\overline{U}}^-_+$)$t_1$
& ($\prescript{g}{}{U}^+_-$, $\prescript{g}{}{U}^-_+$)$t_3$ 
& ($\prescript{g}{}{\overline{U}}^+_+$, $\prescript{g}{}{\overline{U}}^-_-$) $t_3$ 
\\
& &($\prescript{g}{}{A}^-_+$, $\prescript{g}{}{A}^-_-$) 
&($\prescript{g}{}{\overline{A}}^+_-$, $\prescript{g}{}{\overline{A}}^+_+$)  
&($\prescript{g}{}{A}^+_-$, $\prescript{g}{}{A}^+_+$)  
&($\prescript{g}{}{\overline{A}}^-_+$, $\prescript{g}{}{\overline{A}}^-_-$) 
\\[3pt]
\cline{3-6}
& & ($\prescript{c}{}{U}^+_+$, $\prescript{c}{}{U}^-_-$)$t_0$
& ($\prescript{c}{}{\overline{U}}^+_-$, $\prescript{c}{}{\overline{U}}^-_+$)$t_2$
& ($\prescript{c}{}{U}^+_-$, $\prescript{c}{}{U}^-_+$)$t_2$
& ($\prescript{c}{}{\overline{U}}^+_+$, $\prescript{c}{}{\overline{U}}^-_-$)$t_0$ 
\\
& &($\prescript{c}{}{A}^-_+$, $\prescript{c}{}{A}^-_-$) 
&($\prescript{c}{}{\overline{A}}^+_-$, $\prescript{c}{}{\overline{A}}^+_+$) 
&($\prescript{c}{}{A}^+_-$, $\prescript{c}{}{A}^+_+$) 
&($\prescript{c}{}{\overline{A}}^-_+$, $\prescript{c}{}{\overline{A}}^-_-$) 
\\[3pt]
\hline
\multirow{4}{*}{6} 
& \multirow{4}{*}{CII} 
&($\prescript{g}{}{U}^+_{++}$, $\prescript{g}{}{U}^-_{--}$, $\prescript{c}{}{\overline{U}}^+_{++}$, $\prescript{c}{}{\overline{U}}^-_{--}$)$t_1$
&($\prescript{g}{}{U}^+_{+-}$, $\prescript{g}{}{U}^-_{-+}$, $\prescript{c}{}{\overline{U}}^+_{-+}$, $\prescript{c}{}{\overline{U}}^-_{+-}$)$t_2$
&($\prescript{g}{}{U}^+_{--}$, $\prescript{g}{}{U}^-_{++}$, $\prescript{c}{}{\overline{U}}^+_{--}$, $\prescript{c}{}{\overline{U}}^-_{++}$)$t_3$
&($\prescript{g}{}{U}^+_{-+}$, $\prescript{g}{}{U}^-_{+-}$, $\prescript{c}{}{\overline{U}}^+_{+-}$, $\prescript{c}{}{\overline{U}}^-_{-+}$)$t_0$
\\
& &($\prescript{g}{}{A}^-_{++}$, $\prescript{g}{}{A}^-_{--}$, $\prescript{c}{}{\overline{A}}^-_{++}$, $\prescript{c}{}{\overline{A}}^-_{--}$)
& ($\prescript{g}{}{A}^-_{+-}$, $\prescript{g}{}{A}^-_{-+}$, $\prescript{c}{}{\overline{A}}^+_{-+}$, $\prescript{c}{}{\overline{A}}^+_{+-}$)
& ($\prescript{g}{}{A}^+_{--}$, $\prescript{g}{}{A}^+_{++}$, $\prescript{c}{}{\overline{A}}^+_{--}$, $\prescript{c}{}{\overline{A}}^+_{++}$)
& ($\prescript{g}{}{A}^+_{-+}$, $\prescript{g}{}{A}^+_{+-}$, $\prescript{c}{}{\overline{A}}^-_{+-}$, $\prescript{c}{}{\overline{A}}^-_{-+}$)
\\[3pt]
\cline{3-6}
& &($\prescript{c}{}{U}^+_{++}$, $\prescript{c}{}{U}^-_{--}$, $\prescript{g}{}{\overline{U}}^+_{++}$, $\prescript{g}{}{\overline{U}}^-_{--}$)$t_0$
&($\prescript{c}{}{U}^+_{+-}$, $\prescript{c}{}{U}^-_{-+}$, $\prescript{g}{}{\overline{U}}^+_{-+}$, $\prescript{g}{}{\overline{U}}^-_{+-}$)$t_1$
&($\prescript{c}{}{U}^+_{--}$, $\prescript{c}{}{U}^-_{++}$, $\prescript{g}{}{\overline{U}}^+_{--}$, $\prescript{g}{}{\overline{U}}^-_{++}$)$t_2$
&($\prescript{c}{}{U}^+_{-+}$, $\prescript{c}{}{U}^-_{+-}$, $\prescript{g}{}{\overline{U}}^+_{+-}$, $\prescript{g}{}{\overline{U}}^-_{-+}$)$t_3$
\\
& &($\prescript{c}{}{A}^-_{++}$, $\prescript{c}{}{A}^-_{--}$, $\prescript{g}{}{\overline{A}}^-_{++}$, $\prescript{g}{}{\overline{A}}^-_{--}$)
& ($\prescript{c}{}{A}^-_{+-}$, $\prescript{c}{}{A}^-_{-+}$, $\prescript{g}{}{\overline{A}}^+_{-+}$, $\prescript{g}{}{\overline{A}}^+_{+-}$)
& ($\prescript{c}{}{A}^+_{--}$, $\prescript{c}{}{A}^+_{++}$, $\prescript{g}{}{\overline{A}}^+_{--}$, $\prescript{g}{}{\overline{A}}^+_{++}$)
& ($\prescript{c}{}{A}^+_{-+}$, $\prescript{c}{}{A}^+_{+-}$, $\prescript{g}{}{\overline{A}}^-_{+-}$, $\prescript{g}{}{\overline{A}}^-_{-+}$)
\\[3pt]
\hline
\multirow{4}{*}{7}
& \multirow{4}{*}{C}
& ($\prescript{g}{}{U}^+_+$, $\prescript{g}{}{U}^-_-$)$t_1$ 
& ($\prescript{g}{}{\overline{U}}^+_+$, $\prescript{g}{}{\overline{U}}^-_-$)$t_1$ 
& ($\prescript{g}{}{U}^+_-$, $\prescript{g}{}{U}^-_+$)$t_3$
& ($\prescript{g}{}{\overline{U}}^+_-$, $\prescript{g}{}{\overline{U}}^-_+$)$t_3$
\\ 
& &($\prescript{c}{}{\overline{A}}^-_+$, $\prescript{c}{}{\overline{A}}^-_-$) 
&($\prescript{c}{}{A}^-_+$, $\prescript{c}{}{A}^-_-$) 
&($\prescript{c}{}{\overline{A}}^+_-$, $\prescript{c}{}{\overline{A}}^+_+$) 
&($\prescript{c}{}{A}^+_-$, $\prescript{c}{}{A}^+_+$)
\\[3pt] 
\cline{3-6}
& & ($\prescript{c}{}{U}^+_+$, $\prescript{c}{}{U}^-_-$)$t_0$ 
& ($\prescript{c}{}{\overline{U}}^+_+$, $\prescript{c}{}{\overline{U}}^-_-$)$t_2$ 
& ($\prescript{c}{}{U}^+_-$, $\prescript{c}{}{U}^-_+$)$t_2$
& ($\prescript{c}{}{\overline{U}}^+_-$, $\prescript{c}{}{\overline{U}}^-_+$)$t_0$
\\ 
& &($\prescript{g}{}{\overline{A}}^-_+$, $\prescript{g}{}{\overline{A}}^-_-$) 
&($\prescript{g}{}{A}^-_+$, $\prescript{g}{}{A}^-_-$) 
&($\prescript{g}{}{\overline{A}}^+_-$, $\prescript{g}{}{\overline{A}}^+_+$) 
&($\prescript{g}{}{A}^+_-$, $\prescript{g}{}{A}^+_+$)
\\[3pt]
\hline
\multirow{4}{*}{0} 
&\multirow{4}{*}{CI} 
&($\prescript{g}{}{U}^+_{++}$, $\prescript{g}{}{U}^-_{--}$, $\prescript{c}{}{\overline{U}}^-_{++}$, $\prescript{c}{}{\overline{U}}^+_{--}$)$t_1$
&($\prescript{g}{}{U}^+_{-+}$, $\prescript{g}{}{U}^-_{+-}$, $\prescript{c}{}{\overline{U}}^+_{-+}$, $\prescript{c}{}{\overline{U}}^-_{+-}$)$t_2$
&($\prescript{g}{}{U}^+_{--}$, $\prescript{g}{}{U}^-_{++}$, $\prescript{c}{}{\overline{U}}^-_{--}$, $\prescript{c}{}{\overline{U}}^+_{++}$)$t_3$ 
&($\prescript{g}{}{U}^+_{+-}$, $\prescript{g}{}{U}^-_{-+}$, $\prescript{c}{}{\overline{U}}^+_{+-}$, $\prescript{c}{}{\overline{U}}^-_{-+}$)$t_0$
\\ 
& &($\prescript{g}{}{A}^+_{++}$, $\prescript{g}{}{A}^+_{--}$, $\prescript{c}{}{\overline{A}}^-_{++}$, $\prescript{c}{}{\overline{A}}^-_{--}$)
& ($\prescript{g}{}{A}^-_{-+}$, $\prescript{g}{}{A}^-_{+-}$, $\prescript{c}{}{\overline{A}}^-_{-+}$, $\prescript{c}{}{\overline{A}}^-_{+-}$)
& ($\prescript{g}{}{A}^-_{--}$, $\prescript{g}{}{A}^-_{++}$, $\prescript{c}{}{\overline{A}}^+_{--}$, $\prescript{c}{}{\overline{A}}^+_{++}$)
& ($\prescript{g}{}{A}^+_{+-}$, $\prescript{g}{}{A}^+_{-+}$, $\prescript{c}{}{\overline{A}}^+_{+-}$, $\prescript{c}{}{\overline{A}}^+_{-+}$)
\\[3pt]
\cline{3-6}
& &($\prescript{c}{}{U}^+_{++}$, $\prescript{c}{}{U}^-_{--}$, $\prescript{g}{}{\overline{U}}^-_{++}$, $\prescript{g}{}{\overline{U}}^+_{--}$)$t_0$
&($\prescript{c}{}{U}^+_{-+}$, $\prescript{c}{}{U}^-_{+-}$, $\prescript{g}{}{\overline{U}}^+_{-+}$, $\prescript{g}{}{\overline{U}}^-_{+-}$)$t_1$
&($\prescript{c}{}{U}^+_{--}$, $\prescript{c}{}{U}^-_{++}$, $\prescript{g}{}{\overline{U}}^-_{--}$, $\prescript{g}{}{\overline{U}}^+_{++}$)$t_2$ 
&($\prescript{c}{}{U}^+_{+-}$, $\prescript{c}{}{U}^-_{-+}$, $\prescript{g}{}{\overline{U}}^+_{+-}$, $\prescript{g}{}{\overline{U}}^-_{-+}$)$t_3$
\\ 
& &($\prescript{c}{}{A}^+_{++}$, $\prescript{c}{}{A}^+_{--}$, $\prescript{g}{}{\overline{A}}^-_{++}$, $\prescript{g}{}{\overline{A}}^-_{--}$)
& ($\prescript{c}{}{A}^-_{-+}$, $\prescript{c}{}{A}^-_{+-}$, $\prescript{g}{}{\overline{A}}^-_{-+}$, $\prescript{g}{}{\overline{A}}^-_{+-}$)
& ($\prescript{c}{}{A}^-_{--}$, $\prescript{c}{}{A}^-_{++}$, $\prescript{g}{}{\overline{A}}^+_{--}$, $\prescript{g}{}{\overline{A}}^+_{++}$)
& ($\prescript{c}{}{A}^+_{+-}$, $\prescript{c}{}{A}^+_{-+}$, $\prescript{g}{}{\overline{A}}^+_{+-}$, $\prescript{g}{}{\overline{A}}^+_{-+}$)\\[3pt]
\hline \hline
\end{tabular}}
\label{Tab:Symmetry_type AZ}
\end{center}
\end{table*}

In this section, we consider the classification of real AZ classes with order-two spatial symmetries. Throughout this section, all Eqs. are also valid upon exchange $g \leftrightarrow c$; The superscripts denote the squared value of the spatial operators, and the subscripts denote the commutation relationship between spatial operators and $\mathcal{T}_+$ and $\mathcal{C}_-$. Throughout this paper, we take the convention that $[\mathcal{T}_+,\mathcal{C}_-]=0$.

For class A and AII, we have the following equivalence symmetries

\textit{AI and AII} ($\mathcal{T}_+$):
\begin{align}
    &\prescript{g}{}{U}_{\eta_T}^{\epsilon_U}=\textrm{i}  \prescript{g}{}{U}_{-\eta_T}^{-\epsilon_U}=\mathcal{T}_+ \prescript{g}{}{A}_{\eta_T}^{\eta_T \epsilon_T \epsilon_U}=\textrm{i}\mathcal{T}_{+}\prescript{g}{}{A}_{-\eta_T}^{\eta_T \epsilon_T \epsilon_U} \nonumber\\
    &\prescript{g}{}{\overline{U}}_{\eta_T}^{\epsilon_U}=\textrm{i}  \prescript{g}{}{\overline{U}}_{-\eta_T}^{-\epsilon_U}=\mathcal{T}_+ \prescript{g}{}{\overline{A}}_{\eta_T}^{\eta_T \epsilon_T \epsilon_U}=\textrm{i}\mathcal{T}_{+}\prescript{g}{}{\overline{A}}_{-\eta_T}^{\eta_T \epsilon_T \epsilon_U},
\end{align}
where the subscript $\eta_T$ denotes the commutation relationship between the operator and $\mathcal{T}_+$. 

For class D and C, we have the following equivalence symmetries 

\textit{D and C} ($\mathcal{C}_-$):
\begin{align}
    &\prescript{g}{}{U}_{\eta_C}^{\epsilon_U}=\textrm{i}  \prescript{g}{}{U}_{-\eta_C}^{-\epsilon_U}=\mathcal{C}_- \prescript{c}{}{\overline{A}}_{\eta_C}^{\eta_C \epsilon_C \epsilon_U}=\textrm{i}\mathcal{C}_{-}\prescript{c}{}{\overline{A}}_{-\eta_C}^{\eta_C \epsilon_C \epsilon_U}\nonumber\\
    &\prescript{g}{}{\overline{U}}_{\eta_C}^{\epsilon_U}=\textrm{i}  \prescript{g}{}{\overline{U}}_{-\eta_C}^{-\epsilon_U}=\mathcal{C}_- \prescript{c}{}{A}_{\eta_C}^{\eta_C \epsilon_C \epsilon_U}=\textrm{i}\mathcal{C}_{-}\prescript{c}{}{A}_{-\eta_C}^{\eta_C \epsilon_C \epsilon_U},
\end{align}
where the subscript $\eta_C$ denotes the commutation relationship between the operator and $\mathcal{C}_-$. 

Finally, for class BDI, DIII, CII, and CI, we have

\textit{BDI, DIII, CII, and CI} ($\mathcal{C}_-$ and $\mathcal{T}_+$):
\begin{align}
    \prescript{g}{}{U}_{\eta_T,\eta_C}^{\epsilon_U}&=\textrm{i}  \prescript{g}{}{U}_{-\eta_T,-\eta_C}^{-\epsilon_U}=\mathcal{T}_{+} \prescript{g}{}{A}_{\eta_T,\eta_C}^{\eta_T \epsilon_T \epsilon_U}=\textrm{i}\mathcal{T}_{+} \prescript{g}{}{A}_{-\eta_T,-\eta_C}^{\eta_T \epsilon_T \epsilon_U}\nonumber\\
    &=\mathcal{C}_{-} \prescript{c}{}{\overline{A}}_{\eta_T,\eta_C}^{\eta_C \epsilon_C \epsilon_U}=\textrm{i} \mathcal{C}_{-} \prescript{c}{}{\overline{A}}_{-\eta_T,-\eta_C}^{\eta_C \epsilon_C \epsilon_U}\nonumber
\end{align}
\begin{align}
    \prescript{g}{}{\overline{U}}_{\eta_T,\eta_C}^{\epsilon_U}&=\textrm{i}  \prescript{g}{}{\overline{U}}_{-\eta_T,-\eta_C}^{-\epsilon_U}=\mathcal{T}_{+} \prescript{g}{}{\overline{A}}_{\eta_T,\eta_C}^{\eta_T \epsilon_T \epsilon_U}=\textrm{i}\mathcal{T}_{+} \prescript{g}{}{\overline{A}}_{-\eta_T,-\eta_C}^{\eta_T \epsilon_T \epsilon_U}\nonumber\\
    &=\mathcal{C}_{-} \prescript{c}{}{A}_{\eta_T,\eta_C}^{\eta_C \epsilon_C \epsilon_U}=\textrm{i} \mathcal{C}_{-} \prescript{c}{}{A}_{-\eta_T,-\eta_C}^{\eta_C \epsilon_C \epsilon_U},
\end{align}
where the first (second) subscript $\eta_T$ ($\eta_C$) indicates the commutation relationship between the operator and $\mathcal{T}_+$ ($\mathcal{C}_-$).

By going over all possible combinations of $\epsilon_U,\eta_T,\eta_C$, we summarize possible order-two spatial symmetries for real AZ classes in Table~\ref{Tab:Symmetry_type AZ}. 

Next, we discuss the effect of $\Sigma$ symmetry. Similar to the previous case for complex AZ classes, $\Sigma$ symmetry would "unify" some spatial symmetries in Table~\ref{Tab:Symmetry_type AZ}. As an example, consider class AI with spatial symmetry $\prescript{g}{}{U}^+_+$. Recall that the $\Sigma$ operator in Eq.~\eqref{eq:additional CS Sigma} anti-commutes with generalized spatial symmetries Eq.~\eqref{eq:generalized spatial extension}. Therefore, $\Sigma \prescript{g}{}{U}^+_+$ has the same classification as $\prescript{g}{}{\overline{U}}^+_-$. We may repeat this process for all classes in Table~\ref{Tab:Symmetry_type AZ}. For each class, we label symmetries that are independent by $t=0,1,2,3$, where $t$ is defined mod$4$. We abbreviate these symbols into $t_0,t_1,t_2,t_3$ and label them at the end of the circular bracket in Table~\ref{Tab:Symmetry_type AZ}.  

In Appendix~\ref{Sec:DH AZ and AZ dag with spatial}, we show that the $K$ group for real AZ classes with spatial symmetries obey the following relationship
\begin{align}
    &K_{\mathbb{R}}^{U/A}(s,t;d,d_{\parallel},D,D_{\parallel})\nonumber\\
    &=K_{\mathbb{R}}^{U/A}(s-d+D,t-d_{\parallel}+D_{\parallel};0,0,0,0),
    \label{eq:K group real AZ}
\end{align}
where the superscript $U/A$ indicates this relationship works for both unitary and anti-unitary spatial symmetries. Furthermore,
\begin{align}
    &K_{\mathbb{R}}^{U/A}(s,t=0;0,0,0,0)=\pi_0 (\mathcal{R}_s)\times \pi_0 (\mathcal{R}_s)\nonumber\\
    &K_{\mathbb{R}}^{U/A}(s,t=1;0,0,0,0)=\pi_0 (\mathcal{R}_{s-1})\nonumber\\
    &K_{\mathbb{R}}^{U/A}(s,t=2;0,0,0,0)=\pi_0 (\mathcal{C}_{s})\nonumber\\
    &K_{\mathbb{R}}^{U/A}(s,t=3;0,0,0,0)=\pi_0 (\mathcal{R}_{s+1}),
\end{align}
which we show in Appendix~\ref{Appendix:classifying space} using Clifford Algebra.
For both Eqs. $s$ is defined mod$8$ while $t$ is defined mod$4$.
\subsection{Real AZ$^\dag$ classes with order-two symmetry}

\begin{table*}[!]
\begin{center}
\caption{Possible order-two spatial symmetries for real AZ$^\dag$ classes. The superscript of the operators indicates sign the squared value of the operator ; The subscript indicates the commutation relationship between the spatial symmetry operators and TRS$^\dag$ $\mathcal{C}_+$ or PHS$^\dag$ $\mathcal{T}_-$: $+$ for commutation and $-$ for anti-commutation. If both TRS$^\dag$ and PHS$^\dag$ are present, the first (second) subscript indicates the commutation relationship between the spatial symmetry operator and TRS$^\dag$ (PHS$^\dag$). Symmetries inside the same circular bracket are equivalent. For each AZ$^\dag$ class, spatial symmetries that are related by $\Sigma$ symmetry~\eqref{eq:additional CS Sigma} are labeled by the $t_n, n\in\{0,1,2,3\}$ at the end of the circular bracket.}
\resizebox{\textwidth}{!}{
\begin{tabular}[t]{cclllllllll}
\hline \hline
$s^\dag$ & AZ$^\dag$ class & &  &  &  \\
\hline
\multirow{4}{*}{7} 
& \multirow{4}{*}{AI$^\dag$} & ($\prescript{g}{}{U}^+_+$, $\prescript{g}{}{U}^-_-$)$t_1$ 
& ($\prescript{g}{}{\overline{U}}^+_-$, $\prescript{g}{}{\overline{U}}^-_+$)$t_3$
& ($\prescript{g}{}{U}^+_-$, $\prescript{g}{}{U}^-_+$)$t_3$
& ($\prescript{g}{}{\overline{U}}^+_+$, $\prescript{g}{}{\overline{U}}^-_-$)$t_1$
\\ 
& & ($\prescript{c}{}{A}^+_+$, $\prescript{c}{}{A}^+_-$) 
& ($\prescript{c}{}{\overline{A}}^-_+$, $\prescript{c}{}{\overline{A}}^-_-$) 
& ($\prescript{c}{}{A}^-_-$, $\prescript{c}{}{A}^-_+$) 
& ($\prescript{c}{}{\overline{A}}^+_+$, $\prescript{c}{}{\overline{A}}^+_-$)
\\[3pt]
\cline{3-6}
& 
& ($\prescript{c}{}{U}^+_+$, $\prescript{c}{}{U}^-_-$)$t_0$ 
& ($\prescript{c}{}{\overline{U}}^+_-$, $\prescript{c}{}{\overline{U}}^-_+$)$t_0$
& ($\prescript{c}{}{U}^+_-$, $\prescript{c}{}{U}^-_+$)$t_2$
& ($\prescript{c}{}{\overline{U}}^+_+$, $\prescript{c}{}{\overline{U}}^-_-$)$t_2$
\\ 
& & ($\prescript{g}{}{A}^+_+$, $\prescript{g}{}{A}^+_-$) 
& ($\prescript{g}{}{\overline{A}}^-_+$, $\prescript{g}{}{\overline{A}}^-_-$) 
& ($\prescript{g}{}{A}^-_-$, $\prescript{g}{}{A}^-_+$) 
& ($\prescript{g}{}{\overline{A}}^+_+$, $\prescript{g}{}{\overline{A}}^+_-$)
\\ [3pt]
\hline 
\multirow{4}{*}{0} 
& \multirow{4}{*}{BDI$^\dag$} 
& ($\prescript{g}{}{U}_{++}^+$, $\prescript{g}{}{U}_{--}^-$, $\prescript{c}{}{\overline{U}}_{++}^+$, $\prescript{c}{}{\overline{U}}_{--}^-$)$t_1$
& ($\prescript{g}{}{U}^+_{+-}$, $\prescript{g}{}{U}^-_{-+}$, $\prescript{c}{}{\overline{U}}^+_{-+}$, $\prescript{c}{}{\overline{U}}^-_{+-}$)$t_0$
& ($\prescript{g}{}{U}^+_{--}$, $\prescript{g}{}{U}^-_{++}$, $\prescript{c}{}{\overline{U}}^+_{--}$, $\prescript{c}{}{\overline{U}}^-_{++}$)$t_3$
& ($\prescript{g}{}{U}^+_{-+}$, $\prescript{g}{}{U}^-_{+-}$, $\prescript{c}{}{\overline{U}}^+_{+-}$, $\prescript{c}{}{\overline{U}}^-_{-+}$)$t_2$
\\
& & ($\prescript{c}{}{A}^+_{++}$, $\prescript{c}{}{A}^+_{--}$, $\prescript{g}{}{\overline{A}}^+_{++}$, $\prescript{g}{}{\overline{A}}^+_{--}$) 
& ($\prescript{c}{}{A}^+_{+-}$, $\prescript{c}{}{A}^+_{-+}$, $\prescript{g}{}{\overline{A}}^-_{-+}$, $\prescript{g}{}{\overline{A}}^-_{+-}$) 
& ($\prescript{c}{}{A}^-_{--}$, $\prescript{c}{}{A}^-_{++}$, $\prescript{g}{}{\overline{A}}^-_{--}$, $\prescript{g}{}{\overline{A}}^-_{++}$) 
& ($\prescript{c}{}{A}^-_{-+}$, $\prescript{c}{}{A}^-_{+-}$, $\prescript{g}{}{\overline{A}}^+_{+-}$, $\prescript{g}{}{\overline{A}}^+_{-+}$) 
\\[3pt]
\cline{3-6}
&  
& ($\prescript{c}{}{U}_{++}^+$, $\prescript{c}{}{U}_{--}^-$, $\prescript{g}{}{\overline{U}}_{++}^+$, $\prescript{g}{}{\overline{U}}_{--}^-$)$t_0$
& ($\prescript{c}{}{U}^+_{+-}$, $\prescript{c}{}{U}^-_{-+}$, $\prescript{g}{}{\overline{U}}^+_{-+}$, $\prescript{g}{}{\overline{U}}^-_{+-}$)$t_3$
& ($\prescript{c}{}{U}^+_{--}$, $\prescript{c}{}{U}^-_{++}$, $\prescript{g}{}{\overline{U}}^+_{--}$, $\prescript{g}{}{\overline{U}}^-_{++}$)$t_2$
& ($\prescript{c}{}{U}^+_{-+}$, $\prescript{c}{}{U}^-_{+-}$, $\prescript{g}{}{\overline{U}}^+_{+-}$, $\prescript{g}{}{\overline{U}}^-_{-+}$)$t_1$
\\
& & ($\prescript{g}{}{A}^+_{++}$, $\prescript{g}{}{A}^+_{--}$, $\prescript{c}{}{\overline{A}}^+_{++}$, $\prescript{c}{}{\overline{A}}^+_{--}$) 
& ($\prescript{g}{}{A}^+_{+-}$, $\prescript{g}{}{A}^+_{-+}$, $\prescript{c}{}{\overline{A}}^-_{-+}$, $\prescript{c}{}{\overline{A}}^-_{+-}$) 
& ($\prescript{g}{}{A}^-_{--}$, $\prescript{g}{}{A}^-_{++}$, $\prescript{c}{}{\overline{A}}^-_{--}$, $\prescript{c}{}{\overline{A}}^-_{++}$) 
& ($\prescript{g}{}{A}^-_{-+}$, $\prescript{g}{}{A}^-_{+-}$, $\prescript{c}{}{\overline{A}}^+_{+-}$, $\prescript{c}{}{\overline{A}}^+_{-+}$)
\\[3pt]
\hline
\multirow{4}{*}{1}
& \multirow{4}{*}{D$^\dag$}
& ($\prescript{g}{}{U}^+_+$, $\prescript{g}{}{U}^-_-$)$t_1$
& ($\prescript{g}{}{\overline{U}}^+_+$, $\prescript{g}{}{\overline{U}}^-_-$)$t_3$ 
& ($\prescript{g}{}{U}^+_-$, $\prescript{g}{}{U}^-_+$)$t_3$ 
& ($\prescript{g}{}{\overline{U}}^+_-$, $\prescript{g}{}{\overline{U}}^-_+$)$t_1$ 
\\
& & ($\prescript{g}{}{\overline{A}}^+_+$, $\prescript{g}{}{\overline{A}}^+_-$) 
& ($\prescript{g}{}{A}^+_+$, $\prescript{g}{}{A}^+_-$) 
& ($\prescript{g}{}{\overline{A}}^-_-$, $\prescript{g}{}{\overline{A}}^-_+$) 
& ($\prescript{g}{}{A}^-_+$, $\prescript{g}{}{A}^-_-$) 
\\[3pt]
\cline{3-6}
& 
& ($\prescript{c}{}{U}^+_+$, $\prescript{c}{}{U}^-_-$)$t_0$
& ($\prescript{c}{}{\overline{U}}^+_+$, $\prescript{c}{}{\overline{U}}^-_-$)$t_0$ 
& ($\prescript{c}{}{U}^+_-$, $\prescript{c}{}{U}^-_+$)$t_2$ 
& ($\prescript{c}{}{\overline{U}}^+_-$, $\prescript{c}{}{\overline{U}}^-_+$)$t_2$ 
\\
& & ($\prescript{c}{}{\overline{A}}^+_+$, $\prescript{c}{}{\overline{A}}^+_-$) 
& ($\prescript{c}{}{A}^+_+$, $\prescript{c}{}{A}^+_-$) 
& ($\prescript{c}{}{\overline{A}}^-_-$, $\prescript{c}{}{\overline{A}}^-_+$) 
& ($\prescript{c}{}{A}^-_+$, $\prescript{c}{}{A}^-_-$)
\\[3pt] 
\hline
\multirow{4}{*}{2} 
& \multirow{4}{*}{DIII$^\dag$} 
& ($\prescript{g}{}{U}^+_{++}$, $\prescript{g}{}{U}^-_{--}$, $\prescript{c}{}{\overline{U}}^-_{++}$, $\prescript{c}{}{\overline{U}}^+_{--}$)$t_1$
& ($\prescript{g}{}{U}^+_{-+}$, $\prescript{g}{}{U}^-_{+-}$, $\prescript{c}{}{\overline{U}}^+_{-+}$, $\prescript{c}{}{\overline{U}}^-_{+-}$)$t_0$
& ($\prescript{g}{}{U}^+_{--}$, $\prescript{g}{}{U}^-_{++}$, $\prescript{c}{}{\overline{U}}^-_{--}$, $\prescript{c}{}{\overline{U}}^+_{++}$)$t_3$
& ($\prescript{g}{}{U}^+_{+-}$, $\prescript{g}{}{U}^-_{-+}$, $\prescript{c}{}{\overline{U}}^+_{+-}$, $\prescript{c}{}{\overline{U}}^-_{-+}$)$t_2$
\\
& & ($\prescript{c}{}{A}^-_{++}$, $\prescript{c}{}{A}^-_{--}$, $\prescript{g}{}{\overline{A}}^+_{++}$, $\prescript{g}{}{\overline{A}}^+_{--}$)
& ($\prescript{c}{}{A}^+_{-+}$, $\prescript{c}{}{A}^+_{+-}$, $\prescript{g}{}{\overline{A}}^+_{-+}$, $\prescript{g}{}{\overline{A}}^+_{+-}$)
& ($\prescript{c}{}{A}^+_{--}$, $\prescript{c}{}{A}^+_{++}$, $\prescript{g}{}{\overline{A}}^-_{--}$, $\prescript{g}{}{\overline{A}}^-_{++}$)
& ($\prescript{c}{}{A}^-_{+-}$, $\prescript{c}{}{A}^-_{-+}$, $\prescript{g}{}{\overline{A}}^-_{+-}$, $\prescript{g}{}{\overline{A}}^-_{-+}$)
\\[3pt]
\cline{3-6}
&
& ($\prescript{c}{}{U}^+_{++}$, $\prescript{c}{}{U}^-_{--}$, $\prescript{g}{}{\overline{U}}^-_{++}$, $\prescript{g}{}{\overline{U}}^+_{--}$)$t_0$
& ($\prescript{c}{}{U}^+_{-+}$, $\prescript{c}{}{U}^-_{+-}$, $\prescript{g}{}{\overline{U}}^+_{-+}$, $\prescript{g}{}{\overline{U}}^-_{+-}$)$t_3$
& ($\prescript{c}{}{U}^+_{--}$, $\prescript{c}{}{U}^-_{++}$, $\prescript{g}{}{\overline{U}}^-_{--}$, $\prescript{g}{}{\overline{U}}^+_{++}$)$t_2$
& ($\prescript{c}{}{U}^+_{+-}$, $\prescript{c}{}{U}^-_{-+}$, $\prescript{g}{}{\overline{U}}^+_{+-}$, $\prescript{g}{}{\overline{U}}^-_{-+}$)$t_1$
\\
& & ($\prescript{g}{}{A}^-_{++}$, $\prescript{g}{}{A}^-_{--}$, $\prescript{c}{}{\overline{A}}^+_{++}$, $\prescript{c}{}{\overline{A}}^+_{--}$)
& ($\prescript{g}{}{A}^+_{-+}$, $\prescript{g}{}{A}^+_{+-}$, $\prescript{c}{}{\overline{A}}^+_{-+}$, $\prescript{c}{}{\overline{A}}^+_{+-}$)
& ($\prescript{g}{}{A}^+_{--}$, $\prescript{g}{}{A}^+_{++}$, $\prescript{c}{}{\overline{A}}^-_{--}$, $\prescript{c}{}{\overline{A}}^-_{++}$)
& ($\prescript{g}{}{A}^-_{+-}$, $\prescript{g}{}{A}^-_{-+}$, $\prescript{c}{}{\overline{A}}^-_{+-}$, $\prescript{c}{}{\overline{A}}^-_{-+}$)
\\[3pt]
\hline
\multirow{4}{*}{3}
& \multirow{4}{*}{AII$^\dag$}
& ($\prescript{g}{}{U}^+_+$, $\prescript{g}{}{U}^-_-$)$t_1$
& ($\prescript{g}{}{\overline{U}}^+_-$, $\prescript{g}{}{\overline{U}}^-_+$)$t_3$
& ($\prescript{g}{}{U}^+_-$, $\prescript{g}{}{U}^-_+$)$t_3$
& ($\prescript{g}{}{\overline{U}}^+_+$, $\prescript{g}{}{\overline{U}}^-_-$)$t_1$ 
\\
& &($\prescript{c}{}{A}^-_+$, $\prescript{c}{}{A}^-_-$) 
&($\prescript{c}{}{\overline{A}}^+_-$, $\prescript{c}{}{\overline{A}}^+_+$) 
&($\prescript{c}{}{A}^+_-$, $\prescript{c}{}{A}^+_+$) 
&($\prescript{c}{}{\overline{A}}^-_+$, $\prescript{c}{}{\overline{A}}^-_-$) 
\\[3pt]
\cline{3-6}
&
& ($\prescript{c}{}{U}^+_+$, $\prescript{c}{}{U}^-_-$)$t_0$
& ($\prescript{c}{}{\overline{U}}^+_-$, $\prescript{c}{}{\overline{U}}^-_+$)$t_0$
& ($\prescript{c}{}{U}^+_-$, $\prescript{c}{}{U}^-_+$)$t_2$
& ($\prescript{c}{}{\overline{U}}^+_+$, $\prescript{c}{}{\overline{U}}^-_-$)$t_2$ 
\\
& &($\prescript{g}{}{A}^-_+$, $\prescript{g}{}{A}^-_-$) 
&($\prescript{g}{}{\overline{A}}^+_-$, $\prescript{g}{}{\overline{A}}^+_+$) 
&($\prescript{g}{}{A}^+_-$, $\prescript{g}{}{A}^+_+$) 
&($\prescript{g}{}{\overline{A}}^-_+$, $\prescript{g}{}{\overline{A}}^-_-$) 
\\[3pt]
\hline
\multirow{4}{*}{4} 
& \multirow{4}{*}{CII$^\dag$} 
&($\prescript{g}{}{U}^+_{++}$, $\prescript{g}{}{U}^-_{--}$, $\prescript{c}{}{\overline{U}}^+_{++}$, $\prescript{c}{}{\overline{U}}^-_{--}$)$t_1$
&($\prescript{g}{}{U}^+_{+-}$, $\prescript{g}{}{U}^-_{-+}$, $\prescript{c}{}{\overline{U}}^+_{-+}$, $\prescript{c}{}{\overline{U}}^-_{+-}$)$t_0$
&($\prescript{g}{}{U}^+_{--}$, $\prescript{g}{}{U}^-_{++}$, $\prescript{c}{}{\overline{U}}^+_{--}$, $\prescript{c}{}{\overline{U}}^-_{++}$)$t_3$
&($\prescript{g}{}{U}^+_{-+}$, $\prescript{g}{}{U}^-_{+-}$, $\prescript{c}{}{\overline{U}}^+_{+-}$, $\prescript{c}{}{\overline{U}}^-_{-+}$)$t_2$
\\
& &($\prescript{c}{}{A}^-_{++}$, $\prescript{c}{}{A}^-_{--}$, $\prescript{g}{}{\overline{A}}^-_{++}$, $\prescript{g}{}{\overline{A}}^-_{--}$)
& ($\prescript{c}{}{A}^-_{+-}$, $\prescript{c}{}{A}^-_{-+}$, $\prescript{g}{}{\overline{A}}^+_{-+}$, $\prescript{g}{}{\overline{A}}^+_{+-}$)
& ($\prescript{c}{}{A}^+_{--}$, $\prescript{c}{}{A}^+_{++}$, $\prescript{g}{}{\overline{A}}^+_{--}$, $\prescript{g}{}{\overline{A}}^+_{++}$)
& ($\prescript{c}{}{A}^+_{-+}$, $\prescript{c}{}{A}^+_{+-}$, $\prescript{g}{}{\overline{A}}^-_{+-}$, $\prescript{g}{}{\overline{A}}^-_{-+}$)
\\[3pt]
\cline{3-6}
&
&($\prescript{c}{}{U}^+_{++}$, $\prescript{c}{}{U}^-_{--}$, $\prescript{g}{}{\overline{U}}^+_{++}$, $\prescript{g}{}{\overline{U}}^-_{--}$)$t_0$
&($\prescript{c}{}{U}^+_{+-}$, $\prescript{c}{}{U}^-_{-+}$, $\prescript{g}{}{\overline{U}}^+_{-+}$, $\prescript{g}{}{\overline{U}}^-_{+-}$)$t_3$
&($\prescript{c}{}{U}^+_{--}$, $\prescript{c}{}{U}^-_{++}$, $\prescript{g}{}{\overline{U}}^+_{--}$, $\prescript{g}{}{\overline{U}}^-_{++}$)$t_2$
&($\prescript{c}{}{U}^+_{-+}$, $\prescript{c}{}{U}^-_{+-}$, $\prescript{g}{}{\overline{U}}^+_{+-}$, $\prescript{g}{}{\overline{U}}^-_{-+}$)$t_1$
\\
& &($\prescript{g}{}{A}^-_{++}$, $\prescript{g}{}{A}^-_{--}$, $\prescript{c}{}{\overline{A}}^-_{++}$, $\prescript{c}{}{\overline{A}}^-_{--}$)
& ($\prescript{g}{}{A}^-_{+-}$, $\prescript{g}{}{A}^-_{-+}$, $\prescript{c}{}{\overline{A}}^+_{-+}$, $\prescript{c}{}{\overline{A}}^+_{+-}$)
& ($\prescript{g}{}{A}^+_{--}$, $\prescript{g}{}{A}^+_{++}$, $\prescript{c}{}{\overline{A}}^+_{--}$, $\prescript{c}{}{\overline{A}}^+_{++}$)
& ($\prescript{g}{}{A}^+_{-+}$, $\prescript{g}{}{A}^+_{+-}$, $\prescript{c}{}{\overline{A}}^-_{+-}$, $\prescript{c}{}{\overline{A}}^-_{-+}$)
\\[3pt]
\hline
\multirow{4}{*}{5}
& \multirow{4}{*}{C$^\dag$}
& ($\prescript{g}{}{U}^+_+$, $\prescript{g}{}{U}^-_-$)$t_1$ 
& ($\prescript{g}{}{\overline{U}}^+_+$, $\prescript{g}{}{\overline{U}}^-_-$)$t_3$ 
& ($\prescript{g}{}{U}^+_-$, $\prescript{g}{}{U}^-_+$)$t_3$
& ($\prescript{g}{}{\overline{U}}^+_-$, $\prescript{g}{}{\overline{U}}^-_+$)$t_1$
\\ 
& &($\prescript{g}{}{\overline{A}}^-_+$, $\prescript{g}{}{\overline{A}}^-_-$) 
&($\prescript{g}{}{A}^-_+$, $\prescript{g}{}{A}^-_-$) 
&($\prescript{g}{}{\overline{A}}^+_-$, $\prescript{g}{}{\overline{A}}^+_+$)
&($\prescript{g}{}{A}^+_-$, $\prescript{g}{}{A}^+_+$)
\\[3pt] 
\cline{3-6}
& 
& ($\prescript{c}{}{U}^+_+$, $\prescript{c}{}{U}^-_-$)$t_0$ 
& ($\prescript{c}{}{\overline{U}}^+_+$, $\prescript{c}{}{\overline{U}}^-_-$)$t_0$ 
& ($\prescript{c}{}{U}^+_-$, $\prescript{c}{}{U}^-_+$)$t_2$ 
& ($\prescript{c}{}{\overline{U}}^+_-$, $\prescript{c}{}{\overline{U}}^-_+$)$t_2$ 
\\ 
& &($\prescript{c}{}{\overline{A}}^-_+$, $\prescript{c}{}{\overline{A}}^-_-$) 
&($\prescript{c}{}{A}^-_+$, $\prescript{c}{}{A}^-_-$) 
&($\prescript{c}{}{\overline{A}}^+_-$, $\prescript{c}{}{\overline{A}}^+_+$) 
&($\prescript{c}{}{A}^+_-$, $\prescript{c}{}{A}^+_+$)
\\[3pt]
\hline
\multirow{4}{*}{6} 
&\multirow{4}{*}{CI$^\dag$} 
&($\prescript{g}{}{U}^+_{++}$, $\prescript{g}{}{U}^-_{--}$, $\prescript{c}{}{\overline{U}}^-_{++}$, $\prescript{c}{}{\overline{U}}^+_{--}$)$t_1$
&($\prescript{g}{}{U}^+_{-+}$, $\prescript{g}{}{U}^-_{+-}$, $\prescript{c}{}{\overline{U}}^+_{-+}$, $\prescript{c}{}{\overline{U}}^-_{+-}$)$t_0$
&($\prescript{g}{}{U}^+_{--}$, $\prescript{g}{}{U}^-_{++}$, $\prescript{c}{}{\overline{U}}^-_{--}$, $\prescript{c}{}{\overline{U}}^+_{++}$)$t_3$ 
&($\prescript{g}{}{U}^+_{+-}$, $\prescript{g}{}{U}^-_{-+}$, $\prescript{c}{}{\overline{U}}^+_{+-}$, $\prescript{c}{}{\overline{U}}^-_{-+}$)$t_2$
\\ 
& &($\prescript{c}{}{A}^+_{++}$, $\prescript{c}{}{A}^+_{--}$, $\prescript{g}{}{\overline{A}}^-_{++}$, $\prescript{g}{}{\overline{A}}^-_{--}$)
& ($\prescript{c}{}{A}^-_{-+}$, $\prescript{c}{}{A}^-_{+-}$, $\prescript{g}{}{\overline{A}}^-_{-+}$, $\prescript{g}{}{\overline{A}}^-_{+-}$)
& ($\prescript{c}{}{A}^-_{--}$, $\prescript{c}{}{A}^-_{++}$, $\prescript{g}{}{\overline{A}}^+_{--}$, $\prescript{g}{}{\overline{A}}^+_{++}$)
& ($\prescript{c}{}{A}^+_{+-}$, $\prescript{c}{}{A}^+_{-+}$, $\prescript{g}{}{\overline{A}}^+_{+-}$, $\prescript{g}{}{\overline{A}}^+_{-+}$)
\\[3pt]
\cline{3-6}
&
&($\prescript{c}{}{U}^+_{++}$, $\prescript{c}{}{U}^-_{--}$, $\prescript{g}{}{\overline{U}}^-_{++}$, $\prescript{g}{}{\overline{U}}^+_{--}$)$t_0$
&($\prescript{c}{}{U}^+_{-+}$, $\prescript{c}{}{U}^-_{+-}$, $\prescript{g}{}{\overline{U}}^+_{-+}$, $\prescript{g}{}{\overline{U}}^-_{+-}$)$t_3$
&($\prescript{c}{}{U}^+_{--}$, $\prescript{c}{}{U}^-_{++}$, $\prescript{g}{}{\overline{U}}^-_{--}$, $\prescript{g}{}{\overline{U}}^+_{++}$)$t_2$ 
&($\prescript{c}{}{U}^+_{+-}$, $\prescript{c}{}{U}^-_{-+}$, $\prescript{g}{}{\overline{U}}^+_{+-}$, $\prescript{g}{}{\overline{U}}^-_{-+}$)$t_1$
\\ 
& &($\prescript{g}{}{A}^+_{++}$, $\prescript{g}{}{A}^+_{--}$, $\prescript{c}{}{\overline{A}}^-_{++}$, $\prescript{c}{}{\overline{A}}^-_{--}$)
& ($\prescript{g}{}{A}^-_{-+}$, $\prescript{g}{}{A}^-_{+-}$, $\prescript{c}{}{\overline{A}}^-_{-+}$, $\prescript{c}{}{\overline{A}}^-_{+-}$)
& ($\prescript{g}{}{A}^-_{--}$, $\prescript{g}{}{A}^-_{++}$, $\prescript{c}{}{\overline{A}}^+_{--}$, $\prescript{c}{}{\overline{A}}^+_{++}$)
& ($\prescript{g}{}{A}^+_{+-}$, $\prescript{g}{}{A}^+_{-+}$, $\prescript{c}{}{\overline{A}}^+_{+-}$, $\prescript{c}{}{\overline{A}}^+_{-+}$)
\\[3pt]
\hline \hline
\end{tabular}}
\label{Tab:Symmetry_type AZ dag}
\end{center}
\end{table*}

Following similar steps as previous section, we derive equivalent symmetries for real AZ$^\dag$ classes in this section. All Eqs. in this section are also valid upon exchange $g \leftrightarrow c$. Throughout this paper, we take the convention that TRS$^\dag$ operator $\mathcal{C}_+$ and PHS$^\dag$ operator $\mathcal{T}_-$ commute: $[\mathcal{C}_+,\mathcal{T}_-]=0$.

For class AI$^\dag$ and AII$^\dag$, we have the following equivalent symmetries

\textit{AI$^\dag$ and AII$^\dag$} ($\mathcal{C}_{+}$):
\begin{align}
    &\prescript{g}{}{U}_{\eta_C}^{\epsilon_U}=\textrm{i}  \prescript{g}{}{U}_{-\eta_C}^{-\epsilon_U}=\mathcal{C}_{+} \prescript{c}{}{A}_{\eta_C}^{\eta_C \epsilon_C \epsilon_U}=\textrm{i}\mathcal{C}_{+}\prescript{c}{}{A}_{-\eta_C}^{\eta_C \epsilon_C \epsilon_U}\nonumber\\
    &\prescript{g}{}{\overline{U}}_{\eta_C}^{\epsilon_U}=\textrm{i}  \prescript{g}{}{\overline{U}}_{-\eta_C}^{-\epsilon_U}=\mathcal{C}_{+} \prescript{c}{}{\overline{A}}_{\eta_C}^{\eta_C \epsilon_C \epsilon_U}=\textrm{i}\mathcal{C}_{+}\prescript{c}{}{\overline{A}}_{-\eta_C}^{\eta_C \epsilon_C \epsilon_U},
\end{align}
where the subscript indicates the commutation relationship between spatial symmetries and TRS$^\dag$ operator $\mathcal{C}_+$.

For class D$^\dag$ and C$^\dag$, we have

\textit{D$^\dag$ and C$^\dag$} ($\mathcal{T}_-$):
\begin{align}
    &\prescript{g}{}{U}_{\eta_T}^{\epsilon_U}=\textrm{i}  \prescript{g}{}{U}_{-\eta_T}^{-\epsilon_U}=\mathcal{T}_{-} \prescript{g}{}{\overline{A}}_{\eta_T}^{\eta_T \epsilon_T \epsilon_U}=\textrm{i}\mathcal{T}_{-}\prescript{g}{}{\overline{A}}_{-\eta_T}^{\eta_T \epsilon_T \epsilon_U} \nonumber\\
    &\prescript{g}{}{\overline{U}}_{\eta_T}^{\epsilon_U}=\textrm{i}  \prescript{g}{}{\overline{U}}_{-\eta_T}^{-\epsilon_U}=\mathcal{T}_{-} \prescript{g}{}{A}_{\eta_T}^{\eta_T \epsilon_T \epsilon_U}=\textrm{i}\mathcal{T}_{-}\prescript{g}{}{A}_{-\eta_T}^{\eta_T \epsilon_T \epsilon_U},
\end{align}
where the subscript indicates the commutation relationship of PHS$^\dag$ operator $\mathcal{T}_-$.

Finally, for class BDI$^\dag$, DIII$^\dag$, CII$^\dag$, and CI$^\dag$, we have

\textit{BDI$^\dag$, DIII$^\dag$, CII$^\dag$, and CI$^\dag$}:
\begin{align}
    \prescript{g}{}{U}_{\eta_C,\eta_T}^{\epsilon_U}&=\textrm{i}  \prescript{g}{}{U}_{-\eta_C,-\eta_T}^{-\epsilon_U}=\mathcal{C}_{+} \prescript{c}{}{A}_{\eta_C,\eta_T}^{\eta_C \epsilon_C \epsilon_U}=\textrm{i}\mathcal{C}_{+} \prescript{c}{}{A}_{-\eta_C,-\eta_T}^{\eta_C \epsilon_C \epsilon_U}\nonumber\\
    &=\mathcal{T}_{-} \prescript{g}{}{\overline{A}}_{\eta_C,\eta_T}^{\eta_T \epsilon_T \epsilon_U}=\textrm{i} \mathcal{T}_{-} \prescript{g}{}{\overline{A}}_{-\eta_C,-\eta_T}^{\eta_T \epsilon_T \epsilon_U}\nonumber
\end{align}
\begin{align}
  \prescript{g}{}{\overline{U}}_{\eta_C,\eta_T}^{\epsilon_U}&=\textrm{i}  \prescript{g}{}{\overline{U}}_{-\eta_C,-\eta_T}^{-\epsilon_U}=\mathcal{C}_{+} \prescript{c}{}{\overline{A}}_{\eta_C,\eta_T}^{\eta_C \epsilon_C \epsilon_U}=\textrm{i}\mathcal{C}_{+} \prescript{c}{}{\overline{A}}_{-\eta_C,-\eta_T}^{\eta_C \epsilon_C \epsilon_U}\nonumber\\
    &=\mathcal{T}_{-} \prescript{g}{}{A}_{\eta_C,\eta_T}^{\eta_T \epsilon_T \epsilon_U}=\textrm{i} \mathcal{T}_{-} \prescript{g}{}{A}_{-\eta_C,-\eta_T}^{\eta_T \epsilon_T \epsilon_U},
\end{align}
where the first (second) subscript indicates the commutation relationship between spatial symmetries and TRS$^\dag$ (PHS$^\dag$).

By going over all possible combinations of $\epsilon_U,\eta_T,\eta_C$, we summarize the equivalent symmetries in Table~\ref{Tab:Symmetry_type AZ dag}. Similar to real AZ classes, $\Sigma$ would unify certain spatial symmetries for AZ$^\dag$ class. The procedure to classify them is the same as AZ class. We label them by $t_0,t_1,t_2,t_3$ at the end of circular bracket in Table~\ref{Tab:Symmetry_type AZ dag}. 

In Appendix~\ref{Sec:DH AZ and AZ dag with spatial}, we show the following $K$ group relationship for AZ$^\dag$ classes with spatial symmetries
\begin{align}
    &K_{\mathbb{R}}^{\dag U/A}(s^\dag,t;d,d_{\parallel},D,D_{\parallel})\nonumber\\
    &=K_{\mathbb{R}}^{\dag U/A}(s^\dag-d+D,t+d_{\parallel}-D_{\parallel};0,0,0,0),
    \label{eq:K group AZ dag}
\end{align}
where the superscript $U/A$ indicates this relationship works for both unitary and antiunitary spatial symmetries. Furthermore,
\begin{align}
    &K_{\mathbb{R}}^{\dag U/A}(s^\dag,t=0) = \pi_0 (\mathcal{R}_{s^\dag})\times\pi_0 (\mathcal{R}_{s^\dag})\nonumber\\
    &K_{\mathbb{R}}^{\dag U/A}(s^\dag,t=1)=\pi_0 (\mathcal{R}_{s^\dag-1})\nonumber\\
    &K_{\mathbb{R}}^{\dag U/A}(s^\dag,t=2)=\pi_0(\mathcal{C}_{s^\dag})\nonumber\\
    &K_{\mathbb{R}}^{\dag U/A}(s^\dag,t=3)=\pi_0 (\mathcal{R}_{s^\dag+1}),
\end{align}
which we show in Appendix~\ref{Appendix:classifying space} using Clifford Algebra.
For both Eqs, $s^\dag$ is defined mod$8$ while $t$ is defined mod$4$

\section{Periodic table of spatial symmetries for point-gapped Hamiltonian}\label{sec:periodic table}
Having introduced the $K$ group of AZ and AZ$^\dag$ class in the presence of spatial symmetries, we are now ready to present their periodic table under spatial symmetries. 

Each AZ (AZ$^\dag$) class is labeled by $s$ ($s^\dag$). The dimensional hierarchy for these parameters is controlled by the difference between spatial dimensions $d$ and the dimension $D$ of the $D$-sphere that surrounds the defects: $\delta=d-D$. The periodic table is given by ranging $\delta$ from $0$ to $7$ for each class (See Table~\ref{tab: AZ}). In the previous section, for each class, we give them a finer labeling $t$ in the presence of spatial symmetries, where $t$ is defined mod 2 for A and AIII or mod 4 for the rest of the classes. As we can see from $K$ group relationship we introduce at the end of each sub-section [Eqs.~\eqref{eq:K group complex AZ unitary},~\eqref{eq:K group real AZ}, and~\eqref{eq:K group AZ dag}], the dimensional hierarchy for $t$ is controlled by $\delta_{\parallel}=d_{\parallel}-D_{\parallel}$, where $d_{\parallel}$ ($D_{\parallel}$) is the number of spatial dimension (dimension of $D$-sphere that surrounds the defect) that is being flipped under spatial symmetry. In the following, we introduce periodic tables by ranging $\delta_{\parallel}$ from $0$ to $3$. This would give us four distinct tables. We summarize the result in table~\ref{tab:Periodic table delta_parallel=0},~\ref{tab:Periodic table delta_parallel=1},~\ref{tab:Periodic table delta_parallel=2}, and~\ref{tab:Periodic table delta_parallel=3} for $\delta_{\parallel}=0,1,2,3$, respectively. In the following sections, we use specific examples to illustrate these topological invariants.

\begin{table*}[]
    \centering
    \caption{Periodic table for $\delta_{\parallel}=0$.}
    \resizebox{\textwidth}{!}{
    \begin{tabular}{ccccccccccc}\hline \hline
         Symmetry& Class &  Classifying space & $\delta=0$ & $1$ & $2$ & $3$ & $4$ & $5$ & $6$ & $7$ \\ \hline
        \multirow{2}{*}{$t_1$} & A & $\mathcal{C}_1 \times \mathcal{C}_1$ & $0$ & $\mathbb{Z} \oplus \mathbb{Z}$ & $0$ & $\mathbb{Z}\oplus \mathbb{Z}$ & $0$ & $\mathbb{Z}\oplus \mathbb{Z}$ & $0$ & $\mathbb{Z}\oplus \mathbb{Z}$ \\
        & AIII & $\mathcal{C}_0 \times \mathcal{C}_0$ & $\mathbb{Z} \oplus \mathbb{Z}$ & $0$ & $\mathbb{Z}\oplus \mathbb{Z}$ & $0$ & $\mathbb{Z}\oplus \mathbb{Z}$ & $0$ & $\mathbb{Z}\oplus \mathbb{Z}$ & $0$\\ \hline

        \multirow{2}{*}{$t_1$} & A & $\mathcal{C}_0$ & $\mathbb{Z}$ & $0$ & $\mathbb{Z}$ & $0$ & $\mathbb{Z}$ & $0$ & $\mathbb{Z}$ & $0$ \\
        & AIII & $\mathcal{C}_1$ & $0$ & $\mathbb{Z}$ & $0$ & $\mathbb{Z}$ & $0$ & $\mathbb{Z}$ & $0$ & $\mathbb{Z}$\\ \hline
         
         \multirow{8}{*}{$t_0$} & AI,D$^\dag$ & $\mathcal{R}_1 \times \mathcal{R}_1$ & $\mathbb{Z}_2\oplus \mathbb{Z}_2$ & $\mathbb{Z}\oplus \mathbb{Z}$ & $0$ & $0$ & $0$ & $2\mathbb{Z}\oplus 2\mathbb{Z}$ & $0$ & $\mathbb{Z}_2\oplus \mathbb{Z}_2$\\
         & BDI,DIII$^\dag$ & $\mathcal{R}_2 \times \mathcal{R}_2$ & $\mathbb{Z}_2\oplus \mathbb{Z}_2$ & $\mathbb{Z}_2\oplus \mathbb{Z}_2$ & $\mathbb{Z} \oplus \mathbb{Z}$ & $0$ & $0$ & $0$ & $2\mathbb{Z}\oplus 2\mathbb{Z}$& $0$ \\
         & D,AII$^\dag$ & $\mathcal{R}_3 \times \mathcal{R}_3$ & $0$ & $\mathbb{Z}_2\oplus \mathbb{Z}_2$ & $\mathbb{Z}_2\oplus \mathbb{Z}_2$ & $\mathbb{Z} \oplus \mathbb{Z}$ & $0$ & $0$ & $0$ & $2\mathbb{Z}\oplus 2\mathbb{Z}$\\
         & DIII,CII$^\dag$ & $\mathcal{R}_4 \times \mathcal{R}_4$ & $2\mathbb{Z}\oplus 2\mathbb{Z}$ & $0$ & $\mathbb{Z}_2\oplus \mathbb{Z}_2$ & $\mathbb{Z}_2\oplus \mathbb{Z}_2$ & $\mathbb{Z} \oplus \mathbb{Z}$ & $0$ & $0$ & $0$\\
         & AII,C$^\dag$ & $\mathcal{R}_5 \times \mathcal{R}_5$ & $0$ & $2\mathbb{Z}\oplus 2\mathbb{Z}$ & $0$ & $\mathbb{Z}_2\oplus \mathbb{Z}_2$ & $\mathbb{Z}_2\oplus \mathbb{Z}_2$ & $\mathbb{Z} \oplus \mathbb{Z}$ & $0$ & $0$ \\
         & CII,CI$^\dag$ & $\mathcal{R}_6 \times \mathcal{R}_6$ & $0$ & $0$ & $2\mathbb{Z}\oplus 2\mathbb{Z}$ & $0$ & $\mathbb{Z}_2\oplus \mathbb{Z}_2$ & $\mathbb{Z}_2\oplus \mathbb{Z}_2$ & $\mathbb{Z} \oplus \mathbb{Z}$ & $0$ \\
         & C,AI$^\dag$ & $\mathcal{R}_7 \times \mathcal{R}_7$ & $0$ & $0$ & $0$ & $2\mathbb{Z}\oplus 2\mathbb{Z}$ & $0$ & $\mathbb{Z}_2\oplus \mathbb{Z}_2$ & $\mathbb{Z}_2\oplus \mathbb{Z}_2$ & $\mathbb{Z} \oplus \mathbb{Z}$ \\
         & CI,BDI$^\dag$ & $\mathcal{R}_0 \times \mathcal{R}_0$ & $\mathbb{Z} \oplus \mathbb{Z}$ & $0$ & $0$ & $0$ & $2\mathbb{Z}\oplus 2\mathbb{Z}$ & $0$ & $\mathbb{Z}_2\oplus \mathbb{Z}_2$ & $\mathbb{Z}_2\oplus \mathbb{Z}_2$\\ \hline

         \multirow{8}{*}{$t_1$}& AI,AI$^\dag$ & $\mathcal{R}_0$ & $\mathbb{Z}$ & $0$ & $0$ & $0$ & $2\mathbb{Z}$ & $0$ & $\mathbb{Z}_2$ & $\mathbb{Z}_2$\\
         & BDI,BDI$^\dag$ & $\mathcal{R}_1$ & $\mathbb{Z}_2$ & $\mathbb{Z}$ & $0$ & $0$ & $0$ & $2\mathbb{Z}$ & $0$ & $\mathbb{Z}_2$ \\
         & D,D$^\dag$ & $\mathcal{R}_2$ & $\mathbb{Z}_2$ & $\mathbb{Z}_2$ & $\mathbb{Z}$ & $0$ & $0$ & $0$ & $2\mathbb{Z}$ & $0$ \\
         & DIII,DIII$^\dag$ & $\mathcal{R}_3$ & $0$ & $\mathbb{Z}_2$ & $\mathbb{Z}_2$ & $\mathbb{Z}$ & $0$ & $0$ & $0$ & $2\mathbb{Z}$ \\
         & AII,AII$^\dag$ & $\mathcal{R}_4$ & $2\mathbb{Z}$ & $0$ & $\mathbb{Z}_2$ & $\mathbb{Z}_2$ & $\mathbb{Z}$ & $0$ & $0$ & $0$\\
         & CII,CII$^\dag$ & $\mathcal{R}_5$ & $0$ & $2\mathbb{Z}$ & $0$ & $\mathbb{Z}_2$ & $\mathbb{Z}_2$ & $\mathbb{Z}$ & $0$ & $0$\\
         & C,C$^\dag$ & $\mathcal{R}_6$ & $0$ & $0$ & $2\mathbb{Z}$ & $0$ & $\mathbb{Z}_2$ & $\mathbb{Z}_2$ & $\mathbb{Z}$ & $0$ \\
         & CI,CI$^\dag$ & $\mathcal{R}_7$ & $0$ & $0$ & $0$ & $2\mathbb{Z}$ & $0$ & $\mathbb{Z}_2$ & $\mathbb{Z}_2$ & $\mathbb{Z}$ \\ \hline

         \multirow{2}{*}{$t_2$} & AI,D,AII,C,AI$^\dag$,D$^\dag$,AII$^\dag$,C$^\dag$ & $\mathcal{C}_1$ & $0$ & $\mathbb{Z}$ & $0$ & $\mathbb{Z}$ & $0$ & 
        $\mathbb{Z}$ & $0$ & $\mathbb{Z}$\\
        & BDI,DIII,CII,CI,BDI$^\dag$,DIII$^\dag$,CII$^\dag$,CI$^\dag$ & $\mathcal{C}_0$ & $\mathbb{Z}$ & $0$ & $\mathbb{Z}$ & $0$ & $\mathbb{Z}$ & $0$ & $\mathbb{Z}$ & $0$ \\ \hline

        \multirow{8}{*}{$t_3$} & AI,AII$^\dag$ & $\mathcal{R}_2$ & $\mathbb{Z}_2$ & $\mathbb{Z}_2$ & $\mathbb{Z}$ & $0$ & $0$ & $0$ & $2\mathbb{Z}$ & $0$\\
        & BDI,CII$^\dag$ & $\mathcal{R}_3$ & $0$ & $\mathbb{Z}_2$ & $\mathbb{Z}_2$ & $\mathbb{Z}$ & $0$ & $0$ & $0$ & $2\mathbb{Z}$ \\
        & D,C$^\dag$ & $\mathcal{R}_4$ & $2\mathbb{Z}$ & $0$ & $\mathbb{Z}_2$ & $\mathbb{Z}_2$ & $\mathbb{Z}$ & $0$ & $0$ & $0$\\
        & DIII,CI$^\dag$ & $\mathcal{R}_5$ & $0$ & $2\mathbb{Z}$ & $0$ & $\mathbb{Z}_2$ & $\mathbb{Z}_2$ & $\mathbb{Z}$ & $0$ & $0$\\
        & AII,AI$^\dag$ & $\mathcal{R}_6$ & $0$ & $0$ & $2\mathbb{Z}$ & $0$ & $\mathbb{Z}_2$ & $\mathbb{Z}_2$ & $\mathbb{Z}$ & $0$ \\
        & CII,BDI$^\dag$ & $\mathcal{R}_7$ & $0$ & $0$ & $0$ & $2\mathbb{Z}$ & $0$ & $\mathbb{Z}_2$ & $\mathbb{Z}_2$ & $\mathbb{Z}$\\
        & C,D$^\dag$ &  $\mathcal{R}_0$ & $\mathbb{Z}$ & $0$ & $0$ & $0$ & $2\mathbb{Z}$ & $0$ & $\mathbb{Z}_2$ & $\mathbb{Z}_2$\\
        & CI,DIII$^\dag$ & $\mathcal{R}_1$ & $\mathbb{Z}_2$ & $\mathbb{Z}$ & $0$ & $0$ & $0$ & $2\mathbb{Z}$ & $0$ & $\mathbb{Z}_2$\\ \hline\hline
    \end{tabular}}
    \label{tab:Periodic table delta_parallel=0}
\end{table*}

\begin{table*}[]
    \centering
    \caption{Periodic table for $\delta_{\parallel}=1$.}
    \resizebox{\textwidth}{!}{
    \begin{tabular}{ccccccccccc}\hline \hline
         Symmetry& Class &  Classifying space & $\delta=0$ & $1$ & $2$ & $3$ & $4$ & $5$ & $6$ & $7$ \\ \hline
        \multirow{2}{*}{$t_1$} & A & $\mathcal{C}_0 \times \mathcal{C}_0$ & $\mathbb{Z} \oplus \mathbb{Z}$ & $0$ & $\mathbb{Z}\oplus \mathbb{Z}$ & $0$ & $\mathbb{Z}\oplus \mathbb{Z}$ & $0$ & $\mathbb{Z}\oplus \mathbb{Z}$ & $0$ \\
        & AIII & $\mathcal{C}_1 \times \mathcal{C}_1$ & $0$ & $\mathbb{Z} \oplus \mathbb{Z}$ & $0$ & $\mathbb{Z}\oplus \mathbb{Z}$ & $0$ & $\mathbb{Z}\oplus \mathbb{Z}$ & $0$ & $\mathbb{Z}\oplus \mathbb{Z}$ \\ \hline

        \multirow{2}{*}{$t_1$} & A & $\mathcal{C}_1$ & $0$ & $\mathbb{Z}$ & $0$ & $\mathbb{Z}$ & $0$ & $\mathbb{Z}$ & $0$ & $\mathbb{Z}$ \\
        & AIII & $\mathcal{C}_0$ & $\mathbb{Z}$ & $0$ & $\mathbb{Z}$ & $0$ & $\mathbb{Z}$ & $0$ & $\mathbb{Z}$ & $0$\\ \hline
         
         \multirow{8}{*}{
         \begin{tabular}{c}
              AZ:$t_0$\\
              AZ$^\dag$: $t_0$
         \end{tabular}} & AI,D$^\dag$ & $\mathcal{R}_2$ & $\mathbb{Z}_2$ & $\mathbb{Z}_2$ & $\mathbb{Z}$ & $0$ & $0$ & $0$ & $2\mathbb{Z}$ & $0$\\
         & BDI,DIII$^\dag$ & $\mathcal{R}_3$ & $0$ & $\mathbb{Z}_2$ & $\mathbb{Z}_2$ & $\mathbb{Z}$ & $0$ & $0$ & $0$ & $2\mathbb{Z}$ \\
         & D,AII$^\dag$ & $\mathcal{R}_4$ & $2\mathbb{Z}$ & $0$ & $\mathbb{Z}_2$ & $\mathbb{Z}_2$ & $\mathbb{Z}$ & $0$ & $0$ & $0$\\
         & DIII,CII$^\dag$ & $\mathcal{R}_5$& $0$ & $2\mathbb{Z}$ & $0$ & $\mathbb{Z}_2$ & $\mathbb{Z}_2$ & $\mathbb{Z}$ & $0$ & $0$\\
         & AII,C$^\dag$ & $\mathcal{R}_6$ & $0$ & $0$ & $2\mathbb{Z}$ & $0$ & $\mathbb{Z}_2$ & $\mathbb{Z}_2$ & $\mathbb{Z}$ & $0$ \\
         & CII,CI$^\dag$ & $\mathcal{R}_7$ & $0$ & $0$ & $0$ & $2\mathbb{Z}$ & $0$ & $\mathbb{Z}_2$ & $\mathbb{Z}_2$ & $\mathbb{Z}$ \\
         & C,AI$^\dag$ & $\mathcal{R}_0$ & $\mathbb{Z}$ & $0$ & $0$ & $0$ & $2\mathbb{Z}$ & $0$ & $\mathbb{Z}_2$ & $\mathbb{Z}_2$ \\
         & CI,BDI$^\dag$ & $\mathcal{R}_1$& $\mathbb{Z}_2$ & $\mathbb{Z}$ & $0$ & $0$ & $0$ & $2\mathbb{Z}$ & $0$ & $\mathbb{Z}_2$\\ \hline

         \multirow{8}{*}{
         \begin{tabular}{c}
            AZ:$t_1$  \\
              AZ$^\dag$:$t_3$
         \end{tabular}
          }& AI,D$^\dag$  & $\mathcal{R}_1 \times \mathcal{R}_1$ & $\mathbb{Z}_2\oplus \mathbb{Z}_2$ & $\mathbb{Z}\oplus \mathbb{Z}$ & $0$ & $0$ & $0$ & $2\mathbb{Z}\oplus 2\mathbb{Z}$ & $0$ & $\mathbb{Z}_2\oplus \mathbb{Z}_2$\\
         & BDI,DIII$^\dag$ & $\mathcal{R}_2 \times \mathcal{R}_2$ & $\mathbb{Z}_2\oplus \mathbb{Z}_2$ & $\mathbb{Z}_2\oplus \mathbb{Z}_2$ & $\mathbb{Z} \oplus \mathbb{Z}$ & $0$ & $0$ & $0$ & $2\mathbb{Z}\oplus 2\mathbb{Z}$& $0$ \\
         & D,AII$^\dag$ & $\mathcal{R}_3 \times \mathcal{R}_3$ & $0$ & $\mathbb{Z}_2\oplus \mathbb{Z}_2$ & $\mathbb{Z}_2\oplus \mathbb{Z}_2$ & $\mathbb{Z} \oplus \mathbb{Z}$ & $0$ & $0$ & $0$ & $2\mathbb{Z}\oplus 2\mathbb{Z}$\\
         & DIII,CII$^\dag$ & $\mathcal{R}_4 \times \mathcal{R}_4$ & $2\mathbb{Z}\oplus 2\mathbb{Z}$ & $0$ & $\mathbb{Z}_2\oplus \mathbb{Z}_2$ & $\mathbb{Z}_2\oplus \mathbb{Z}_2$ & $\mathbb{Z} \oplus \mathbb{Z}$ & $0$ & $0$ & $0$\\
         & AII,C$^\dag$ & $\mathcal{R}_5 \times \mathcal{R}_5$ & $0$ & $2\mathbb{Z}\oplus 2\mathbb{Z}$ & $0$ & $\mathbb{Z}_2\oplus \mathbb{Z}_2$ & $\mathbb{Z}_2\oplus \mathbb{Z}_2$ & $\mathbb{Z} \oplus \mathbb{Z}$ & $0$ & $0$\\
         & CII,CI$^\dag$ & $\mathcal{R}_6 \times \mathcal{R}_6$ & $0$ & $0$ & $2\mathbb{Z}\oplus 2\mathbb{Z}$ & $0$ & $\mathbb{Z}_2\oplus \mathbb{Z}_2$ & $\mathbb{Z}_2\oplus \mathbb{Z}_2$ & $\mathbb{Z} \oplus \mathbb{Z}$ & $0$\\
         & C,AI$^\dag$ & $\mathcal{R}_7 \times \mathcal{R}_7$  & $0$ & $0$ & $0$ & $2\mathbb{Z}\oplus 2\mathbb{Z}$ & $0$ & $\mathbb{Z}_2\oplus \mathbb{Z}_2$ & $\mathbb{Z}_2\oplus \mathbb{Z}_2$ & $\mathbb{Z} \oplus \mathbb{Z}$  \\
         & CI,BDI$^\dag$ & $\mathcal{R}_0 \times \mathcal{R}_0$ & $\mathbb{Z} \oplus \mathbb{Z}$ & $0$ & $0$ & $0$ & $2\mathbb{Z}\oplus 2\mathbb{Z}$ & $0$ & $\mathbb{Z}_2\oplus \mathbb{Z}_2$ & $\mathbb{Z}_2\oplus \mathbb{Z}_2$\\ \hline

        \multirow{8}{*}{
        \begin{tabular}{c}
             AZ:$t_2$\\
             AZ$^\dag$:$t_2$
        \end{tabular}
        } & AI,D$^\dag$ & $\mathcal{R}_0$ & $\mathbb{Z}$ & $0$ & $0$ & $0$ & $2\mathbb{Z}$ & $0$ & $\mathbb{Z}_2$ & $\mathbb{Z}_2$\\
        & BDI,DIII$^\dag$ & $\mathcal{R}_1$ & $\mathbb{Z}_2$ & $\mathbb{Z}$ & $0$ & $0$ & $0$ & $2\mathbb{Z}$ & $0$ & $\mathbb{Z}_2$ \\
        & D,AII$^\dag$ & $\mathcal{R}_2$ & $\mathbb{Z}_2$ & $\mathbb{Z}_2$ & $\mathbb{Z}$ & $0$ & $0$ & $0$ & $2\mathbb{Z}$ & $0$\\
        & DIII,CII$^\dag$& $\mathcal{R}_3$ & $0$ & $\mathbb{Z}_2$ & $\mathbb{Z}_2$ & $\mathbb{Z}$ & $0$ & $0$ & $0$ & $2\mathbb{Z}$ \\
        & AII,C$^\dag$ & $\mathcal{R}_4$ & $2\mathbb{Z}$ & $0$ & $\mathbb{Z}_2$ & $\mathbb{Z}_2$ & $\mathbb{Z}$ & $0$ & $0$ & $0$ \\
        & CII,CI$^\dag$ & $\mathcal{R}_5$ & $0$ & $2\mathbb{Z}$ & $0$ & $\mathbb{Z}_2$ & $\mathbb{Z}_2$ & $\mathbb{Z}$ & $0$ & $0$\\
        & C,AI$^\dag$ & $\mathcal{R}_6$ & $0$ & $0$ & $2\mathbb{Z}$ & $0$ & $\mathbb{Z}_2$ & $\mathbb{Z}_2$ & $\mathbb{Z}$ & $0$ \\
        & CI,BDI$^\dag$ & $\mathcal{R}_7$ & $0$ & $0$ & $0$ & $2\mathbb{Z}$ & $0$ & $\mathbb{Z}_2$ & $\mathbb{Z}_2$ & $\mathbb{Z}$\\ \hline
        
         \multirow{2}{*}{\begin{tabular}{c}
             AZ:$t_3$\\
             AZ$^\dag$:$t_1$
        \end{tabular}} & AI,D,AII,C,AI$^\dag$,D$^\dag$,AII$^\dag$,C$^\dag$ & $\mathcal{C}_1$ & $0$ & $\mathbb{Z}$ & $0$ & $\mathbb{Z}$ & $0$ & 
        $\mathbb{Z}$ & $0$ & $\mathbb{Z}$\\
        & BDI,DIII,CII,CI,BDI$^\dag$,DIII$^\dag$,CII$^\dag$,CI$^\dag$ & $\mathcal{C}_0$ & $\mathbb{Z}$ & $0$ & $\mathbb{Z}$ & $0$ & $\mathbb{Z}$ & $0$ & $\mathbb{Z}$ & $0$ \\ \hline

 \hline\hline
    \end{tabular}}
    \label{tab:Periodic table delta_parallel=1}
\end{table*}

\begin{table*}[]
    \centering
    \caption{Periodic table for $\delta_{\parallel}=2$.}
    \resizebox{\textwidth}{!}{
    \begin{tabular}{ccccccccccc}\hline \hline
         Symmetry& Class &  Classifying space & $\delta=0$ & $1$ & $2$ & $3$ & $4$ & $5$ & $6$ & $7$ \\ \hline
                
         \multirow{2}{*}{\begin{tabular}{c}
             AZ:$t_0$\\
             AZ$^\dag$:$t_0$
        \end{tabular}} & AI,D,AII,C,AI$^\dag$,D$^\dag$,AII$^\dag$,C$^\dag$ & $\mathcal{C}_1$ & $0$ & $\mathbb{Z}$ & $0$ & $\mathbb{Z}$ & $0$ & 
        $\mathbb{Z}$ & $0$ & $\mathbb{Z}$\\
        & BDI,DIII,CII,CI,BDI$^\dag$,DIII$^\dag$,CII$^\dag$,CI$^\dag$ & $\mathcal{C}_0$ & $\mathbb{Z}$ & $0$ & $\mathbb{Z}$ & $0$ & $\mathbb{Z}$ & $0$ & $\mathbb{Z}$ & $0$ \\ \hline
        
         \multirow{8}{*}{
         \begin{tabular}{c}
              AZ:$t_1$\\
              AZ$^\dag$: $t_3$
         \end{tabular}} & AI,D$^\dag$ & $\mathcal{R}_2$ & $\mathbb{Z}_2$ & $\mathbb{Z}_2$ & $\mathbb{Z}$ & $0$ & $0$ & $0$ & $2\mathbb{Z}$ & $0$\\
         & BDI,DIII$^\dag$ & $\mathcal{R}_3$ & $0$ & $\mathbb{Z}_2$ & $\mathbb{Z}_2$ & $\mathbb{Z}$ & $0$ & $0$ & $0$ & $2\mathbb{Z}$ \\
         & D,AII$^\dag$ & $\mathcal{R}_4$ & $2\mathbb{Z}$ & $0$ & $\mathbb{Z}_2$ & $\mathbb{Z}_2$ & $\mathbb{Z}$ & $0$ & $0$ & $0$\\
         & DIII,CII$^\dag$ & $\mathcal{R}_5$& $0$ & $2\mathbb{Z}$ & $0$ & $\mathbb{Z}_2$ & $\mathbb{Z}_2$ & $\mathbb{Z}$ & $0$ & $0$\\
         & AII,C$^\dag$ & $\mathcal{R}_6$ & $0$ & $0$ & $2\mathbb{Z}$ & $0$ & $\mathbb{Z}_2$ & $\mathbb{Z}_2$ & $\mathbb{Z}$ & $0$ \\
         & CII,CI$^\dag$ & $\mathcal{R}_7$ & $0$ & $0$ & $0$ & $2\mathbb{Z}$ & $0$ & $\mathbb{Z}_2$ & $\mathbb{Z}_2$ & $\mathbb{Z}$ \\
         & C,AI$^\dag$ & $\mathcal{R}_0$ & $\mathbb{Z}$ & $0$ & $0$ & $0$ & $2\mathbb{Z}$ & $0$ & $\mathbb{Z}_2$ & $\mathbb{Z}_2$ \\
         & CI,BDI$^\dag$ & $\mathcal{R}_1$& $\mathbb{Z}_2$ & $\mathbb{Z}$ & $0$ & $0$ & $0$ & $2\mathbb{Z}$ & $0$ & $\mathbb{Z}_2$\\ \hline

         \multirow{8}{*}{
         \begin{tabular}{c}
            AZ:$t_2$  \\
              AZ$^\dag$:$t_2$
         \end{tabular}
          }& AI,D$^\dag$  & $\mathcal{R}_1 \times \mathcal{R}_1$ & $\mathbb{Z}_2\oplus \mathbb{Z}_2$ & $\mathbb{Z}\oplus \mathbb{Z}$ & $0$ & $0$ & $0$ & $2\mathbb{Z}\oplus 2\mathbb{Z}$ & $0$ & $\mathbb{Z}_2\oplus \mathbb{Z}_2$\\
         & BDI,DIII$^\dag$ & $\mathcal{R}_2 \times \mathcal{R}_2$ & $\mathbb{Z}_2\oplus \mathbb{Z}_2$ & $\mathbb{Z}_2\oplus \mathbb{Z}_2$ & $\mathbb{Z} \oplus \mathbb{Z}$ & $0$ & $0$ & $0$ & $2\mathbb{Z}\oplus 2\mathbb{Z}$& $0$ \\
         & D,AII$^\dag$ & $\mathcal{R}_3 \times \mathcal{R}_3$ & $0$ & $\mathbb{Z}_2\oplus \mathbb{Z}_2$ & $\mathbb{Z}_2\oplus \mathbb{Z}_2$ & $\mathbb{Z} \oplus \mathbb{Z}$ & $0$ & $0$ & $0$ & $2\mathbb{Z}\oplus 2\mathbb{Z}$\\
         & DIII,CII$^\dag$ & $\mathcal{R}_4 \times \mathcal{R}_4$ & $2\mathbb{Z}\oplus 2\mathbb{Z}$ & $0$ & $\mathbb{Z}_2\oplus \mathbb{Z}_2$ & $\mathbb{Z}_2\oplus \mathbb{Z}_2$ & $\mathbb{Z} \oplus \mathbb{Z}$ & $0$ & $0$ & $0$\\
         & AII,C$^\dag$ & $\mathcal{R}_5 \times \mathcal{R}_5$ & $0$ & $2\mathbb{Z}\oplus 2\mathbb{Z}$ & $0$ & $\mathbb{Z}_2\oplus \mathbb{Z}_2$ & $\mathbb{Z}_2\oplus \mathbb{Z}_2$ & $\mathbb{Z} \oplus \mathbb{Z}$ & $0$ & $0$\\
         & CII,CI$^\dag$ & $\mathcal{R}_6 \times \mathcal{R}_6$ & $0$ & $0$ & $2\mathbb{Z}\oplus 2\mathbb{Z}$ & $0$ & $\mathbb{Z}_2\oplus \mathbb{Z}_2$ & $\mathbb{Z}_2\oplus \mathbb{Z}_2$ & $\mathbb{Z} \oplus \mathbb{Z}$ & $0$\\
         & C,AI$^\dag$ & $\mathcal{R}_7 \times \mathcal{R}_7$  & $0$ & $0$ & $0$ & $2\mathbb{Z}\oplus 2\mathbb{Z}$ & $0$ & $\mathbb{Z}_2\oplus \mathbb{Z}_2$ & $\mathbb{Z}_2\oplus \mathbb{Z}_2$ & $\mathbb{Z} \oplus \mathbb{Z}$  \\
         & CI,BDI$^\dag$ & $\mathcal{R}_0 \times \mathcal{R}_0$ & $\mathbb{Z} \oplus \mathbb{Z}$ & $0$ & $0$ & $0$ & $2\mathbb{Z}\oplus 2\mathbb{Z}$ & $0$ & $\mathbb{Z}_2\oplus \mathbb{Z}_2$ & $\mathbb{Z}_2\oplus \mathbb{Z}_2$\\ \hline

        \multirow{8}{*}{
        \begin{tabular}{c}
             AZ:$t_3$\\
             AZ$^\dag$:$t_1$
        \end{tabular}
        } & AI,D$^\dag$ & $\mathcal{R}_0$ & $\mathbb{Z}$ & $0$ & $0$ & $0$ & $2\mathbb{Z}$ & $0$ & $\mathbb{Z}_2$ & $\mathbb{Z}_2$\\
        & BDI,DIII$^\dag$ & $\mathcal{R}_1$ & $\mathbb{Z}_2$ & $\mathbb{Z}$ & $0$ & $0$ & $0$ & $2\mathbb{Z}$ & $0$ & $\mathbb{Z}_2$ \\
        & D,AII$^\dag$ & $\mathcal{R}_2$ & $\mathbb{Z}_2$ & $\mathbb{Z}_2$ & $\mathbb{Z}$ & $0$ & $0$ & $0$ & $2\mathbb{Z}$ & $0$\\
        & DIII,CII$^\dag$& $\mathcal{R}_3$ & $0$ & $\mathbb{Z}_2$ & $\mathbb{Z}_2$ & $\mathbb{Z}$ & $0$ & $0$ & $0$ & $2\mathbb{Z}$ \\
        & AII,C$^\dag$ & $\mathcal{R}_4$ & $2\mathbb{Z}$ & $0$ & $\mathbb{Z}_2$ & $\mathbb{Z}_2$ & $\mathbb{Z}$ & $0$ & $0$ & $0$ \\
        & CII,CI$^\dag$ & $\mathcal{R}_5$ & $0$ & $2\mathbb{Z}$ & $0$ & $\mathbb{Z}_2$ & $\mathbb{Z}_2$ & $\mathbb{Z}$ & $0$ & $0$\\
        & C,AI$^\dag$ & $\mathcal{R}_6$ & $0$ & $0$ & $2\mathbb{Z}$ & $0$ & $\mathbb{Z}_2$ & $\mathbb{Z}_2$ & $\mathbb{Z}$ & $0$ \\
        & CI,BDI$^\dag$ & $\mathcal{R}_7$ & $0$ & $0$ & $0$ & $2\mathbb{Z}$ & $0$ & $\mathbb{Z}_2$ & $\mathbb{Z}_2$ & $\mathbb{Z}$\\ \hline

 \hline\hline
    \end{tabular}}
    \label{tab:Periodic table delta_parallel=2}
\end{table*}

\begin{table*}[]
    \centering
    \caption{Periodic table for $\delta_{\parallel}=3$.}
    \resizebox{\textwidth}{!}{
    \begin{tabular}{ccccccccccc}\hline \hline
         Symmetry& Class &  Classifying space & $\delta=0$ & $1$ & $2$ & $3$ & $4$ & $5$ & $6$ & $7$ \\ \hline

        \multirow{8}{*}{
        \begin{tabular}{c}
             AZ:$t_0$\\
             AZ$^\dag$:$t_0$
        \end{tabular}
        } & AI,D$^\dag$ & $\mathcal{R}_0$ & $\mathbb{Z}$ & $0$ & $0$ & $0$ & $2\mathbb{Z}$ & $0$ & $\mathbb{Z}_2$ & $\mathbb{Z}_2$\\
        & BDI,DIII$^\dag$ & $\mathcal{R}_1$ & $\mathbb{Z}_2$ & $\mathbb{Z}$ & $0$ & $0$ & $0$ & $2\mathbb{Z}$ & $0$ & $\mathbb{Z}_2$ \\
        & D,AII$^\dag$ & $\mathcal{R}_2$ & $\mathbb{Z}_2$ & $\mathbb{Z}_2$ & $\mathbb{Z}$ & $0$ & $0$ & $0$ & $2\mathbb{Z}$ & $0$\\
        & DIII,CII$^\dag$& $\mathcal{R}_3$ & $0$ & $\mathbb{Z}_2$ & $\mathbb{Z}_2$ & $\mathbb{Z}$ & $0$ & $0$ & $0$ & $2\mathbb{Z}$ \\
        & AII,C$^\dag$ & $\mathcal{R}_4$ & $2\mathbb{Z}$ & $0$ & $\mathbb{Z}_2$ & $\mathbb{Z}_2$ & $\mathbb{Z}$ & $0$ & $0$ & $0$ \\
        & CII,CI$^\dag$ & $\mathcal{R}_5$ & $0$ & $2\mathbb{Z}$ & $0$ & $\mathbb{Z}_2$ & $\mathbb{Z}_2$ & $\mathbb{Z}$ & $0$ & $0$\\
        & C,AI$^\dag$ & $\mathcal{R}_6$ & $0$ & $0$ & $2\mathbb{Z}$ & $0$ & $\mathbb{Z}_2$ & $\mathbb{Z}_2$ & $\mathbb{Z}$ & $0$ \\
        & CI,BDI$^\dag$ & $\mathcal{R}_7$ & $0$ & $0$ & $0$ & $2\mathbb{Z}$ & $0$ & $\mathbb{Z}_2$ & $\mathbb{Z}_2$ & $\mathbb{Z}$\\ \hline

        \multirow{2}{*}{\begin{tabular}{c}
             AZ:$t_1$\\
             AZ$^\dag$:$t_3$
        \end{tabular}} & AI,D,AII,C,AI$^\dag$,D$^\dag$,AII$^\dag$,C$^\dag$ & $\mathcal{C}_1$ & $0$ & $\mathbb{Z}$ & $0$ & $\mathbb{Z}$ & $0$ & 
        $\mathbb{Z}$ & $0$ & $\mathbb{Z}$\\
        & BDI,DIII,CII,CI,BDI$^\dag$,DIII$^\dag$,CII$^\dag$,CI$^\dag$ & $\mathcal{C}_0$ & $\mathbb{Z}$ & $0$ & $\mathbb{Z}$ & $0$ & $\mathbb{Z}$ & $0$ & $\mathbb{Z}$ & $0$ \\ \hline

         \multirow{8}{*}{
         \begin{tabular}{c}
              AZ:$t_2$\\
              AZ$^\dag$: $t_2$
         \end{tabular}} & AI,D$^\dag$ & $\mathcal{R}_2$ & $\mathbb{Z}_2$ & $\mathbb{Z}_2$ & $\mathbb{Z}$ & $0$ & $0$ & $0$ & $2\mathbb{Z}$ & $0$\\
         & BDI,DIII$^\dag$ & $\mathcal{R}_3$ & $0$ & $\mathbb{Z}_2$ & $\mathbb{Z}_2$ & $\mathbb{Z}$ & $0$ & $0$ & $0$ & $2\mathbb{Z}$ \\
         & D,AII$^\dag$ & $\mathcal{R}_4$ & $2\mathbb{Z}$ & $0$ & $\mathbb{Z}_2$ & $\mathbb{Z}_2$ & $\mathbb{Z}$ & $0$ & $0$ & $0$\\
         & DIII,CII$^\dag$ & $\mathcal{R}_5$& $0$ & $2\mathbb{Z}$ & $0$ & $\mathbb{Z}_2$ & $\mathbb{Z}_2$ & $\mathbb{Z}$ & $0$ & $0$\\
         & AII,C$^\dag$ & $\mathcal{R}_6$ & $0$ & $0$ & $2\mathbb{Z}$ & $0$ & $\mathbb{Z}_2$ & $\mathbb{Z}_2$ & $\mathbb{Z}$ & $0$ \\
         & CII,CI$^\dag$ & $\mathcal{R}_7$ & $0$ & $0$ & $0$ & $2\mathbb{Z}$ & $0$ & $\mathbb{Z}_2$ & $\mathbb{Z}_2$ & $\mathbb{Z}$ \\
         & C,AI$^\dag$ & $\mathcal{R}_0$ & $\mathbb{Z}$ & $0$ & $0$ & $0$ & $2\mathbb{Z}$ & $0$ & $\mathbb{Z}_2$ & $\mathbb{Z}_2$ \\
         & CI,BDI$^\dag$ & $\mathcal{R}_1$& $\mathbb{Z}_2$ & $\mathbb{Z}$ & $0$ & $0$ & $0$ & $2\mathbb{Z}$ & $0$ & $\mathbb{Z}_2$\\ \hline

        \multirow{8}{*}{
         \begin{tabular}{c}
            AZ:$t_3$  \\
              AZ$^\dag$:$t_1$
         \end{tabular}
          }& AI,D$^\dag$  & $\mathcal{R}_1 \times \mathcal{R}_1$ & $\mathbb{Z}_2\oplus \mathbb{Z}_2$ & $\mathbb{Z}\oplus \mathbb{Z}$ & $0$ & $0$ & $0$ & $2\mathbb{Z}\oplus 2\mathbb{Z}$ & $0$ & $\mathbb{Z}_2\oplus \mathbb{Z}_2$\\
         & BDI,DIII$^\dag$ & $\mathcal{R}_2 \times \mathcal{R}_2$ & $\mathbb{Z}_2\oplus \mathbb{Z}_2$ & $\mathbb{Z}_2\oplus \mathbb{Z}_2$ & $\mathbb{Z} \oplus \mathbb{Z}$ & $0$ & $0$ & $0$ & $2\mathbb{Z}\oplus 2\mathbb{Z}$& $0$ \\
         & D,AII$^\dag$ & $\mathcal{R}_3 \times \mathcal{R}_3$ & $0$ & $\mathbb{Z}_2\oplus \mathbb{Z}_2$ & $\mathbb{Z}_2\oplus \mathbb{Z}_2$ & $\mathbb{Z} \oplus \mathbb{Z}$ & $0$ & $0$ & $0$ & $2\mathbb{Z}\oplus 2\mathbb{Z}$\\
         & DIII,CII$^\dag$ & $\mathcal{R}_4 \times \mathcal{R}_4$ & $2\mathbb{Z}\oplus 2\mathbb{Z}$ & $0$ & $\mathbb{Z}_2\oplus \mathbb{Z}_2$ & $\mathbb{Z}_2\oplus \mathbb{Z}_2$ & $\mathbb{Z} \oplus \mathbb{Z}$ & $0$ & $0$ & $0$\\
         & AII,C$^\dag$ & $\mathcal{R}_5 \times \mathcal{R}_5$ & $0$ & $2\mathbb{Z}\oplus 2\mathbb{Z}$ & $0$ & $\mathbb{Z}_2\oplus \mathbb{Z}_2$ & $\mathbb{Z}_2\oplus \mathbb{Z}_2$ & $\mathbb{Z} \oplus \mathbb{Z}$ & $0$ & $0$\\
         & CII,CI$^\dag$ & $\mathcal{R}_6 \times \mathcal{R}_6$ & $0$ & $0$ & $2\mathbb{Z}\oplus 2\mathbb{Z}$ & $0$ & $\mathbb{Z}_2\oplus \mathbb{Z}_2$ & $\mathbb{Z}_2\oplus \mathbb{Z}_2$ & $\mathbb{Z} \oplus \mathbb{Z}$ & $0$\\
         & C,AI$^\dag$ & $\mathcal{R}_7 \times \mathcal{R}_7$  & $0$ & $0$ & $0$ & $2\mathbb{Z}\oplus 2\mathbb{Z}$ & $0$ & $\mathbb{Z}_2\oplus \mathbb{Z}_2$ & $\mathbb{Z}_2\oplus \mathbb{Z}_2$ & $\mathbb{Z} \oplus \mathbb{Z}$  \\
         & CI,BDI$^\dag$ & $\mathcal{R}_0 \times \mathcal{R}_0$ & $\mathbb{Z} \oplus \mathbb{Z}$ & $0$ & $0$ & $0$ & $2\mathbb{Z}\oplus 2\mathbb{Z}$ & $0$ & $\mathbb{Z}_2\oplus \mathbb{Z}_2$ & $\mathbb{Z}_2\oplus \mathbb{Z}_2$\\ \hline

 \hline\hline
    \end{tabular}}
    \label{tab:Periodic table delta_parallel=3}
\end{table*}

\section{Examples}\label{Sec:examples}
\begin{figure}
    \centering
    \includegraphics[width=\columnwidth]{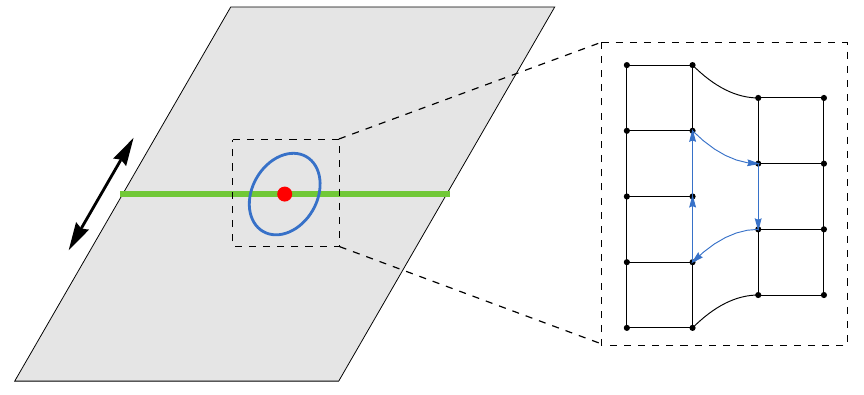}
    \caption{Point defect with reflection symmetry in $y$ direction. Red dot indicates the location of the defect. Blue circle is the $S^1$ sphere that surrounds the defect. Green line is the reflection symmetry line i.e. the lattice is reflection symmetric with respect to the green line. In this case, Burgers vector $\textbf{B}=\hat{y}$ which is shown by an additional upwards arrow after completing a loop surround the defect. }
    \label{fig:point Defect}
\end{figure}
In this section, we give several concrete examples of toy models that demonstrate how to use our periodic table introduced in the previous section. Before introducing concrete models, we first discuss several preliminary considerations.

To simplify our discussion and also to be aligned with current interest within the field of non-Hermitian defects, we focus on the case of point defect in 2D [Fig.~\ref{fig:point Defect} (a)] protected by reflection symmetry. In this case, $\delta_{\parallel}=0$ and the periodic table is table~\ref{tab:Periodic table delta_parallel=0}. 
Without loss of generality, we assume the model obeys reflection symmetry in $y$ direction. For a complete enumeration of defects for different $\delta_{\parallel}=0$, see Ref.~\cite{Shiozaki_2014}.

Physically, a point defect on lattice is obtained by removing a site and reintroducing hopping for sites after the defect (Fig.~\ref{fig:point Defect} (b)). For a point-gapped non-Hermitian system in 2D, point defect is known to generate skin effects, which is referred to as dislocation non-Hermitian skin effect in previous works~\cite{Schindler_2021,Bhargava_2021,Panigrahi_2022}. We see that, by this construction of point defects, the circle surrounds the point would give us a Burgers vector $\textbf{B}=\hat{y}$. In order to preserve periodic boundary condition (PBC), another defect with burgers vector pointing in the opposite direction must be introduced at the other end hoppings chains that are reintroduced. Since the defect is probing the topological property of sub-manifold $\textbf{B}\cdot \textbf{k}\mod{2\pi}=\pi$, we see that for $\textbf{B}=\hat{y}$, the topological property of the defect is given by the sub-manifold $k_y=\pi$. 

Finally, to give a sharp contrast between models with and without reflection symmetries, we are investigating the reflection-symmetry-protected phases. By comparing table~\ref{tab:Periodic table delta_parallel=0} and~\ref{tab: AZ}, we see that class AI$^\dag$ with $t_2$ class symmetry and class C with $t_2$ class symmetry are protected by the extra spatial symmetry with invariant $\mathbb{Z}$; Class DIII with $t_1$ class symmetry is also protected by the extra spatial symmetry with a $\mathbb{Z}_2$ invariant. If we break the relevant spatial symmetry (reflection in this case), the topological invariant will vanish since their original classification in Table~\ref{tab: AZ} is trivial. By doing so, there are two possible ways the system will change: (i) the modes that are originally localized around the defect will be delocalized, or (ii) a line gap will open, in which case the topological property is governed by line gap instead of point gap. In this sense, the defects mode for point-gapped spectrum is protected by the reflection symmetry. On the other hand, for class D$^\dag$ with $t_2$ class symmetry, if the spatial symmetry is broken, we expect the defect mode and point gap remain the same since D$^\dag$ has a $\mathbb{Z}$ invariant for $\delta=1$ without spatial symmetry. In this case, the reflection symmetry does not protect the point gap or the defect modes. 

In the following, we introduce four models in $\delta_{\parallel}=0$ (Table~\ref{tab:Periodic table delta_parallel=0}) that would realize scenarios that we described above. They belong to class AI$^\dag$, C, and D$^\dag$ with $t_2$ class reflection symmetry, and class DIII with $t_1$ class symmetry. 

\subsection{Point defect in AI$^\dag$ protected by reflection symmetry $\prescript{c}{}{U}^+_-$}
\begin{figure}
    \centering
    \includegraphics[width=\columnwidth]{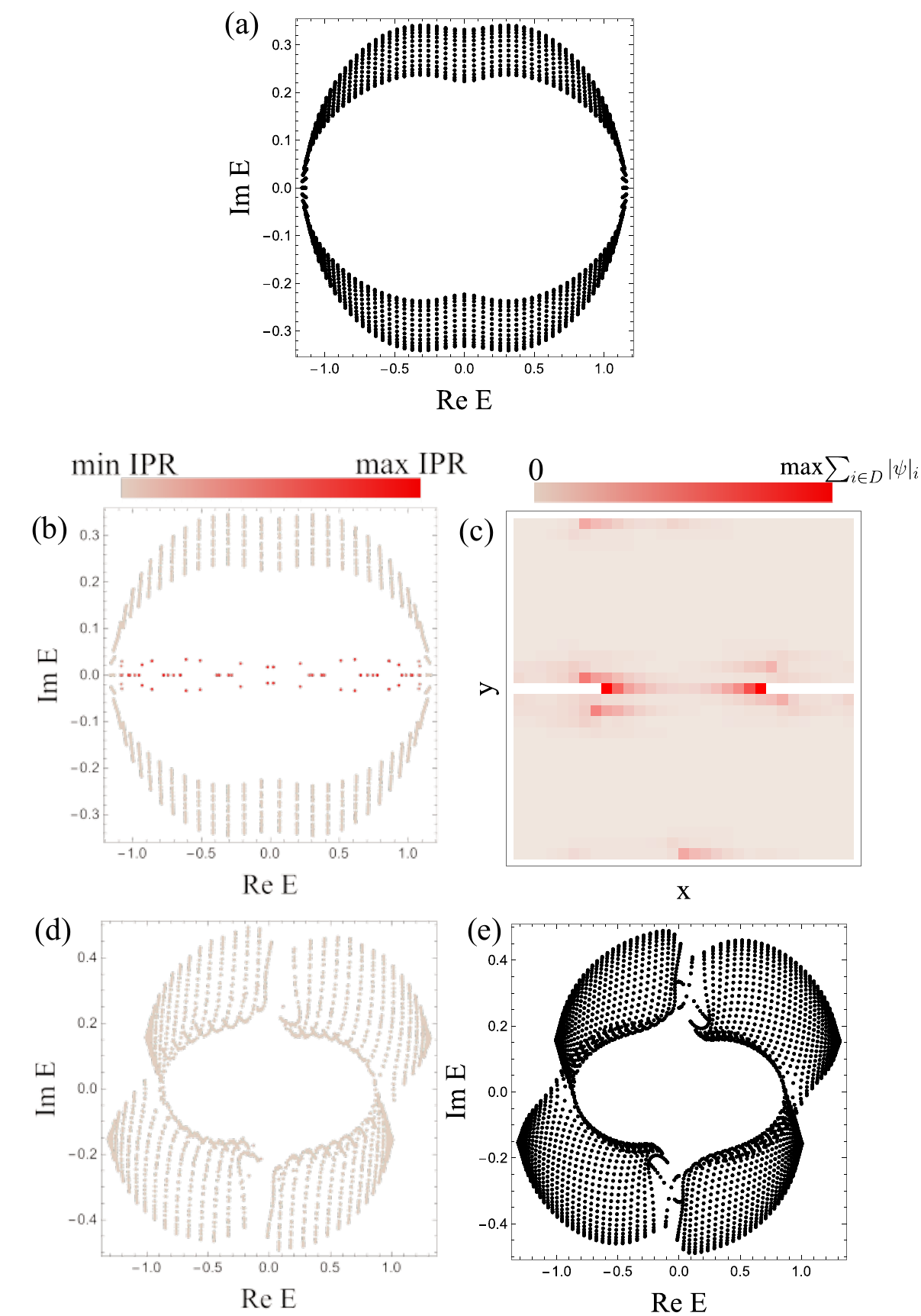}
    \caption{Spectrum and eigenstates distribution of model~\eqref{eq:AI dag heartsuit model}. The parameters are set at $t=0.5$, $g_1=0.4$, $g_2=1$, $\mu_1=0.3$, $\mu_2=0.2$. (a) Spectrum on complex plane plotted under full PBC. (b) Spectrum when two point defects are introduced in the $x$ direction. In-gap states that are localized at the defect appear. (c) Distribution of in-gap in real space. $D$ on the legend is referred to defect states inside the point gap [i.e. eigenstates of red dots in (b)]. White space indicates sites that are removed. (d), (e) Spectrum of ~\eqref{eq:AI dag heartsuit model} after Introducing a reflection-symmetry-breaking perturbation $0.15 (1+\textrm{i}) \sigma_x\tau_0$ for (d) with two point defects, for which the in-gap states vanish, and (e) full PBC.}
    \label{fig:AI dag model}
\end{figure}
We build a two-dimensional AI$^\dag$ model that obeys reflection symmetry of the type $\prescript{c}{}{U}^+_-$. Following the construction of $Z_2$ NHSE model~\cite{Okuma_2019}, we stack an HN model with its AI$^\dag$ pair in $x$ direction with symmetry-preserving coupling in $y$ direction. The resulting model is
\begin{align}
    &h_{\text{AI}^\dag}(k_x,k_y)=\begin{pmatrix}
        h_{HN}(k_x) & \Delta_1(k_y)\\
        \Delta_2(k_y) & h_{HN}^T(-k_x)
    \end{pmatrix}\nonumber\\
    &=t \sigma_0 \tau_x\nonumber\\
    &+\frac{1}{2}(\sigma_0 \tau_x-\textrm{i}\sigma_z \tau_y)g_1\cos k_x-\frac{\textrm{i}}{2}(\sigma_z\tau_x-\textrm{i}\sigma_0\tau_y)g_1 \sin k_x\nonumber\\
    &+\frac{1}{2}(\sigma_0 \tau_x+\textrm{i}\sigma_z \tau_y)g_2\cos k_x+
    \frac{\textrm{i}}{2}(\sigma_z\tau_x+\textrm{i}\sigma_0\tau_y)g_2 \sin k_x\nonumber\\
    &-\frac{\textrm{i}}{2}(\sigma_x-\textrm{i}\sigma_y)\tau_y\mu_1\sin k_y-\frac{\textrm{i}}{2}(\sigma_x+\textrm{i}\sigma_y)\tau_y\mu_2\sin k_y,
    \label{eq:AI dag heartsuit model}
\end{align}
where $h_{HN}(k_x)$ is the two-band HN model
\begin{equation}
    h_{HN}(k_x)=\begin{pmatrix}
        0 & t+g_2 \exp(\textrm{i}k_x)\\
        t+g_1 \exp(\textrm{i}k_x) & 0
    \end{pmatrix}.
    \label{eq:two bands HN model}
\end{equation}
$\Delta_{1(2)}(k_y)=-\frac{\textrm{i}}{2}(\sigma_x\mp \textrm{i}\sigma_y)\tau_y\mu_{1(2)}\sin k_y$ is the symmetry-preserving coupling in $y$ direction. $t,g_1,g_2,\mu_1,\mu_2$ are all real parameters. 
Model~\eqref{eq:AI dag heartsuit model} obeys TRS$^\dag$ with $\mathcal{C}_+=\sigma_x \tau_0\mathcal{K}$. It also obeys the reflection $\prescript{c}{}{U}^+_-=\sigma_z\tau_0$ such that $\prescript{c}{}{U}^+_- h_{\text{AI}^\dag}(k_x,k_y) \prescript{c}{}{U}^+_-=h_{\text{AI}^\dag}(k_x,-k_y)$. According to the classification table~\ref{Tab:Symmetry_type AZ dag} and~\ref{tab:Periodic table delta_parallel=0}, this model belongs to the subclass $t_2$ of AI$^\dag$. In the case of 2D with a point defect such that $\delta=2-1=1$, this topological phases has a $\mathbb{Z}$ invariant which we will later show is given by the mirror winding number. 

We plot the spectrum of model~\eqref{eq:AI dag heartsuit model} in Fig.~\ref{fig:AI dag model} (a) under full PBC. After introducing two defects for a HN chain in $x$ direction, in-gap states that are localized at the defect appear [Fig.~\ref{fig:AI dag model} (b) and (c)]. As mentioned at the beginning of the section, in this construction, the Burgers vector is given by $\textbf{B}=\pm \hat{y}$, which is probing the topological property of the sub-manifold $k_y=\pi$. At $k_y=\pi$, $\Delta_1(\pi)=\Delta_2(\pi)=0$, Eq.~\eqref{eq:AI dag heartsuit model} decoupled into two independent HN models with opposite winding numbers. We can diagonalize the resulting Hamiltonian into sectors that correspond to the eigenvalues $\pm1$ of $\prescript{c}{}{U}^+_-=\sigma_z\tau_0$. This was automatically done for model~\eqref{eq:AI dag heartsuit model}. We are now left with
\begin{equation}
    h_{\text{AI}^\dag}(k_x,\pi)=\begin{pmatrix}
        h_{HN}(k_x) & 0\\
        0 & h_{HN}^T(-k_x)
        \end{pmatrix}.
\end{equation}
For each sectors of the above Hamiltonian, we can define a winding number
\begin{align}
    &W_+=\frac{1}{2\pi \textrm{i}}\int_{k_x=-\pi}^{k_x=\pi} d k_x \frac{d}{d k_x}\log \det h_{HN}(k_x)\nonumber\\
    &W_-=\frac{1}{2\pi \textrm{i}}\int_{k_x=-\pi}^{k_x=\pi} d k_x \frac{d}{d k_x}\log \det h_{HN}^T(-k_x).
\end{align}
The $\mathbb{Z}$ invariant is given by the \emph{mirror winding number}~\cite{Yoshida_2020}
\begin{equation}
    W_m=\frac{W_+-W_-}{2}.
    \label{eq:mirror winding number}
\end{equation}
For model~\eqref{eq:AI dag heartsuit model}, we have $W_m=1$ since $W_{\pm}=\pm1$. In turn, this nontrivial mirror winding number gives rise to the localization of in-gap state around the defect. Notice that now the total winding number $W_{\text{tot}}=W_+ +W_-$ vanishes. Since $W_m$ is only well-defined under reflection symmetry while $W_{\text{tot}}$ is always well-defined, we see that the topological phase of model~\eqref{eq:AI dag heartsuit model} is protected by reflection symmetry. Mirror winding number is first proposed in Ref.~\cite{Yoshida_2020} for NHSE protected by mirror symmetry (which is a different name of reflection symmetry). The mirror winding number is defined in a similar way as mirror Chern number~\cite{Shiozaki_2014}. 

The protection of reflection symmetry can also be seen by noticing that in the classification without reflection symmetry (Table~\ref{tab: AZ}), class AI$^\dag$ is trivial in the case $\delta=1$. Therefore, by breaking reflection symmetry, we expect either the in-gap state to loss localization and become trivial or the system opens a line gap such that the classification changes. To show this explicitly, we introduce a reflection-symmetry-breaking term $\epsilon(1+\textrm{i}) \sigma_x \tau_0$. As shown in Fig.~\ref{fig:AI dag model} (d) and (e), the in-gap states vanish after the addition of a symmetry-breaking term for $\epsilon=0.15$. Notice that for smaller $\epsilon$, the in-gap states still exist. However, by increasing the value of $\epsilon$ the in-gap states would vanish \emph{without} going through a phase transition. Thus, in-gap states after the addition of reflection-symmetry-breaking term are not topological and can be gapped out by continuous deformation. 

\subsection{Point defect in C protected by reflection symmetry $\prescript{c}{}{U}^+_-$}
\begin{figure}
    \centering
    \includegraphics[width=\columnwidth]{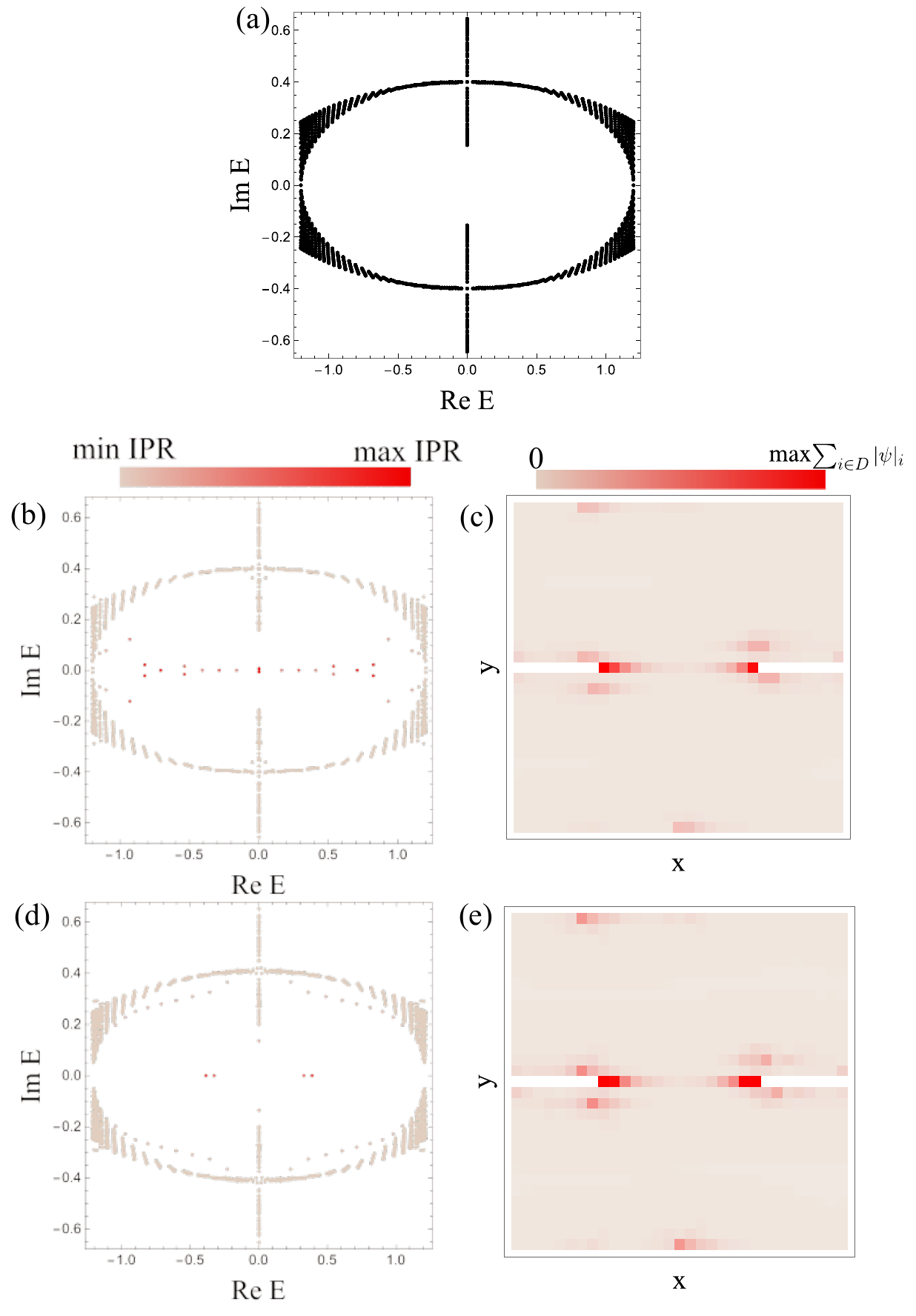}
    \caption{Spectrum and eigenstates distribution of the model~\eqref{eq:C with reflection}. Parameters are set at $t=0.5$, $g_1=0.3$, $g_2=1.3$, $\mu_1=0.2$, $\mu_2=0.3$. (a)-(c) are the same contents as Fig.~\ref{fig:AI dag model} but for model~\eqref{eq:C with reflection}. (d) The spectrum of the model~\eqref{eq:C with reflection} after the addition of reflection-symmetry-breaking term $0.12 \sigma_y \tau_0$ with two point defects. (e) Distribution of in-gap states after the addition of reflection-symmetry-breaking term.}
    \label{fig:Class C model}
\end{figure}
To strengthen our findings, we now turn our attention to class C. Consider the model
\begin{align}
    &h_{C}(k_x,k_y)=\begin{pmatrix}
        h_{HN}(k_x)& \Delta_1(k_y)\\
        \Delta_2(k_y)& -h_{HN}^T (-k_x)
    \end{pmatrix}\nonumber\\
    &=-t \sigma_z \tau_X\nonumber\\
    &+\frac{1}{2}(\sigma_z \tau_x-\textrm{i}\sigma_0 \tau_y)g_1\cos k_x-\frac{\textrm{i}}{2}(\sigma_0\tau_x+\textrm{i}\sigma_z\tau_y)g_1 \sin k_x\nonumber\\
    &+\frac{1}{2}(\sigma_z \tau_x+\textrm{i}\sigma_0\tau_y)g_2 \cos k_x+\frac{\textrm{i}}{2}(\sigma_0\tau_x-\textrm{i}\sigma_z\tau_y)g_2 \sin k_x\nonumber\\
    &-\frac{\textrm{i}}{2}(\sigma_x-\textrm{i}\sigma_y)\tau_y\mu_1\sin k_y-\frac{\textrm{i}}{2}(\sigma_x+\textrm{i}\sigma_y)\tau_y\mu_2\sin k_y,
    \label{eq:C with reflection}
\end{align}
where $h_{HN}$ is the two bands HN model defined in Eq.~\eqref{eq:two bands HN model}. $\Delta_1$ and $\Delta_2$ are the same as the previous section. This model obeys PHS with $\mathcal{C}_-=\textrm{i}\sigma_y \tau_0 \mathcal{K}$. It also obeys reflection symmetry in $y$ direction with $\prescript{c}{}{U}^+_-=\sigma_z\tau_0$ such that $\sigma_z h_{C}(k_x,k_y) \sigma_z=h_{C}(k_x,-k_y)$. According to the classification Table~\ref{Tab:Symmetry_type AZ} and~\ref{tab:Periodic table delta_parallel=0}, this model belongs to class C with $t_2$ symmetries, which possess a $\mathbb{Z}$ invariant. 

In the sub-manifold $k_y=\pi$, we have $\Delta_1(\pi)=\Delta_2(\pi)=0$. Then we can define the winding number $W_+$ ($W_-$) for $h_{HN}(k_x)$ ($-h_{HN}^T(-k_x)$), which lives in the $+1$($-1$) sector of reflection symmetry operator $\prescript{c}{}{U}^+_-=\sigma_z\tau_0$. Notice that since $h_{HN}$ is a $2$ by $2$ matrix, $\det[ -h_{HN}^T(-k_x)]=\det h_{HN}^T(-k_x)$. Then we see that $W_{\pm}=\pm 1$. Hence, similar to model~\eqref{eq:AI dag heartsuit model}, mirror winding number~\eqref{eq:mirror winding number} $W_m=1$ while the total winding number vanish. Thus, we see that the topological phase of model~\eqref{eq:C with reflection} is indeed protected by reflection symmetry. 

We plot the spectrum of the model~\eqref{eq:C with reflection} in Fig.~\ref{fig:Class C model} (a) under full periodic boundary conditions (PBC). After the introduction of point defect, in-gap states that are localized around the defect begin to show (Fig.~\ref{fig:Class C model} (b) and (c)). Since the classification of class C without spatial symmetries is trivial in dimension $\delta=1$ (See Table~\ref{tab: AZ}), model~\eqref{eq:C with reflection} is another example of reflection-symmetry protected topological phases. We now try to break reflection symmetry while preserving PHS by introducing an on-site perturbation $0.12 \sigma_y \tau_0$ to the Hamiltonian. The resulting spectrum and localization of in-gap states are shown in Fig.~\ref{fig:Class C model} (d) and (e). Notably, instead of vanishing defect-localized in-gap states like class AI$^\dag$, an imaginary-line gap is now open. The classification is now given by line gaps instead of point gaps. In this sense, the reflection symmetry is protecting point gaps for model~\eqref{eq:C with reflection}. 

\subsection{Point defect in D$^\dag$ with reflection symmetry $\prescript{c}{}{U}^+_-$}
\begin{figure}
    \centering
    \includegraphics[width=\columnwidth]{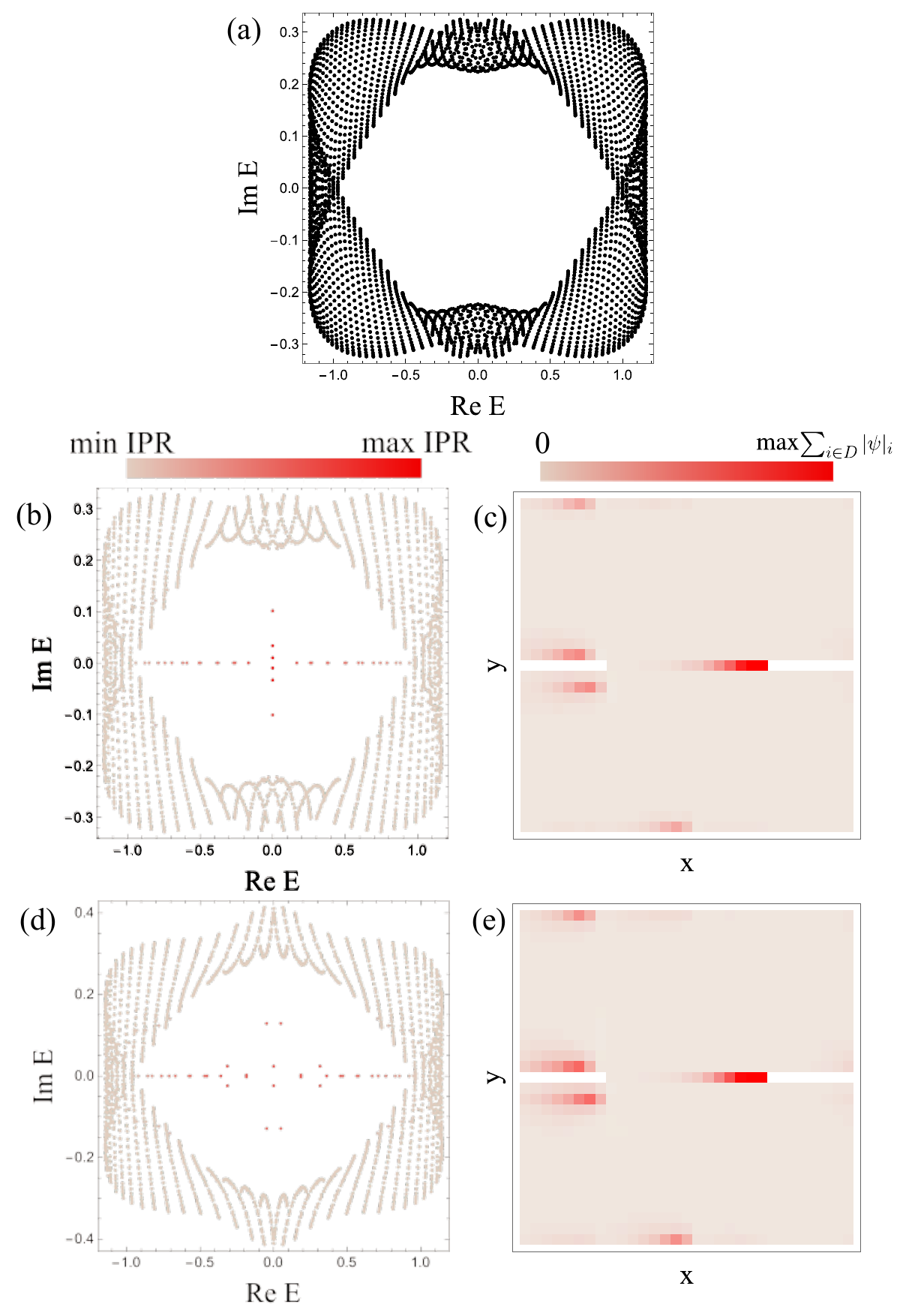}
    \caption{Spectrum and eigenstates distribution of model~\eqref{eq:D dag model}. Parameters are set at $t=0.5$, $g_1=0.4$, $g_2=1$, $\mu=0.2$. The contents of this plot are the same as Fig.~\ref{fig:Class C model}. }
    \label{fig:D dag model}
\end{figure}
Now we introduce a model that is not protected by reflection symmetry. First notice, for class D$^\dag$ in dimension $\delta=1$, it has the same topological invariant $\mathbb{Z}$ in the absence of spatial symmetry (Table~\ref{tab: AZ}) and with spatial symmetry in $t_2$ class (Table~\ref{tab:Periodic table delta_parallel=0}). We then consider the following model
\begin{align}
    &h_{D^\dag}(k_x,k_y)=\begin{pmatrix}
        h_{HN}(k_x) & \Delta(k_y)\\
        \Delta(k_y) & -h_{HN}^* (-k_x)
    \end{pmatrix}\nonumber\\
    &=t \sigma_z \tau_x\nonumber\\
    &+\frac{1}{2}(\sigma_z\tau_x-\textrm{i} \sigma_z\tau_y)g_1 \cos k_x-\frac{\textrm{i}}{2}(\sigma_z\tau_x-\textrm{i} \sigma_z\tau_y)g_1 \sin k_x\nonumber\\
    &+\frac{1}{2}(\sigma_z\tau_x+\textrm{i} \sigma_z\tau_y)g_2 \cos k_x+\frac{\textrm{i}}{2}(\sigma_z\tau_x+\textrm{i} \sigma_z\tau_y)g_2 \sin k_x\nonumber\\
    &-\frac{\textrm{i}}{2}(\sigma_x-\textrm{i}\sigma_y)\tau_y\mu\sin k_y-\frac{\textrm{i}}{2}(\sigma_x+\textrm{i}\sigma_y)\tau_y\mu\sin k_y,
    \label{eq:D dag model}
\end{align}
where $h_{HN}$ is two-bands HN model defined in Eq.~\eqref{eq:two bands HN model}; $\Delta(k_y)=-\frac{\textrm{i}}{2}\tau_y\mu\sin k_y$. This model obeys PHS$^\dag$ with $\mathcal{T}_-=\sigma_x \tau_0 \mathcal{K}$. It also obeys the reflection symmetry in the $y$ direction with operator $\prescript{c}{}{U}^+_-=\sigma_z\tau_0$. 

Unlike previous models~\eqref{eq:AI dag heartsuit model} and~\eqref{eq:C with reflection}, for model~\eqref{eq:D dag model}, $W_\pm$ defined on the sub-manifold $k_y=\pi$ for $h_{HN}(k_x)$ and $-h_{HN}^* (-k_x)$ have the same sign. More specifically, $W_+=W_-=1$. Now, the mirror winding number~\eqref{eq:mirror winding number} vanishes while the total winding number $W_{\text{tot}}=W_+ +W_-=2$ is nontrivial for $k_y=\pi$. This shows that the topological phase of model~\eqref{eq:D dag model} is not protected by reflection symmetry. 

We plot its spectrum under full PBC in Fig.~\ref{fig:D dag model} (a). Similar to the previous two models, we introduce defects with Burgers vector $\textbf{B}=\pm \hat{y}$. After the introduction of defects, as shown in Fig.~\ref{fig:D dag model} (b) and (c), we see that in-gap states that are localized around the defect appears. We now consider a reflection-symmetry-breaking term $0.2 \textrm{i}\sigma_1\tau_0$ added to Hamiltonian. Since the topological invariant is not protected by reflection symmetry $\prescript{c}{}{U}^+_-=\sigma_z\tau_0$, we see that both point gap and localization of in-gap states are not affected by the introduction of additional term [Fig.~\ref{fig:D dag model} (d) and (e)]. 

\subsection{Point defect in DIII protected by reflection symmetry $\prescript{c}{}{U}^+_{-+}$}
\begin{figure}
    \centering
    \includegraphics[width=\columnwidth]{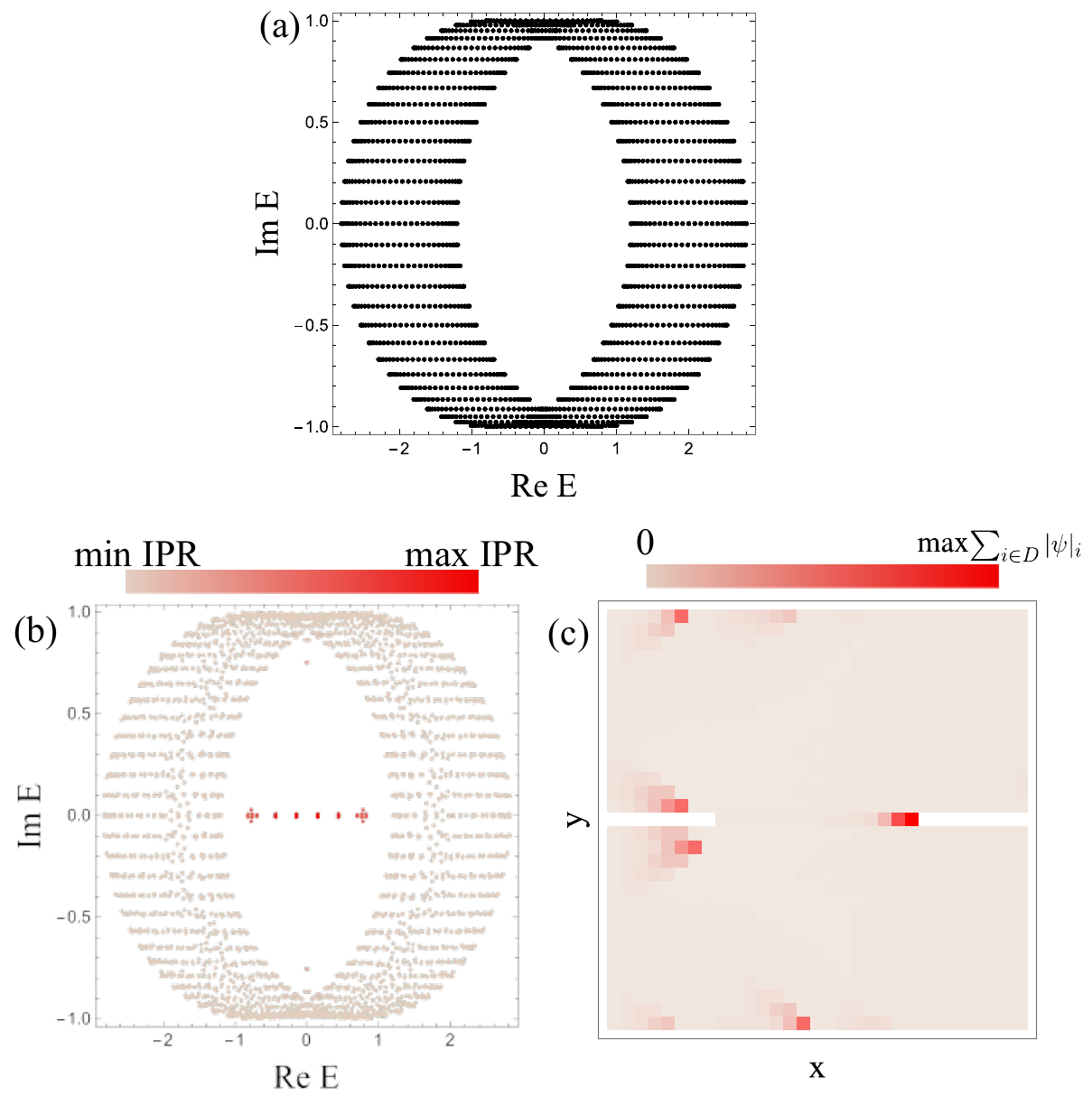}
    \caption{Spectrum and eigenstates distribution of model~\eqref{eq:model DIII} with pint defects. Parameters are chosen at $t=0.5,g=1.5,\mu_1=0.4,\mu_2=0.8$. The contents of the plot are the same as Fig.~\ref{fig:AI dag model}-\ref{fig:D dag model} (a)-(c). }
    \label{fig:DIII model}
\end{figure}
Finally, we consider the following model
\begin{align}
    &h_{DIII}(k_x,k_y)=\begin{pmatrix}
        h_1(k_x) & \Delta(k_y)\\
        \Delta(k_y) & h_1^*(-k_x)
    \end{pmatrix}\nonumber\\
    &=\sigma_x \tau_0 \mu_2 \sin k_y+\frac{1}{2}\sigma_z (\tau_x+\textrm{i}\tau_y)\mu_1\sin k_x\nonumber\\
    &+\sigma_0 \tau_z  (g+t)\cos k_x+\sigma_0\tau_0 \textrm{i}(g-t)\sin k_x,
    \label{eq:model DIII}
\end{align}
where $\Delta(k_y)=\sigma_x \tau_0 \mu_2 \sin k_y$, $h_1(k_x)=\frac{1}{2}(\tau_x+\textrm{i}\tau_y)\mu_1 \sin k_x+\tau_z  (g+t)\cos k_x+\tau_0 \textrm{i}(g-t)\sin k_x$.
Model~\eqref{eq:model DIII} obeys TRS with $\mathcal{T}_+=\textrm{i}\sigma_y \tau_0 \mathcal{K}$, PHS with $\mathcal{C}_-=\sigma_0\tau_x\mathcal{K}$, and reflection symmetry in $y$ direction with $\prescript{c}{}{U}^+_{-+}=\sigma_z\tau_0$. Since $\mathcal{T}_+^2=-1$, $\mathcal{C}_-^2=1$, this place model~\eqref{eq:model DIII} in class DIII with symmetry $t_1$. According to Table~\ref{tab:Periodic table delta_parallel=0}, the model~\eqref{eq:model DIII} has a $\mathbb{Z}_2$ invariant in the presence of a point defect. Furthermore, in the absence of spatial symmetries, class DIII has trivial classification for $\delta=1$ (Table~\ref{tab: AZ}). 

We plot its spectrum under full PBC in Fig.~\ref{fig:DIII model} (a). Upon the introduction of point defects, in-gap states localized at the defects begin to appear [Fig.~\ref{fig:DIII model} (b) and (c)]. In-gap states are protected by a $\mathbb{Z}_2$ topological invariant. Notice that in the sub-manifold $k_y=0$, reflection symmetry $\prescript{c}{}{U}^+_{-+}=\sigma_z\tau_0$ ensures that the coupling term would vanish. To see this, we consider the most general $4$ by $4$ Hamiltonian
\begin{equation}
    h_g(k_x,k_y)=\begin{pmatrix}
        h_+(k_x,k_y) & \Delta_1 (k_x,k_y)\\
        \Delta_2(k_x,k_y) & h_-(k_x,k_y)
    \end{pmatrix}
\end{equation}
The the reflection symmetry gives
\begin{align}
   &\sigma_z\tau_0 h_g(k_x,k_y) \sigma_z\tau_0=\begin{pmatrix}
        h_+(k_x,k_y) & -\Delta_1 (k_x,k_y)\\
        -\Delta_2(k_x,k_y) & h_-(k_x,k_y)
    \end{pmatrix}\nonumber\\
    &=\begin{pmatrix}
        h_+(k_x,-k_y) & \Delta_1 (k_x,-k_y)\\
        \Delta_2(k_x,-k_y) & h_-(k_x,-k_y)
    \end{pmatrix}=h_g(k_x,-k_y).
\end{align}
Then we see that $-\Delta_{1,2} (k_x,k_y)=\Delta_{1,2} (k_x,-k_y)$, which ensures that they must vanish at $k_y=\pi$. In this sense, the reflection symmetry ensures that $h_+$ and $h_-$ are decoupled at $k_y=\pi$. Return to model~\eqref{eq:model DIII}, we can define the $\mathbb{Z}_2$ topological invariant $\nu\in\{0,1\}$ for each $\pm 1$ sector of $\prescript{c}{}{U}^+_{-+}=\sigma_z\tau_0$ defined for sub-manifold $k_y=\pi$, where
\begin{align}
    (-1)^\nu=&\text{sgn}\{\frac{\text{Pf}[h_1(\pi)U_{\mathcal{C}}]}{\text{Pf}[h_1(0)U_{\mathcal{C}}]}\nonumber\\
    &\times\exp[-\frac{1}{2}\int_{k_x=0}^{k_x=\pi}d k_x \frac{d}{dk_x}\log \det (h_1(k_x)U_{\mathcal{C}})]\},
\end{align}
where we also defined $\mathcal{C}_-=U_{\mathcal{C}} \mathcal{K}$. A similar invariant can also be defined for $h_1^*(-k_x)$ Since the diagonal Hamiltonian $h_1$ and $h_1^*(-k_x)$ are \emph{decoupled} at $k_y=\pi$, even if they both have $\nu=1$, the overall Hamiltonian is still topological. 

To further confirm our claims that reflection symmetry protects $\mathbb{Z}_2$ topological invariant, we check the following two things: (i) By breaking reflection symmetries, the model~\eqref{eq:model DIII} would either loss localization of in-gap states or open a line gap; (ii) To demonstrate the $\mathbb{Z}_2$ nature of this model, we stack two copies of~\eqref{eq:model DIII} together with symmetry-preserving couplings. We expect the in-gap states would vanish in latter case. As we will see, these checking are indeed valid. 

\begin{figure}
    \centering
    \includegraphics[width=\columnwidth]{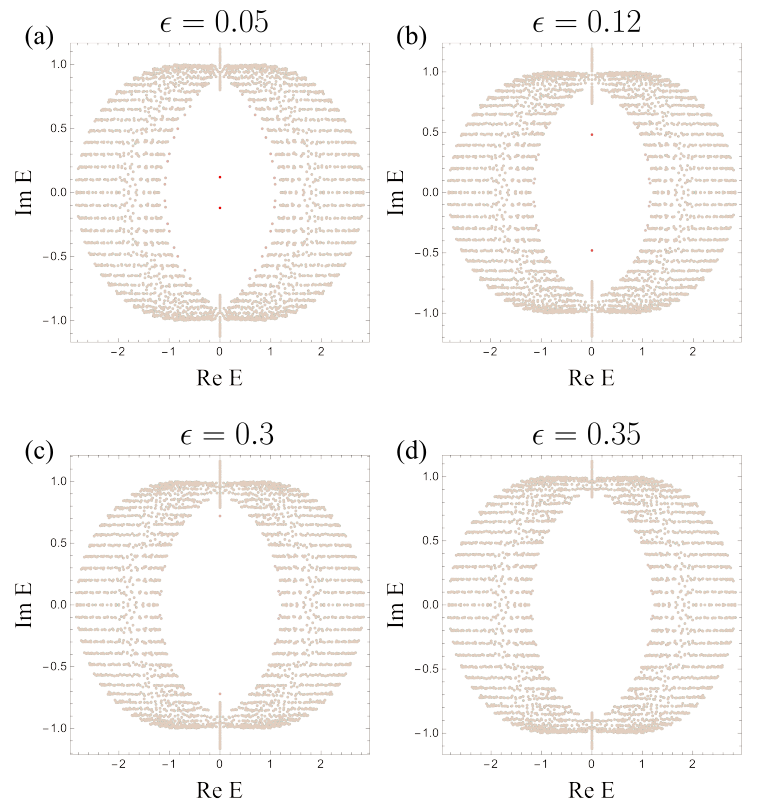}
    \caption{Adding on-site perturbation term $\epsilon \sigma_y\tau_y$ to Hamiltonian~\eqref{eq:model DIII}. The parameter $\epsilon$ increases from (a)-(d) which causes the in-gap states vanish without going through a phase transition. }
    \label{fig:Deformation DIII 1}
\end{figure}

To confirm (i), we consider the introduction of a reflection-symmetry-breaking term $\epsilon \sigma_y\tau_y$. We first tune $\epsilon=0.05$. As shown in Fig.~\ref{fig:Deformation DIII 1} (a), there still exist two in-gap states. However, these two states are not topological. Indeed, increasing the strength $\epsilon$ of added term, we see that these two states shift into the bulk \emph{without} closing the band gap of bulk bands [Fig.~\ref{fig:Deformation DIII 1} (b)-(d)]. This means that these two states can be gaped out through continuous deformation. Hence, we see that the reflection symmetry is protecting the topological in-gap states. 

\begin{figure*}
    \centering
    \includegraphics[width=\linewidth]{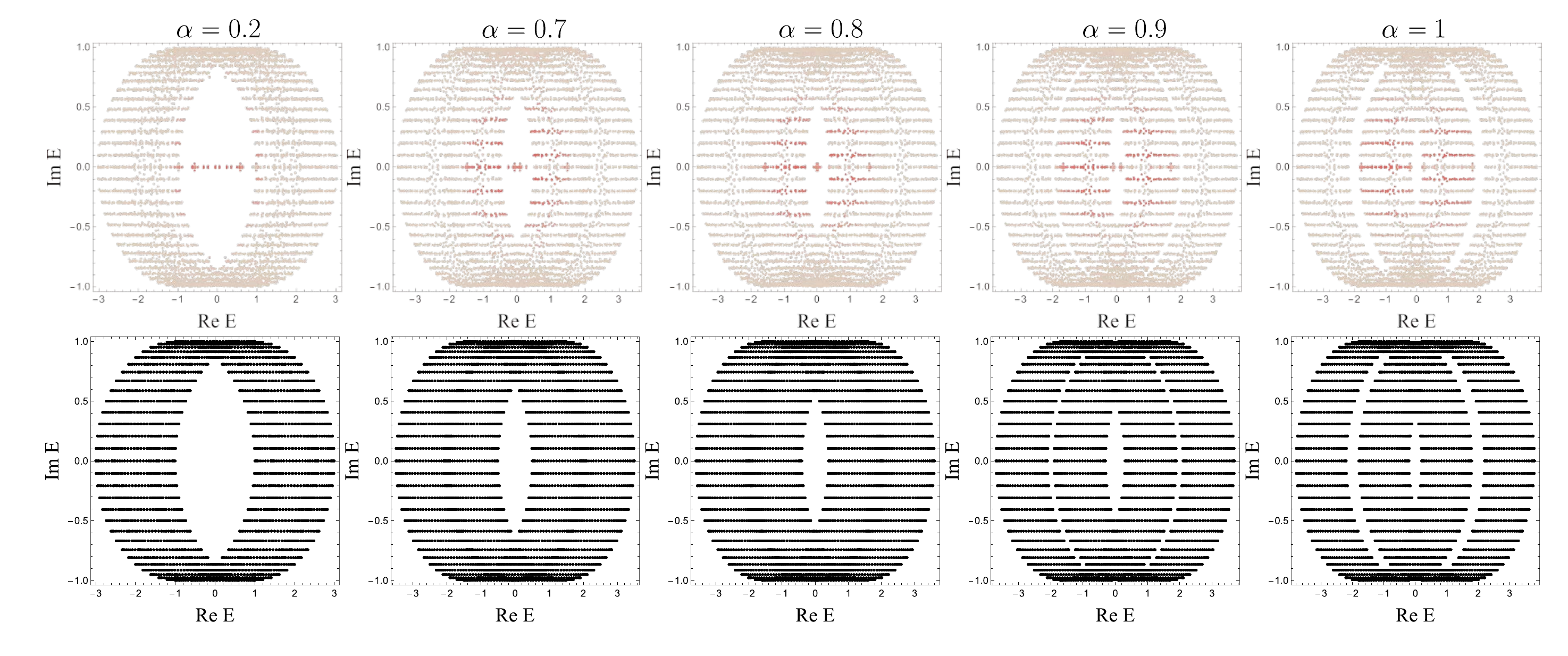}
    \caption{The spectrum of stacked Hamiltonian~\eqref{eq:stacked Hamiltonnian} by increasing the value of coupling strength $\alpha$. The in-gap states merge into the bulk without going throgh a phase transition.}
    \label{fig:DIII deformation 2}
\end{figure*}

To confirm (ii), we consider the stacked Hamiltonian
\begin{equation}
    h_{\text{stacked}}(k_x,k_y)=\begin{pmatrix}
        h_{DIII}(k_x,k_y)& \alpha \sigma_0\tau_0 \\
        \alpha \sigma_0\tau_0 & h_{DIII}(k_x,k_y)
    \end{pmatrix},
    \label{eq:stacked Hamiltonnian}
\end{equation}
where the coupling is controlled by the parameter $\alpha$. The stacked Hamiltonian has all the symmetries that $h_{DIII}$ obeys. Although in-gap states exist for small value of $\alpha$, these states would merge into the bulk if we increase the value of $\alpha$ \emph{without} going through a phase transition (Fig.~\ref{fig:DIII deformation 2}). Therefore, these states are not topological. Hence, we see that model~\eqref{eq:model DIII} indeed has a $\mathbb{Z}_2$ invariant. 

\section{Discussions and Conclusions}\label{sec:discussion}
In this paper, we give topological classification for all order-two spatial symmetries within AZ and AZ$^\dag$ class. The classification of crystalline symmetry for non-Hermitian system is a field that has been rarely studied. The classification table~\ref{tab:Periodic table delta_parallel=0},~\ref{tab:Periodic table delta_parallel=1},~\ref{tab:Periodic table delta_parallel=2},~\ref{tab:Periodic table delta_parallel=3} that we derived in the paper give a systematic way of understanding dislocation non-Hermitian skin effect proposed in~\cite{Schindler_2021,Bhargava_2021,Panigrahi_2022}. In this paper, we generalized the classification results from Ref.~\cite{Shiozaki_2014} to non-Hermitian systems. Our classification is also more generalized than Ref.~\cite{Liu_2019_reflection}, which only considered reflection symmetry for non-Hermitian Hamiltonian without defects, and Ref.~\cite{Liu_2019_defects}, which only considered defects with no spatial symmetries. Ref.~\cite{Shiozaku_indicator_2021} considered more generalized crystalline symmetries in non-Hermitian systems. However, Ref.~\cite{Shiozaku_indicator_2021} did not include internal symmetries in their classifications. In the grand scheme, our result paved the way for discovering novel symmetry-protected topological phases of non-Hermitian systems. As we have demonstrated in this paper, the existence of novel symmetry-protected topological phases also reveals that it is crucial to reexamine the effect of crystalline symmetries on non-Hermitian systems. 

In this paper, we also demonstrate our findings through several toy models. Furthermore, we generalized the "clock" shaped dimensional hierarchy of Hermitian Hamiltonian proposed in Ref.~\cite{Teo_2010} to NH Hamiltonian within AZ and AZ$^\dag$ class. However, since AZ and AZ$^\dag$ constitute only $16$ out of $38$-fold classification of NH matrices, it still remains a problem to find the classification for the rest of the $22$ classes. Since the $22$-fold SLS classes are rarely studied in current literature, their classifications are left for future works.

\pagebreak

\appendix
\section{Dimensional Hierarchy of AZ and AZ$^\dag$ without spatial symmetries}\label{Sec:dimensional Hierarchy AZ and AZ dag}
\begin{table*}[]
    \centering
    \caption{Dimensional Hierarchy maps of AZ and AZ$^\dag$ classes.}
    \begin{tabular}{ccccccc}\hline \hline
        AZ class & Mapping & Type of $\theta$ & Mapped AZ class & TRS & PHS & CS  \\ \hline

        A&$H_c(\textbf{k},\textbf{r},\theta)=\cos \theta \Tilde{H}_{nc}(\textbf{k},\textbf{r})\otimes \tau_z+\sin \theta \textrm{i}\Sigma \otimes \tau_x$&$\textbf{k}/\textbf{r}$&AIII&&&$1\otimes \tau_x$ \\

        AIII&$\Tilde{H}_{nc}(\textbf{k},\textbf{r},\theta)=\cos\theta \Tilde{H}_{c}(\textbf{k},\textbf{r})+\sin \theta \Tilde{\Gamma}$&$\textbf{k}/\textbf{r}$&A&&&\\ \hline
                
         \multirow{2}{*}{AI/AII}& $H_c(\textbf{k},\textbf{r},\theta)=\cos\theta \Tilde{H}_{nc}(\textbf{k},\textbf{r})\otimes\tau_z+\sin \theta 
         \textrm{i}\otimes\tau_0$ & $\textbf{k}$ & BDI/CII & $\Tilde{\mathcal{T}}_+ \otimes \tau_0$ & $\Tilde{\mathcal{T}}_+ \otimes \tau_x$ & $1\otimes\tau_x$ \\
         & $H_c(\textbf{k},\textbf{r},\theta)=\cos\theta \Tilde{H}_{nc}(\textbf{k},\textbf{r})\otimes\tau_z+\sin \theta 
         \textrm{i}\Sigma\otimes\tau_x$ & $\textbf{r}$ & CI/DIII & $\Tilde{\mathcal{T}}_+\otimes \tau_z$ & $\Tilde{\mathcal{T}}_+\otimes \tau_y$ & $1\otimes\tau_x$ \\ \hline

        \multirow{2}{*}{BDI/CII} & \multirow{2}{*}{$\Tilde{H}_{nc}(\textbf{k},\textbf{r},\theta)=\cos \theta \Tilde{H}_c(\textbf{k},\textbf{r})+\sin\theta \Tilde{\mathcal{T}}_+ \Tilde{\mathcal{C}}_-$} & $\textbf{k}$ & D/C & & $\Tilde{\mathcal{C}}_-$ & \\
        & & $\textbf{r}$ & AI/AII & $\Tilde{\mathcal{T}}_+$ & & \\ \hline

         \multirow{2}{*}{D/C} & \multirow{2}{*}{$H_c(\textbf{k},\textbf{r},\theta)=\textrm{i}\cos \theta \Tilde{H}_{nc}(\textbf{k},\textbf{r})\otimes \tau_0+\sin \theta 1 \otimes \tau_z$} & $\textbf{k}$ & DIII/CI &$\Tilde{\mathcal{C}}_-\otimes  \tau_y$ &$\Tilde{\mathcal{C}}_-\otimes \tau_z$  & $1\otimes\tau_x$\\
         & & $\textbf{r}$ & BDI/CII &$\Tilde{\mathcal{C}}_-\otimes \tau_0$ &$\Tilde{\mathcal{C}}_-\otimes \tau_x$  & $1\otimes \tau_x$ \\ \hline

         \multirow{2}{*}{DIII/CI} & \multirow{2}{*}{$\Tilde{H}_{nc}(\textbf{k},\textbf{r},\theta)=\cos \theta \Tilde{H}_c(\textbf{k},\textbf{r})+\sin \theta \textrm{i} \Tilde{\mathcal{T}}_+ \Tilde{\mathcal{C}}_-$} & $\textbf{k}$ & AII/AI & $\Tilde{\mathcal{T}}_+$ & & \\
         & & $\textbf{r}$ & D/C & & $\Tilde{\mathcal{C}}_-$ & \\ \hline\hline

         AZ$^\dag$ class & & & Mapped AZ$^\dag$ class & TRS$^\dag$ & PHS$^\dag$ & \\ \hline

         \multirow{2}{*}{AI$^\dag$/AII$^\dag$}& \multirow{2}{*}{$H_c(\textbf{k},\textbf{r},\theta)=\textrm{i}\cos \theta \Tilde{H}_{nc}(\textbf{k},\textbf{r})\otimes\tau_0+\sin \theta 1\otimes \tau_z$} & $\textbf{k}$ & BDI$^\dag$/CII$^\dag$ & $\Tilde{\mathcal{C}}_+ \otimes \tau_x$ & $\Tilde{\mathcal{C}}_+ \otimes \tau_0$ & $1\otimes\tau_x$ \\
        &  & $\textbf{r}$ & CI$^\dag$/DIII$^\dag$ &$\Tilde{\mathcal{C}}_+\otimes\tau_z$ & $\Tilde{\mathcal{C}}_+\otimes \tau_y$  & $1\otimes \tau_x$ \\ \hline

         \multirow{2}{*}{BDI$^\dag$/CII$^\dag$} & \multirow{2}{*}{$\Tilde{H}_{nc}(\textbf{k},\textbf{r},\theta)=\cos \theta \Tilde{H}_c(\textbf{k},\textbf{r})+\sin\theta \Tilde{\mathcal{C}}_+ \Tilde{\mathcal{T}}_-$} & $\textbf{k}$ & D$^\dag$/C$^\dag$ & & $\Tilde{\mathcal{T}}_-$ & \\
        & & $\textbf{r}$ & AI$^\dag$/AII$^\dag$ & $\Tilde{\mathcal{C}}_+$ & & \\ \hline
        
        \multirow{2}{*}{D$^\dag$/C$^\dag$} & $H_c(\textbf{k},\textbf{r},\theta)=\cos \theta \Tilde{H}_{nc}(\textbf{k},\textbf{r})\otimes\tau_z+\sin \theta \textrm{i}\Sigma \otimes \tau_x$ & $\textbf{k}$ & DIII$^\dag$/CI$^\dag$  &  $\Tilde{\mathcal{T}}_-\otimes\tau_y$ & $\Tilde{\mathcal{T}}_-\otimes \tau_z$& $1\otimes\tau_x$ \\
        & $H_c(\textbf{k},\textbf{r},\theta)=\cos\theta \Tilde{H}_{nc}(\textbf{k},\textbf{r})\otimes \tau_z+\sin \theta \textrm{i} \otimes \tau_0$ & $\textbf{r}$ & BDI$^\dag$/CII$^\dag$ &  $\Tilde{\mathcal{T}}_-\otimes\tau_x$ & $\Tilde{\mathcal{T}}_-\otimes\tau_0$ & $1\otimes\tau_x$\\ \hline

         \multirow{2}{*}{DIII$^\dag$/CI$^\dag$} & \multirow{2}{*}{$\Tilde{H}_{nc}(\textbf{k},\textbf{r},\theta)=\cos \theta \Tilde{H}_c(\textbf{k},\textbf{r})+\sin\theta \textrm{i}\Tilde{\mathcal{C}}_+ \Tilde{\mathcal{T}}_-$} & $\textbf{k}$ & AII$^\dag$/AI$^\dag$ & $\Tilde{\mathcal{C}}_+$ & & \\
         & & $\textbf{r}$ & D$^\dag$/C$^\dag$ & & $\Tilde{\mathcal{T}}_-$ & \\ \hline\hline
    \end{tabular}
    \label{tab:dimensional hierarchy mappings}
\end{table*}

\begin{table*}[]
    \centering
    \caption{Homomorphism from $K_{\mathbb{C}}^U(s,t;d+D,d_{\parallel}+D_{\parallel},0,0)$ to $K_{\mathbb{C}}^U(s+1,t;d+D+1,d_{\parallel}+D_{\parallel},0,0)$ and $K_{\mathbb{C}}^U(s+1,t+1;d+D+1,d_{\parallel}+D_{\parallel}+1,0,0)$. }
    \begin{tabular}{cccccccccc}\hline \hline
        AZ class & $t$ & Symmetry & Type of $\theta$ & Mapped to & TRS & PHS & CS & Mapped symmetry & Mapped $t$\\ \hline
         \multirow{4}{*}{A}&0&$\prescript{c}{}{\overline{U}}$&$\textbf{k}_{\perp}/\textbf{r}_{\perp}$&\multirow{4}{*}{AIII}& & &\multirow{4}{*}{$1\otimes \tau_x$}&$\prescript{g}{}{\overline{U}}_+=\prescript{c}{}{\Tilde{\overline{U}}}\otimes \tau_0$&0\\
         &1&$\prescript{g}{}{\overline{U}}$&$\textbf{k}_{\perp}/\textbf{r}_{\perp}$&&&&&$\prescript{g}{}{\overline{U}}_-=\prescript{g}{}{\Tilde{\overline{U}}}\otimes \tau_z$&1\\
         &0&$\prescript{c}{}{U}$&$\textbf{k}_{\parallel}/\textbf{r}_{\parallel}$&&&&&$\prescript{c}{}{\overline{U}}_+=\prescript{c}{}{\Tilde{U}}\otimes \tau_x$&1\\
         &1&$\prescript{g}{}{U}$&$\textbf{k}_{\parallel}/\textbf{r}_{\parallel}$&&&&&$\prescript{c}{}{\overline{U}}_-=\prescript{g}{}{\Tilde{U}}\otimes \tau_y$&0\\ \hline

         \multirow{4}{*}{AIII}&0&$\prescript{c}{}{U}_+$&$\textbf{k}_{\perp}/\textbf{r}_{\perp}$&\multirow{4}{*}{A}&&&&$\prescript{c}{}{U}=\prescript{c}{}{U}_+$&0\\
         &1&$\prescript{g}{}{U}_+$&$\textbf{k}_{\perp}/\textbf{r}_{\perp}$&&&&&$\prescript{g}{}{U}=\prescript{g}{}{U}_+$&1\\
         &0&$\prescript{c}{}{U}_+$&$\textbf{k}_{\parallel}/\textbf{r}_{\parallel}$&&&&&$\prescript{g}{}{\overline{U}}=\Gamma \prescript{c}{}{U}_+$&1\\
         &1&$\prescript{g}{}{U}_+$&$\textbf{k}_{\parallel}/\textbf{r}_{\parallel}$&&&&&$\prescript{c}{}{\overline{U}}=\Gamma \prescript{g}{}{U}_+$&0\\ \hline \hline
    \end{tabular}
    \label{tab:DH complex AZ}
\end{table*}

\begin{table*}
    \centering
    \caption{Homomorphism from $K_{\mathbb{R}}^{U/A}(s,t;d,d_{\parallel},D,D_{\parallel})$ to $K_{\mathbb{R}}^{U/A}(s+1,t;d+1,d_{\parallel},D,D_{\parallel})$, $K_{\mathbb{R}}^{U/A}(s+1,t+1;d+1,d_{\parallel}+1,D,D_{\parallel})$, $K_{\mathbb{R}}^{U/A}(s-1,t;d,d_{\parallel},D+1,D_{\parallel})$, and $K_{\mathbb{R}}^{U/A}(s-1,t-1;d,d_{\parallel},D+1,D_{\parallel}+1)$ for AZ classes without chiral symmetry.}
    \begin{tabular}{cccccccccc}\hline \hline
        AZ class & $t$ & Symmetry & Type of $\theta$ & Mapped to & TRS & PHS & CS & Mapped symmetry & Mapped $t$\\ \hline
         \multirow{4}{*}{AI/AII}& 0 & $\prescript{c}{}{U}^+_+$ & \multirow{4}{*}{$\textbf{k}_{\perp}$} & \multirow{4}{*}{BDI/CII} & \multirow{4}{*}{$\Tilde{\mathcal{T}}_+\otimes\tau_0$} & \multirow{4}{*}{$\Tilde{\mathcal{T}}_+\otimes\tau_x$} & & $\prescript{c}{}{U}^+_{++}=\prescript{c}{}{\Tilde{U}}^+_+\otimes \tau_0$  & 0\\
         & 1 & $\prescript{g}{}{\overline{U}}^+_-$ &&&&&& $\prescript{c}{}{U}^+_{+-}=\prescript{g}{}{\Tilde{\overline{U}}}^+_- \otimes \tau_y$ & 1 \\
         & 2 & $\prescript{c}{}{U}^-_+$ &&&&&& $\prescript{c}{}{U}^-_{++}=\prescript{c}{}{\Tilde{U}}^-_+ \otimes \tau_0$ &2\\
         & 3 &  $\prescript{g}{}{\overline{U}}^+_+$ &&&&&& $\prescript{c}{}{U}^+_{-+}=\prescript{g}{}{\Tilde{\overline{U}}}^+_+\otimes \tau_y$ &3\\ \hline

         \multirow{4}{*}{AI/AII}& 0 & $\prescript{c}{}{U}^+_+$& \multirow{4}{*}{$\textbf{k}_{\parallel}$} & \multirow{4}{*}{BDI/CII} & \multirow{4}{*}{$\Tilde{\mathcal{T}}_+ \otimes\tau_0$} &  \multirow{4}{*}{$\Tilde{\mathcal{T}}_+ \otimes\tau_x$} & &$\prescript{g}{}{U}^+_{++}=\prescript{c}{}{\Tilde{U}}^+_+\otimes\tau_0$& 1\\
         & 1 &  $\prescript{g}{}{\overline{U}}^+_-$ &&&&&& $\prescript{g}{}{U}^+_{+-}=\prescript{g}{}{\Tilde{\overline{U}}}^+_- \otimes \tau_y$ & 2\\
         & 2 & $\prescript{c}{}{U}^-_+$ &&&&&& $\prescript{g}{}{U}^-_{++}=\prescript{c}{}{\Tilde{U}}^-_+\otimes \tau_0$ & 3\\
         & 3& $\prescript{g}{}{\overline{U}}^+_+$ &&&&&& $\prescript{g}{}{U}^+_{-+}=\prescript{g}{}{\Tilde{\overline{U}}}^+_+\otimes \tau_y$ & 0 \\ \hline

         \multirow{4}{*}{AI/AII}& 0 & $\prescript{c}{}{U}^+_+$ & \multirow{4}{*}{$\textbf{r}_{\perp}$} & \multirow{4}{*}{CI/DIII} & \multirow{4}{*}{$\Tilde{\mathcal{T}}_+\otimes \tau_z$} & \multirow{4}{*}{$\Tilde{\mathcal{T}}_+\otimes \tau_y$} &  & $\prescript{c}{}{U}^+_{++}=\prescript{c}{}{\Tilde{U}}^+_+\otimes \tau_0$ & 0 \\
         & 1 & $\prescript{g}{}{\overline{U}}^+_-$ &&&&&& $\prescript{c}{}{U}^+_{-+}=\prescript{g}{}{\Tilde{\overline{U}}}^+_- \otimes \tau_y$ & 1\\
         &2&$\prescript{c}{}{U}^-_+$&&&&&&$\prescript{c}{}{U}^-_{++}=\prescript{c}{}{\Tilde{U}}^-_+\otimes \tau_0$&2\\
         &3&$\prescript{g}{}{\overline{U}}^+_+$&&&&&&$\prescript{c}{}{U}^+_{+-}=\prescript{g}{}{\Tilde{\overline{U}}}^+_+\otimes\tau_y$&3\\ \hline

        \multirow{4}{*}{AI/AII}&0&$\prescript{c}{}{U}^+_+$&\multirow{4}{*}{$\textbf{r}_{\parallel}$}& \multirow{4}{*}{CI/DIII} &\multirow{4}{*}{$\Tilde{\mathcal{T}}_+\otimes \tau_z$} &\multirow{4}{*}{$\Tilde{\mathcal{T}}_+\otimes \tau_y$}&& $\prescript{g}{}{\overline{U}}^+_{+-}=\prescript{c}{}{\Tilde{U}}^+_+\otimes \tau_y$ &3\\
        &1&$\prescript{g}{}{\overline{U}}^+_-$&&&&&&$\prescript{g}{}{\overline{U}}^+_{--}=\prescript{g}{}{\Tilde{\overline{U}}}^+_-\otimes \tau_0$&0\\
        &2&$\prescript{c}{}{U}^-_+$&&&&&&$\prescript{g}{}{\overline{U}}^-_{+-}=\prescript{c}{}{\Tilde{U}}^-_+\otimes\tau_y$&1\\
        &3&$\prescript{g}{}{\overline{U}}^+_+$&&&&&&$\prescript{g}{}{\overline{U}}^+_{++}=\prescript{g}{}{\overline{U}}^+_+\otimes \tau_0$&2\\ \hline

        \multirow{4}{*}{D/C}&0& $\prescript{c}{}{U}^+_+$ &\multirow{4}{*}{$\textbf{k}_{\perp}$} &\multirow{4}{*}{DIII/CI}&\multirow{4}{*}{$\Tilde{\mathcal{C}}_- \otimes \tau_y$}& \multirow{4}{*}{$\Tilde{\mathcal{C}}_-\otimes \tau_z$} &&$\prescript{c}{}{U}^+_{++}=\prescript{c}{}{U}^+_+\otimes \tau_0$&0\\
        &1&$\prescript{g}{}{\overline{U}}^+_+$&&&&&&$\prescript{c}{}{\overline{U}}^+_{--}=\prescript{g}{}{\Tilde{\overline{U}}}^+_+\otimes \tau_x$&1\\
        &2&$\prescript{c}{}{U}^-_+$&&&&&&$\prescript{c}{}{U}^-_{++}=\prescript{c}{}{\Tilde{U}}^-_+\otimes\tau_0$&2 \\
        &3&$\prescript{g}{}{\overline{U}}^+_-$&&&&&&$\prescript{c}{}{\overline{U}}^+_{++}=\prescript{g}{}{\Tilde{\overline{U}}}^+_-\otimes \tau_x$&3\\ \hline

        \multirow{4}{*}{D/C}&0& $\prescript{c}{}{U}^+_+$ &\multirow{4}{*}{$\textbf{k}_{\parallel}$} &\multirow{4}{*}{DIII/CI}&\multirow{4}{*}{$\Tilde{\mathcal{C}}_- \otimes \tau_y$}& \multirow{4}{*}{$\Tilde{\mathcal{C}}_-\otimes \tau_z$} &&$\prescript{c}{}{U}^+_{-+}=\prescript{c}{}{U}^+_+\otimes \tau_y$&1\\
        &1&$\prescript{g}{}{\overline{U}}^+_+$&&&&&&$\prescript{c}{}{\overline{U}}^+_{-+}=\prescript{g}{}{\Tilde{\overline{U}}}^+_+\otimes \tau_z$&2\\
        &2&$\prescript{c}{}{U}^-_+$&&&&&&$\prescript{c}{}{U}^-_{-+}=\prescript{c}{}{\Tilde{U}}^-_+\otimes\tau_y$&3 \\
        &3&$\prescript{g}{}{\overline{U}}^+_-$&&&&&&$\prescript{c}{}{\overline{U}}^+_{+-}=\prescript{g}{}{\Tilde{\overline{U}}}^+_-\otimes \tau_z$&0\\ \hline

        \multirow{4}{*}{D/C}&0& $\prescript{c}{}{U}^+_+$ &\multirow{4}{*}{$\textbf{r}_{\perp}$} &\multirow{4}{*}{BDI/CII}&\multirow{4}{*}{$\Tilde{\mathcal{C}}_- \otimes \tau_0$}& \multirow{4}{*}{$\Tilde{\mathcal{C}}_-\otimes \tau_x$} &&$\prescript{c}{}{U}^+_{++}=\prescript{c}{}{U}^+_+\otimes \tau_0$&0\\
        &1&$\prescript{g}{}{\overline{U}}^+_+$&&&&&&$\prescript{c}{}{\overline{U}}^+_{++}=\prescript{g}{}{\Tilde{\overline{U}}}^+_+\otimes \tau_x$&1\\
        &2&$\prescript{c}{}{U}^-_+$&&&&&&$\prescript{c}{}{U}^-_{++}=\prescript{c}{}{\Tilde{U}}^-_+\otimes\tau_0$&2 \\
        &3&$\prescript{g}{}{\overline{U}}^+_-$&&&&&&$\prescript{c}{}{\overline{U}}^+_{--}=\prescript{g}{}{\Tilde{\overline{U}}}^+_-\otimes \tau_x$&3\\ \hline

        \multirow{4}{*}{D/C}&0& $\prescript{c}{}{U}^+_+$ &\multirow{4}{*}{$\textbf{r}_{\parallel}$} &\multirow{4}{*}{BDI/CII}&\multirow{4}{*}{$\Tilde{\mathcal{C}}_- \otimes \tau_0$}& \multirow{4}{*}{$\Tilde{\mathcal{C}}_-\otimes \tau_x$} &&$\prescript{c}{}{U}^+_{-+}=\prescript{c}{}{U}^+_+\otimes \tau_y$&3\\
        &1&$\prescript{g}{}{\overline{U}}^+_+$&&&&&&$\prescript{c}{}{\overline{U}}^+_{+-}=\prescript{g}{}{\Tilde{\overline{U}}}^+_+\otimes \tau_z$&0\\
        &2&$\prescript{c}{}{U}^-_+$&&&&&&$\prescript{c}{}{U}^-_{-+}=\prescript{c}{}{\Tilde{U}}^-_+\otimes\tau_y$&1 \\
        &3&$\prescript{g}{}{\overline{U}}^+_-$&&&&&&$\prescript{c}{}{\overline{U}}^+_{-+}=\prescript{g}{}{\Tilde{\overline{U}}}^+_-\otimes \tau_z$&2\\ \hline \hline
    \end{tabular}
    \label{tab:DH AZ nc to c}
\end{table*}

\begin{table*}[]
    \centering
    \caption{Homomorphism from $K_{\mathbb{R}}^{U/A}(s,t;d,d_{\parallel},D,D_{\parallel})$ to $K_{\mathbb{R}}^{U/A}(s+1,t;d+1,d_{\parallel},D,D_{\parallel})$, $K_{\mathbb{R}}^{U/A}(s+1,t+1;d+1,d_{\parallel}+1,D,D_{\parallel})$, $K_{\mathbb{R}}^{U/A}(s-1,t;d,d_{\parallel},D+1,D_{\parallel})$, and $K_{\mathbb{R}}^{U/A}(s-1,t-1;d,d_{\parallel},D+1,D_{\parallel}+1)$ for AZ classes with chiral symmetry.}
    \begin{tabular}{cccccccccc}\hline\hline
         AZ class & $t$ & Symmetry & Type of $\theta$ & Mapped to & TRS & PHS & CS & Mapped symmetry & Mapped $t$\\ \hline
         \multirow{4}{*}{BDI/CII}&0&$\prescript{c}{}{U}^+_{++}$&\multirow{4}{*}{$\textbf{k}_{\perp}$}&\multirow{4}{*}{D/C}&&\multirow{4}{*}{$\Tilde{\mathcal{C}}_-$}&&$\prescript{c}{}{U}^+_+=\prescript{c}{}{U}^+_{++}$&0\\
         &1&$\prescript{g}{}{\overline{U}}^+_{-+}$&&&&&&$\prescript{g}{}{\overline{U}}^+_+=\prescript{g}{}{\overline{U}}^+_{-+}$&1\\
         &2&$\prescript{c}{}{U}^+_{--}$&&&&&&$\prescript{c}{}{U}^+_-=\prescript{c}{}{U}^+_{--}$&2\\
         &3&$\prescript{g}{}{\overline{U}}^+_{+-}$&&&&&&$\prescript{g}{}{\overline{U}}^+_-=\prescript{g}{}{\overline{U}}^+_{+-}$&3\\ \hline

        \multirow{4}{*}{BDI/CII}&0&$\prescript{c}{}{U}^+_{++}$&\multirow{4}{*}{$\textbf{k}_{\parallel}$}&\multirow{4}{*}{D/C}&&\multirow{4}{*}{$\Tilde{\mathcal{C}}_-$}&&$\prescript{g}{}{\overline{U}}^+_+=\Gamma\prescript{c}{}{U}^+_{++}$&1\\ 
        &1&$\prescript{g}{}{\overline{U}}^+_{-+}$&&&&&&$\prescript{c}{}{U}^+_-=\textrm{i}\Gamma\prescript{g}{}{\overline{U}}^+_{-+}$&2\\
        &2&$\prescript{c}{}{U}^+_{--}$&&&&&&$\prescript{g}{}{\overline{U}}^+_-=\Gamma \prescript{c}{}{U}^+_{--}$&3\\
        &3&$\prescript{g}{}{\overline{U}}^+_{+-}$&&&&&&$\prescript{c}{}{U}^+_+=\textrm{i}\Gamma\prescript{g}{}{\overline{U}}^+_{+-}$&0\\ \hline

        \multirow{4}{*}{BDI/CII}&0&$\prescript{c}{}{U}^+_{++}$&\multirow{4}{*}{$\textbf{r}_{\perp}$}&\multirow{4}{*}{AI/AII}&\multirow{4}{*}{$\Tilde{\mathcal{T}}_+$}&&&$\prescript{c}{}{U}^+_+=\prescript{c}{}{U}^+_{++}$&0\\
         &1&$\prescript{g}{}{\overline{U}}^+_{-+}$&&&&&&$\prescript{g}{}{\overline{U}}^+_-=\prescript{g}{}{\overline{U}}^+_{-+}$&1\\
         &2&$\prescript{c}{}{U}^+_{--}$&&&&&&$\prescript{c}{}{U}^+_-=\prescript{c}{}{U}^+_{--}$&2\\
         &3&$\prescript{g}{}{\overline{U}}^+_{+-}$&&&&&&$\prescript{g}{}{\overline{U}}^+_+=\prescript{g}{}{\overline{U}}^+_{+-}$&3\\ \hline

         \multirow{4}{*}{BDI/CII}&0&$\prescript{c}{}{U}^+_{++}$&\multirow{4}{*}{$\textbf{r}_{\parallel}$}&\multirow{4}{*}{AI/AII}&\multirow{4}{*}{$\Tilde{\mathcal{T}}_+$}&&&$\prescript{g}{}{\overline{U}}^+_+=\Gamma\prescript{c}{}{U}^+_{++}$&3\\
         &1&$\prescript{g}{}{\overline{U}}^+_{-+}$&&&&&&$\prescript{c}{}{U}^+_+=\textrm{i}\Gamma\prescript{g}{}{\overline{U}}^+_{-+}$&0\\
         &2&$\prescript{c}{}{U}^+_{--}$&&&&&&$\prescript{g}{}{\overline{U}}^+_-=\Gamma\prescript{c}{}{U}^+_{--}$&1\\
         &3&$\prescript{g}{}{\overline{U}}^+_{+-}$&&&&&&$\prescript{c}{}{U}^+_-=\textrm{i}\Gamma\prescript{g}{}{\overline{U}}^+_{+-}$&2\\ \hline

         \multirow{4}{*}{DIII/CI}&0&$\prescript{c}{}{U}^+_{++}$&\multirow{4}{*}{$\textbf{k}_{\perp}$}&\multirow{4}{*}{AII/AI}&\multirow{4}{*}{$\Tilde{\mathcal{T}}_+$}&&&$\prescript{c}{}{U}^+_+=\prescript{c}{}{U}^+_{++}$&0\\
         &1&$\prescript{g}{}{\overline{U}}^+_{-+}$&&&&&&$\prescript{g}{}{\overline{U}}^+_-=\prescript{g}{}{\overline{U}}^+_{-+}$&1\\
         &2&$\prescript{c}{}{U}^+_{--}$&&&&&&$\prescript{c}{}{U}^+_-=\prescript{c}{}{U}^+_{--}$&2\\
         &3&$\prescript{g}{}{\overline{U}}^+_{+-}$&&&&&&$\prescript{g}{}{\overline{U}}^+_+=\prescript{g}{}{\overline{U}}^+_{+-}$&3\\ \hline

         \multirow{4}{*}{DIII/CI}&0&$\prescript{c}{}{U}^+_{++}$&\multirow{4}{*}{$\textbf{k}_{\parallel}$}&\multirow{4}{*}{AII/AI}&\multirow{4}{*}{$\Tilde{\mathcal{T}}_+$}&&&$\prescript{g}{}{\overline{U}}^+_-=\Gamma\prescript{c}{}{U}^+_{++}$&1\\
         &1&$\prescript{g}{}{\overline{U}}^+_{-+}$&&&&&&$\prescript{c}{}{U}^+_-=\textrm{i}\Gamma\prescript{g}{}{\overline{U}}^+_{-+}$&2\\
         &2&$\prescript{c}{}{U}^+_{--}$&&&&&&$\prescript{g}{}{\overline{U}}^+_+=\Gamma\prescript{c}{}{U}^+_{--}$&3\\
         &3&$\prescript{g}{}{\overline{U}}^+_{+-}$&&&&&&$\prescript{c}{}{U}^+_+=\textrm{i}\Gamma\prescript{g}{}{\overline{U}}^+_{+-}$&0\\ \hline

         \multirow{4}{*}{DIII/CI}&0&$\prescript{c}{}{U}^+_{++}$&\multirow{4}{*}{$\textbf{r}_{\perp}$}&\multirow{4}{*}{D/C}&&\multirow{4}{*}{$\Tilde{\mathcal{C}}_-$}&&$\prescript{c}{}{U}^+_+=\prescript{c}{}{U}^+_{++}$&0\\
         &1&$\prescript{g}{}{\overline{U}}^+_{-+}$&&&&&&$\prescript{g}{}{\overline{U}}^+_+=\prescript{g}{}{\overline{U}}^+_{-+}$&1\\
         &2&$\prescript{c}{}{U}^+_{--}$&&&&&&$\prescript{c}{}{U}^+_-=\prescript{c}{}{U}^+_{--}$&2\\
         &3&$\prescript{g}{}{\overline{U}}^+_{+-}$&&&&&&$\prescript{g}{}{\overline{U}}^+_-=\prescript{g}{}{\overline{U}}^+_{+-}$&3\\ \hline

         \multirow{4}{*}{DIII/CI}&0&$\prescript{c}{}{U}^+_{++}$&\multirow{4}{*}{$\textbf{r}_{\parallel}$}&\multirow{4}{*}{D/C}&&\multirow{4}{*}{$\Tilde{\mathcal{C}}_-$}&&$\prescript{g}{}{\overline{U}}^+_-=\Gamma\prescript{c}{}{U}^+_{++}$&3\\
         &1&$\prescript{g}{}{\overline{U}}^+_{-+}$&&&&&&$\prescript{c}{}{U}^+_+=\textrm{i}\Gamma\prescript{g}{}{\overline{U}}^+_{-+}$&0\\
         &2&$\prescript{c}{}{U}^+_{--}$&&&&&&$\prescript{g}{}{\overline{U}}^+_+=\Gamma\prescript{c}{}{U}^+_{--}$&1\\
         &3&$\prescript{g}{}{\overline{U}}^+_{+-}$&&&&&&$\prescript{c}{}{U}^+_-=\textrm{i}\Gamma\prescript{g}{}{\overline{U}}^+_{+-}$&2\\ \hline\hline
    \end{tabular}
    \label{tab:DH AZ c to nc}
\end{table*}

\begin{table*}[]
    \centering
    \caption{Homomorphism from $K_{\mathbb{R}}^{\dag U/A}(s,t;d,d_{\parallel},D,D_{\parallel})$ to $K_{\mathbb{R}}^{\dag U/A}(s+1,t;d+1,d_{\parallel},D,D_{\parallel})$, $K_{\mathbb{R}}^{\dag U/A}(s+1,t-1;d+1,d_{\parallel}+1,D,D_{\parallel})$, $K_{\mathbb{R}}^{\dag U/A}(s-1,t;d,d_{\parallel},D+1,D_{\parallel})$, and $K_{\mathbb{R}}^{\dag U/A}(s-1,t+1;d,d_{\parallel},D+1,D_{\parallel}+1)$ for AZ$^\dag$ classes without chiral symmetry.}
    \begin{tabular}{cccccccccc}\hline \hline
         AZ$^\dag$ class & $t$ & Symmetry & Type of $\theta$ & Mapped to & TRS$^\dag$ & PHS$^\dag$ & CS & Mapped symmetry & Mapped $t$\\ \hline
         \multirow{4}{*}{AI$^\dag$/AII$^\dag$}&0&$\prescript{c}{}{U}^+_+$&\multirow{4}{*}{$\textbf{k}_{\perp}$}&\multirow{4}{*}{BDI$^\dag$/CII$^\dag$}&\multirow{4}{*}{$\Tilde{\mathcal{C}}_+\otimes \tau_x$}&\multirow{4}{*}{$\Tilde{\mathcal{C}}_+\otimes \tau_0$}&&$\prescript{c}{}{U}^+_{++}=\prescript{c}{}{U}^+_+\otimes \tau_0$&0\\
         &1&$\prescript{g}{}{\overline{U}}^+_+$&&&&&&$\prescript{c}{}{\overline{U}}^+_{++}=\prescript{g}{}{\Tilde{\overline{U}}}^+_+\otimes \tau_x$&1\\
         &2&$\prescript{c}{}{U}^-_+$&&&&&&$\prescript{c}{}{U}^-_{++}=\prescript{c}{}{\Tilde{U}}^-_+\otimes \tau_0$&2\\
         &3&$\prescript{g}{}{\overline{U}}^+_-$&&&&&&$\prescript{c}{}{\overline{U}}^+_{--}=\prescript{g}{}{\Tilde{\overline{U}}}^+_-\otimes \tau_x$&3\\ \hline

         \multirow{4}{*}{AI$^\dag$/AII$^\dag$}&0&$\prescript{c}{}{U}^+_+$&\multirow{4}{*}{$\textbf{k}_{\parallel}$}& \multirow{4}{*}{BDI$^\dag$/CII$^\dag$} &\multirow{4}{*}{$\Tilde{\mathcal{C}}_+\otimes \tau_x$}&\multirow{4}{*}{$\Tilde{\mathcal{C}}_+\otimes \tau_0$}&&$\prescript{c}{}{U}^+_{+-}=\prescript{c}{}{\Tilde{U}}^+_+\otimes \tau_y$&3\\
         &1&$\prescript{g}{}{\overline{U}}^+_+$&&&&&&$\prescript{c}{}{\overline{U}}^+_{-+}=\prescript{g}{}{\Tilde{\overline{U}}}^+_+\otimes \tau_z$&0\\
         &2&$\prescript{c}{}{U}^-_+$&&&&&&$\prescript{c}{}{U}^-_{+-}=\prescript{c}{}{U}^-_+\otimes \tau_y$&1\\
         &3&$\prescript{g}{}{\overline{U}}^+_-$&&&&&&$\prescript{c}{}{\overline{U}}^+_{+-}=\prescript{g}{}{\Tilde{\overline{U}}}^+_-\otimes \tau_z$&2\\ \hline

         \multirow{4}{*}{AI$^\dag$/AII$^\dag$}&0&$\prescript{c}{}{U}^+_+$&\multirow{4}{*}{$\textbf{r}_{\perp}$}& \multirow{4}{*}{CI$^\dag$/DIII$^\dag$}&\multirow{4}{*}{$\Tilde{\mathcal{C}}_+\otimes \tau_z$}&\multirow{4}{*}{$\Tilde{\mathcal{C}}_+\otimes \tau_y$}&&$\prescript{c}{}{U}^+_{++}=\prescript{c}{}{U}^+_+\otimes \tau_0$&0\\
         &1&$\prescript{g}{}{\overline{U}}^+_+$&&&&&&$\prescript{c}{}{\overline{U}}^+_{--}=\prescript{g}{}{\Tilde{\overline{U}}}^+_+\otimes\tau_x$&1\\
         &2&$\prescript{c}{}{U}^-_+$&&&&&&$\prescript{c}{}{U}^-_{++}=\prescript{c}{}{\Tilde{U}}^-_+\otimes \tau_0$&2\\
         &3&$\prescript{g}{}{\overline{U}}^+_-$&&&&&&$\prescript{c}{}{\overline{U}}^+_{++}=\prescript{g}{}{\Tilde{\overline{U}}}^+_-\otimes \tau_x$&3\\ \hline

         \multirow{4}{*}{AI$^\dag$/AII$^\dag$}&0&$\prescript{c}{}{U}^+_+$&\multirow{4}{*}{$\textbf{r}_{\parallel}$}& \multirow{4}{*}{CI$^\dag$/DIII$^\dag$}&\multirow{4}{*}{$\Tilde{\mathcal{C}}_+\otimes \tau_z$}&\multirow{4}{*}{$\Tilde{\mathcal{C}}_+\otimes \tau_y$}&&$\prescript{c}{}{U}^+_{+-}=\prescript{c}{}{U}^+_+\otimes \tau_y$&1\\
         &1&$\prescript{g}{}{\overline{U}}^+_+$&&&&&&$\prescript{c}{}{\overline{U}}^+_{+-}=\prescript{g}{}{\Tilde{\overline{U}}}^+_+\otimes\tau_z$&2\\
         &2&$\prescript{c}{}{U}^-_+$&&&&&&$\prescript{c}{}{U}^-_{+-}=\prescript{c}{}{\Tilde{U}}^-_+\otimes \tau_y$&3\\
         &3&$\prescript{g}{}{\overline{U}}^+_-$&&&&&&$\prescript{c}{}{\overline{U}}^+_{-+}=\prescript{g}{}{\Tilde{\overline{U}}}^+_-\otimes \tau_y$&0\\ \hline

         \multirow{4}{*}{D$^\dag$/C$^\dag$}&0&$\prescript{c}{}{U}^+_+$&\multirow{4}{*}{$\textbf{k}_{\perp}$}&\multirow{4}{*}{DIII$^\dag$/CI$^\dag$}&\multirow{4}{*}{$\Tilde{\mathcal{T}}_- \otimes \tau_y$}&\multirow{4}{*}{$\Tilde{\mathcal{T}}_- \otimes \tau_z$}&&$\prescript{c}{}{U}^+_{++}=\prescript{c}{}{\Tilde{U}}^+_+\otimes \tau_0$&0\\
         &1&$\prescript{g}{}{\overline{U}}^+_-$&&&&&&$\prescript{c}{}{U}^+_{+-}=\prescript{g}{}{\Tilde{\overline{U}}}^+_-\otimes \tau_y$&1\\
         &2&$\prescript{c}{}{U}^-_+$&&&&&&$\prescript{c}{}{U}^-_{++}=\prescript{c}{}{\Tilde{U}}^-_+\otimes \tau_0$&2\\
         &3&$\prescript{g}{}{\overline{U}}^+_+$&&&&&&$\prescript{c}{}{U}^+_{-+}=\prescript{g}{}{\Tilde{\overline{U}}}^+_+\otimes \tau_y$&3\\ \hline

         \multirow{4}{*}{D$^\dag$/C$^\dag$}&0&$\prescript{c}{}{U}^+_+$&\multirow{4}{*}{$\textbf{k}_{\parallel}$}&\multirow{4}{*}{DIII$^\dag$/CI$^\dag$}&\multirow{4}{*}{$\Tilde{\mathcal{T}}_- \otimes \tau_y$}&\multirow{4}{*}{$\Tilde{\mathcal{T}}_- \otimes \tau_z$}&&$\prescript{g}{}{\overline{U}}^+_{-+}=\prescript{c}{}{\Tilde{U}}^+_+\otimes \tau_y$&3\\
         &1&$\prescript{g}{}{\overline{U}}^+_-$&&&&&&$\prescript{g}{}{\overline{U}}^+_{--}=\prescript{g}{}{\Tilde{\overline{U}}}^+_-\otimes \tau_0$&0\\
         &2&$\prescript{c}{}{U}^-_+$&&&&&&$\prescript{g}{}{\overline{U}}^-_{-+}=\prescript{c}{}{\Tilde{U}}^-_+\otimes \tau_y$&1\\
         &3&$\prescript{g}{}{\overline{U}}^+_+$&&&&&&$\prescript{g}{}{\overline{U}}^+_{++}=\prescript{g}{}{\Tilde{\overline{U}}}^+_+\otimes \tau_0$&2\\ \hline

         \multirow{4}{*}{D$^\dag$/C$^\dag$}&0&$\prescript{c}{}{U}^+_+$&\multirow{4}{*}{$\textbf{r}_{\perp}$}&\multirow{4}{*}{BDI$^\dag$/CII$^\dag$}&\multirow{4}{*}{$\Tilde{\mathcal{T}}_- \otimes \tau_x$}&\multirow{4}{*}{$\Tilde{\mathcal{T}}_- \otimes \tau_0$}&&$\prescript{c}{}{U}^+_{++}=\prescript{c}{}{\Tilde{U}}^+_+\otimes \tau_0$&0\\
         &1&$\prescript{g}{}{\overline{U}}^+_-$&&&&&&$\prescript{c}{}{U}^+_{-+}=\prescript{g}{}{\Tilde{\overline{U}}}^+_-\otimes \tau_y$&1\\
         &2&$\prescript{c}{}{U}^-_+$&&&&&&$\prescript{c}{}{U}^-_{++}=\prescript{c}{}{\Tilde{U}}^-_+\otimes \tau_0$&2\\
         &3&$\prescript{g}{}{\overline{U}}^+_+$&&&&&&$\prescript{c}{}{U}^+_{+-}=\prescript{g}{}{\Tilde{\overline{U}}}^+_+\otimes \tau_y$&3\\ \hline

         \multirow{4}{*}{D$^\dag$/C$^\dag$}&0&$\prescript{c}{}{U}^+_+$&\multirow{4}{*}{$\textbf{r}_{\parallel}$}&\multirow{4}{*}{BDI$^\dag$/CII$^\dag$}&\multirow{4}{*}{$\Tilde{\mathcal{T}}_- \otimes \tau_x$}&\multirow{4}{*}{$\Tilde{\mathcal{T}}_- \otimes \tau_0$}&&$\prescript{g}{}{U}^+_{++}=\prescript{c}{}{\Tilde{U}}^+_+\otimes \tau_0$&1\\
         &1&$\prescript{g}{}{\overline{U}}^+_-$&&&&&&$\prescript{g}{}{U}^+_{-+}=\prescript{g}{}{\Tilde{\overline{U}}}^+_-\otimes \tau_y$&2\\
         &2&$\prescript{c}{}{U}^-_+$&&&&&&$\prescript{g}{}{U}^-_{++}=\prescript{c}{}{\Tilde{U}}^-_+\otimes \tau_0$&3\\
         &3&$\prescript{g}{}{\overline{U}}^+_+$&&&&&&$\prescript{g}{}{U}^+_{+-}=\prescript{g}{}{\Tilde{\overline{U}}}^+_+\otimes \tau_y$&0\\ \hline\hline

    \end{tabular}
    \label{tab:DH AZ dag nc to c}
\end{table*}

\begin{table*}[]
    \centering
    \caption{Homomorphism from $K_{\mathbb{R}}^{\dag U/A}(s,t;d,d_{\parallel},D,D_{\parallel})$ to $K_{\mathbb{R}}^{\dag U/A}(s+1,t;d+1,d_{\parallel},D,D_{\parallel})$, $K_{\mathbb{R}}^{\dag U/A}(s+1,t-1;d+1,d_{\parallel}+1,D,D_{\parallel})$, $K_{\mathbb{R}}^{\dag U/A}(s-1,t;d,d_{\parallel},D+1,D_{\parallel})$, and $K_{\mathbb{R}}^{\dag U/A}(s-1,t+1;d,d_{\parallel},D+1,D_{\parallel}+1)$ for AZ$^\dag$ classes with chiral symmetry.}
    \begin{tabular}{cccccccccc}\hline\hline
         AZ$^\dag$ class & $t$ & Symmetry & Type of $\theta$ & Mapped to & TRS$^\dag$ & PHS$^\dag$ & CS & Mapped symmetry & Mapped $t$\\ \hline
         \multirow{4}{*}{BDI$^\dag$/CII$^\dag$}&0&$\prescript{c}{}{U}^+_{++}$&\multirow{4}{*}{$\textbf{k}_{\perp}$}&\multirow{4}{*}{D$^\dag$/C$^\dag$}&&\multirow{4}{*}{$\Tilde{\mathcal{T}}_-$}&&$\prescript{c}{}{U}^+_+=\prescript{c}{}{U}^+_{++}$&0\\
         &1&$\prescript{g}{}{\overline{U}}^+_{+-}$&&&&&&$\prescript{g}{}{\overline{U}}^+_-=\prescript{g}{}{\overline{U}}^+_{+-}$&1\\
         &2&$\prescript{c}{}{U}^+_{--}$&&&&&&$\prescript{c}{}{U}^+_-=\prescript{c}{}{U}^+_{--}$&2\\
         &3&$\prescript{g}{}{\overline{U}}^+_{-+}$&&&&&&$\prescript{g}{}{\overline{U}}^+_+=\prescript{g}{}{\overline{U}}^+_{-+}$&3\\ \hline

        \multirow{4}{*}{BDI$^\dag$/CII$^\dag$}&0&$\prescript{c}{}{U}^+_{++}$&\multirow{4}{*}{$\textbf{k}_{\parallel}$}&\multirow{4}{*}{D$^\dag$/C$^\dag$}&&\multirow{4}{*}{$\Tilde{\mathcal{T}}_-$}&&$\prescript{g}{}{\overline{U}}^+_+=\Gamma\prescript{c}{}{U}^+_{++}$&3\\ 
        &1&$\prescript{g}{}{\overline{U}}^+_{+-}$&&&&&&$\prescript{c}{}{U}^+_+=\textrm{i}\Gamma\prescript{g}{}{\overline{U}}^+_{+-}$&0\\
        &2&$\prescript{c}{}{U}^+_{--}$&&&&&&$\prescript{g}{}{\overline{U}}^+_-=\Gamma \prescript{c}{}{U}^+_{--}$&1\\
        &3&$\prescript{g}{}{\overline{U}}^+_{-+}$&&&&&&$\prescript{c}{}{U}^+_-=\textrm{i}\Gamma\prescript{g}{}{\overline{U}}^+_{-+}$&2\\ \hline

        \multirow{4}{*}{BDI$^\dag$/CII$^\dag$}&0&$\prescript{c}{}{U}^+_{++}$&\multirow{4}{*}{$\textbf{r}_{\perp}$}&\multirow{4}{*}{AI$^\dag$/AII$^\dag$}&\multirow{4}{*}{$\Tilde{\mathcal{C}}_+$}&&&$\prescript{c}{}{U}^+_+=\prescript{c}{}{U}^+_{++}$&0\\
         &1&$\prescript{g}{}{\overline{U}}^+_{+-}$&&&&&&$\prescript{g}{}{\overline{U}}^+_+=\prescript{g}{}{\overline{U}}^+_{+-}$&1\\
         &2&$\prescript{c}{}{U}^+_{--}$&&&&&&$\prescript{c}{}{U}^+_-=\prescript{c}{}{U}^+_{--}$&2\\
         &3&$\prescript{g}{}{\overline{U}}^+_{-+}$&&&&&&$\prescript{g}{}{\overline{U}}^+_-=\prescript{g}{}{\overline{U}}^+_{-+}$&3\\ \hline

         \multirow{4}{*}{BDI$^\dag$/CII$^\dag$}&0&$\prescript{c}{}{U}^+_{++}$&\multirow{4}{*}{$\textbf{r}_{\parallel}$}&\multirow{4}{*}{AI$^\dag$/AII$^\dag$}&\multirow{4}{*}{$\Tilde{\mathcal{C}}_+$}&&&$\prescript{g}{}{\overline{U}}^+_+=\Gamma\prescript{c}{}{U}^+_{++}$&1\\
         &1&$\prescript{g}{}{\overline{U}}^+_{+-}$&&&&&&$\prescript{c}{}{U}^+_-=\textrm{i}\Gamma\prescript{g}{}{\overline{U}}^+_{+-}$&2\\
         &2&$\prescript{c}{}{U}^+_{--}$&&&&&&$\prescript{g}{}{\overline{U}}^+_-=\Gamma\prescript{c}{}{U}^+_{--}$&3\\
         &3&$\prescript{g}{}{\overline{U}}^+_{-+}$&&&&&&$\prescript{c}{}{U}^+_+=\textrm{i}\Gamma\prescript{g}{}{\overline{U}}^+_{-+}$&0\\ \hline

         \multirow{4}{*}{DIII$^\dag$/CI$^\dag$}&0&$\prescript{c}{}{U}^+_{++}$&\multirow{4}{*}{$\textbf{k}_{\perp}$}&\multirow{4}{*}{AII$^\dag$/AI$^\dag$}&\multirow{4}{*}{$\Tilde{\mathcal{C}}_+$}&&&$\prescript{c}{}{U}^+_+=\prescript{c}{}{U}^+_{++}$&0\\
         &1&$\prescript{g}{}{\overline{U}}^+_{+-}$&&&&&&$\prescript{g}{}{\overline{U}}^+_+=\prescript{g}{}{\overline{U}}^+_{+-}$&1\\
         &2&$\prescript{c}{}{U}^+_{--}$&&&&&&$\prescript{c}{}{U}^+_-=\prescript{c}{}{U}^+_{--}$&2\\
         &3&$\prescript{g}{}{\overline{U}}^+_{-+}$&&&&&&$\prescript{g}{}{\overline{U}}^+_-=\prescript{g}{}{\overline{U}}^+_{-+}$&3\\ \hline

         \multirow{4}{*}{DIII$^\dag$/CI$^\dag$}&0&$\prescript{c}{}{U}^+_{++}$&\multirow{4}{*}{$\textbf{k}_{\parallel}$}&\multirow{4}{*}{AII$^\dag$/AI$^\dag$}&\multirow{4}{*}{$\Tilde{\mathcal{C}}_+$}&&&$\prescript{g}{}{\overline{U}}^+_-=\Gamma\prescript{c}{}{U}^+_{++}$&3\\
         &1&$\prescript{g}{}{\overline{U}}^+_{+-}$&&&&&&$\prescript{c}{}{U}^+_+=\textrm{i}\Gamma\prescript{g}{}{\overline{U}}^+_{+-}$&0\\
         &2&$\prescript{c}{}{U}^+_{--}$&&&&&&$\prescript{g}{}{\overline{U}}^+_+=\Gamma\prescript{c}{}{U}^+_{--}$&1\\
         &3&$\prescript{g}{}{\overline{U}}^+_{-+}$&&&&&&$\prescript{c}{}{U}^+_-=\textrm{i}\Gamma\prescript{g}{}{\overline{U}}^+_{-+}$&2\\ \hline

         \multirow{4}{*}{DIII$^\dag$/CI$^\dag$}&0&$\prescript{c}{}{U}^+_{++}$&\multirow{4}{*}{$\textbf{r}_{\perp}$}&\multirow{4}{*}{D$^\dag$/C$^\dag$}&&\multirow{4}{*}{$\Tilde{\mathcal{T}}_-$}&&$\prescript{c}{}{U}^+_+=\prescript{c}{}{U}^+_{++}$&0\\
         &1&$\prescript{g}{}{\overline{U}}^+_{+-}$&&&&&&$\prescript{g}{}{\overline{U}}^+_-=\prescript{g}{}{\overline{U}}^+_{+-}$&1\\
         &2&$\prescript{c}{}{U}^+_{--}$&&&&&&$\prescript{c}{}{U}^+_-=\prescript{c}{}{U}^+_{--}$&2\\
         &3&$\prescript{g}{}{\overline{U}}^+_{-+}$&&&&&&$\prescript{g}{}{\overline{U}}^+_+=\prescript{g}{}{\overline{U}}^+_{-+}$&3\\ \hline

         \multirow{4}{*}{DIII$^\dag$/CI$^\dag$}&0&$\prescript{c}{}{U}^+_{++}$&\multirow{4}{*}{$\textbf{r}_{\parallel}$}&\multirow{4}{*}{D$^\dag$/C$^\dag$}&&\multirow{4}{*}{$\Tilde{\mathcal{T}}_-$}&&$\prescript{g}{}{\overline{U}}^+_-=\Gamma\prescript{c}{}{U}^+_{++}$&1\\
         &1&$\prescript{g}{}{\overline{U}}^+_{+-}$&&&&&&$\prescript{c}{}{U}^+_-=\textrm{i}\Gamma\prescript{g}{}{\overline{U}}^+_{+-}$&2\\
         &2&$\prescript{c}{}{U}^+_{--}$&&&&&&$\prescript{g}{}{\overline{U}}^+_+=\Gamma\prescript{c}{}{U}^+_{--}$&3\\
         &3&$\prescript{g}{}{\overline{U}}^+_{-+}$&&&&&&$\prescript{c}{}{U}^+_+=\textrm{i}\Gamma\prescript{g}{}{\overline{U}}^+_{-+}$&0\\ \hline
    \end{tabular}
    \label{tab:DH AZ dag c to nc}
\end{table*}
Similar to Hermitian case, we wish to derive a dimensional hierarchy for AZ and AZ$^\dag$ classes. In this section, we focus on the dimensional hierarchy for point-gapped Hamiltonian without any spatial symmetries. By dimensional hierachy, we mean the following relationship (isomorphism)
\begin{align}
    &K_\mathbb{C}(s;d,D)=K_\mathbb{C}(s-d+D;0,0), \nonumber\\
    &s=0,1 \mod{2}
    \label{eq:Hierarchy complex AZ}
\end{align}
for complex AZ classes and
\begin{align}
    &K_{\mathbb{R}}(s;d,D)=K_{\mathbb{R}}(s-d+D;0,0),\nonumber\\
    &s=0,\cdots,7 \mod{8}
\end{align}
for real AZ classes and finally
\begin{align}
    &K_{\mathbb{R}}^\dag(s^\dag;d,D)=K_{\mathbb{R}}^\dag(s^\dag-d+D;0,0),\nonumber\\
    &s^\dag=0,\cdots,7 \mod{8}
    \label{eq:Hierarchy AZ dag}
\end{align}
for AZ$^\dag$ classes. 

\subsection{CS to non-CS classes}
First we consider AZ class. We first consider a mapping that send a chiral symmetry Hamiltonian $H_c$ to a Hamiltonian without chiral symmetry $H_{nc}$. 

Before discussing specific mappings, we specify the following requirements for the mappings. As mentioned in Sec.~\ref{Sec:struture of classification}, through continuous deformation, we can deform a point-gapped Hamiltonian into a unitary matrix $\mathcal{U}_{c}(\textbf{k},\textbf{r})$, which obeys $\mathcal{U}_{c}^\dag \mathcal{U}_{c}=1$. The classification problem then becomes the classification of extended Hermitian Hamiltonian $\Tilde{H}_{c}(\textbf{k},\textbf{r})$, which obeys $\Tilde{H}_{c}^2=1$ [See Eq.~\eqref{eq:extended Hermitian Hamiltonian}]. We introduce an extra dimension $\theta$ to $\Tilde{H}_{c}(\textbf{k},\textbf{r})$ which allows us to map the extended Hamiltonian to different symmetry classes. Let us call the mapping $\Tilde{H}_{nc}(\textbf{k},\textbf{r},\theta)$, which satisfy
\begin{equation}
    \Tilde{H}_{nc}(\textbf{k},\textbf{r},0)=const,\Tilde{H}_{nc}(\textbf{k},\textbf{r},\pi)=const'.
\end{equation}
The parameter $\theta$ can either be momentum-like or position-like. For a momentum-like (position-like) $\theta$, it transforms the same way as $\textbf{k}$ ($\textbf{r}$) under symmetries. In this case, the dimension $\delta$ of the Hamiltonian will increase (decrease) by $1$ [Recall that $\delta=d-D$].

 We may write the mapping as
\begin{equation}
    \Tilde{H}_{nc}(\textbf{k},\textbf{r},\theta)=\cos \theta \Tilde{H}_c (\textbf{k},\textbf{r})+\sin \theta H_2,
\end{equation}
where $H_2$ is the term that would make the mapped Hamiltonian $\Tilde{H}_{nc}(\textbf{k},\textbf{r},\theta)$ violate CS. Since $\Tilde{H}(\textbf{k},\textbf{r})$ is already in the form of an extended Hamiltonian, this further requires that $H_2$ is anti-diagonal and Hermitian. Finally, the condition $\Tilde{H}^2=1$ requires that $\{\Tilde{H}_c,H_2\}=0$. Since the CS has the form $\Gamma H_c(\textbf{k},\textbf{r},\theta) \Gamma^{-1}=-H^\dag_c(\textbf{k},\textbf{r},\theta)$, we see that by setting $H_2=\Tilde{\Gamma}$ could satisfy all the conditions mentioned above. We thus define the following mapping
\begin{align}
    \Tilde{H}_{nc} (\textbf{k},\textbf{r},\theta)=\cos \theta \Tilde{H}_c (\textbf{k},\textbf{r})+\sin \theta \Tilde{\Gamma},
    \label{eq:c to nc}
\end{align}
where $\theta \in [0,\pi]$. Notice that the mapping is defined for the extended Hermitian Hamiltonian~\eqref{eq:extended Hermitian Hamiltonian}. Due to the one to one relation between point-gapped Hamiltonian and its Hermitian extension, Eq.~\eqref{eq:c to nc} is a valid mapping to send a chiral symmetric point-gapped Hamiltonian to a non-chiral one. Now, by varying types of $\theta$ between momentum-like or position-like, Eq.~\eqref{eq:c to nc} provides a mapping between AZ classes with CS and without CS. In the following, we use some examples to illustrate this process. 

For complex AZ class, notice that since there is no anti-unitary symmetries, we cannot distinguish momentum-like or position-like $\theta$. Therefore, for both types of $\theta$, Eq.~\eqref{eq:c to nc} would send a Hamiltonian from class AIII ($s=0$) to class A ($s=1$).

For real AZ classes with CS, both TRS and PHS symmetries may present. We make the requirement that $[\mathcal{T}_+,\mathcal{C}_-]=0$. We may write the CS operator as
\begin{equation}
    \Gamma=\textrm{i}^{(s-2)/2}\mathcal{T}_+ \mathcal{C}_-,
\end{equation}
where $s$ is defined $s\mod{4}$. 

All the above derivations are also applicable to AZ$^\dag$ class. The mapping for AZ$^\dag$ class from a CS class to a non-CS one is the same as Eq.~\eqref{eq:c to nc}. However, the $\Gamma$ operator is now defined as
\begin{equation}
    \Gamma=\textrm{i}^{(s^\dag)/2}\mathcal{C}_+ \mathcal{T}_-,
\end{equation}
where we used the convention $[\mathcal{C}_+,\mathcal{T}_-]=0$ and $s^\dag$ is defined $s^\dag \mod{4}$. 

As an example, we consider $\theta$ is momentum-like. For AZ class, TRS is violated when $s=2 \mod{4}$ while PHS is violated when $s=0 \mod{4}$. This corresponds to a clockwise rotation $s\rightarrow s+1$. Similarly, if $\theta$ is position-like, TRS is violated when $s=0 \mod{4}$. This corresponds to a counter-clockwise rotation $s\rightarrow s-1$. A similar conclusion also apply to AZ$^\dag$ class. All the mappings are summarized in Table~\ref{tab:dimensional hierarchy mappings}.


To sum up this section, the mappings discussed above allow us to establish a one way mapping (homomorphism) from CS classes to non-CS classes i.e. 
\begin{align}
    &K_\mathbb{C}(s;d,D)\rightarrow K_\mathbb{C}(s-d+D;0,0), \nonumber\\
    &s=0
\end{align}
for complex AZ classes and
\begin{align}
    &K_{\mathbb{R}}(s;d,D)\rightarrow K_{\mathbb{R}}(s-d+D;0,0),\nonumber\\
    &s=0,2,4,6
\end{align}
for real AZ classes and finally
\begin{align}
    &K_{\mathbb{R}}^\dag(s^\dag;d,D)\rightarrow K_{\mathbb{R}}^\dag(s^\dag-d+D;0,0),\nonumber\\
    &s^\dag=0,2,4,6
\end{align}
for AZ$^\dag$ classes. In the next section, we consider one way mappings that would send a non-CS Hamiltonian to CS one such that the equal sign in Eq.~\eqref{eq:Hierarchy complex AZ} to~\eqref{eq:Hierarchy AZ dag} is justified. 
\subsection{Non-CS to CS classes}
We focus on the mapping from non-CS to CS classes for AZ classes in this section ($H_{nc}$ to $H_c$). We first list the process of constructing mappings. Similar to the previous section, we introduce an extra dimension $\theta$ to $\Tilde{H}_{nc}(\textbf{k},\textbf{r})$ such that the mapped Hamiltonian obeys CS. We define the mappings
\begin{equation}
    H_c (\textbf{k},\textbf{r},\theta)=\cos \theta H_1 (\textbf{k},\textbf{r})+\sin \theta H_2,
    \label{eq:nc to c}
\end{equation}
where $H_1 (\textbf{k},\textbf{r})$ is $\Tilde{H}_{nc}(\textbf{k},\textbf{r})$ or $\textbf{i}\Tilde{H}_{nc}(\textbf{k},\textbf{r})$in a larger Clifford algebra dimension (See Table~\ref{tab:dimensional hierarchy mappings}); $H_2$ is an extra term that made $H_c$ to obeys CS. Notice that unlike Eq.~\eqref{eq:c to nc}, the mapped Hamiltonian in Eq.~\eqref{eq:nc to c} is \emph{not} an extended Hamiltonian. Instead, $H_c$ is a point-gapped NH Hamiltonian that obeys unitary condition $H_c^\dag H_c=1$. Finally, we need to make sure that PHS and TRS of the mapped Hamiltonian $H_c$ commute with each other, as this is the convention we use in this papaer.

Bear these requirements in mind, we are now ready to discuss specific mappings.

First we consider a mapping from AI/AII to BDI/CII:
\begin{equation}
    H_c(\textbf{k},\textbf{r},\theta)=\cos\theta \Tilde{H}_{nc}(\textbf{k},\textbf{r})\otimes\tau_z+\sin \theta \textrm{i}\otimes\tau_0,
\end{equation}
where $ \Tilde{H}_{nc}$ obeys TRS with operator $\Tilde{\mathcal{T}}_+$.
If $\theta$ is a momentum-like parameter, the mapped Hamiltonian $H_c$ obeys TRS with $\Tilde{\mathcal{T}}_+\otimes \tau_0$; PHS with $\Tilde{\mathcal{T}}_+\otimes \tau_x$. These two symmetries combined, $H_c$ obeys CS with $1\otimes \tau_x$. Furthermore, since the extended Hamiltonian obeys $\Tilde{H}_{nc}^2=1$, we see that $H_c^\dag H_c=1$. This corresponds to a shift $s\rightarrow s+1$ where $s=1$ (AI) or $5$ (AII).

Next, we consider a mapping that would map AI/AII to CI/DIII:
\begin{equation}
    H_c(\textbf{k},\textbf{r},\theta)=\cos\theta \Tilde{H}_{nc}(\textbf{k},\textbf{r})\otimes\tau_z+\sin \theta\textrm{i}\Sigma\otimes\tau_x.
\end{equation}
The $\Sigma$ is the $\Sigma$ symmetry operator of $\Tilde{H}_{nc}(\textbf{k},\textbf{r})$. If $\theta$ is position-like, the mapped Hamiltonian $H_c$ obeys TRS with $\Tilde{\mathcal{T}}_+ \otimes \tau_z$ and PHS with $\Tilde{\mathcal{T}}_+\otimes \tau_y$. Combining these two symmetries, $H_c$ obeys CS with $1\otimes \tau_x$. This corresponds to a shift $s\rightarrow s-1$ where $s=1$ or $5$. 

By following similar process, we can find a mapping for class D/C and A into adjacent CS classes. For AZ$^\dag$ classes, the process of finding mappings into CS classes is essentially the same which we do not repeat here. The mapping and resulting symmetries are summarized in Table~\ref{tab:dimensional hierarchy mappings}. 

Based on the above discussion, we have homomorphisms that map
\begin{align}
    &K_\mathbb{C}(s;d,D)\rightarrow K_\mathbb{C}(s-d+D;0,0), \nonumber\\
    &s=1
\end{align}
for complex AZ classes and
\begin{align}
    &K_{\mathbb{R}}(s;d,D)\rightarrow K_{\mathbb{R}}(s-d+D;0,0),\nonumber\\
    &s=1,3,5,7
\end{align}
for real AZ classes and finally
\begin{align}
    &K_{\mathbb{R}}^\dag(s^\dag;d,D)\rightarrow K_{\mathbb{R}}^\dag(s^\dag-d+D;0,0),\nonumber\\
    &s^\dag=1,3,5,7.
\end{align}
This allows us to conclude the isomorphism Eq.~\eqref{eq:Hierarchy complex AZ} to~\eqref{eq:Hierarchy AZ dag}. 


\section{Dimensional Hierarchy of AZ and AZ$^\dag$ with order-two spatial symmetries}\label{Sec:DH AZ and AZ dag with spatial}
In this section, we proof the dimensional hierarchy of AZ and AZ$^\dag$ in the presence of spatial symmetries. 
\subsection{Complex AZ with order-two unitary symmetries}
Consider a Hamiltonian $H_{nc}$ in class A with an additional symmetry $\prescript{c}{}{\overline{U}}$. According to Table~\ref{Tab:Unitary+Complex Classes A and AIII}, this belongs to class $(s=0,t=0)$. We consider the extended version $\prescript{c}{}{\Tilde{\overline{U}}}$ on the mapped Hamiltonian in class AIII
\begin{equation}
    H_c(\textbf{k},\textbf{r},\theta)=\cos \theta \Tilde{H}_{nc}(\textbf{k},\textbf{r})\otimes \tau_z+\sin \theta \textrm{i}\Sigma \otimes \tau_x
\end{equation}
as listed in Table~\ref{tab:dimensional hierarchy mappings}. Once again, due to the absence of anti-unitary symmetries, we cannot distinguish whether $\theta$ is momentum-like or position-like. Therefore, we only need to consider whether $\theta$ will change under the spatial symmetries. If $\theta$ is $\textbf{k}_{\perp}/\textbf{r}_{\perp}$ i.e. unchanged under the spatial symmetries, we can define operator $\prescript{g}{}{\overline{U}}_+=\prescript{c}{}{\Tilde{\overline{U}}}\otimes \tau_0$, since
\begin{equation}
    \prescript{c}{}{\Tilde{\overline{U}}}\otimes \tau_0 H_c (\textbf{k},\textbf{r},\theta)\prescript{c}{}{\Tilde{\overline{U}}}\otimes \tau_0=-H^\dag (-\textbf{k}_{\parallel},\textbf{k}_{\perp},-\textbf{r}_{\parallel},\textbf{r}_{\perp},\theta).
\end{equation}
Then the mapped Hamiltonian is now in $(s=1,t=0)$. This process provide a homomorphism $K_{\mathbb{C}}^U(s=0,t=0;d+D,d_{\parallel}+D_{\parallel})\rightarrow K_{\mathbb{C}}^U(s=1,t=0;d+D+1,d_{\parallel}+D_{\parallel})$. Following similar process for other spatial symmetries, we arrive at Table~\ref{tab:DH complex AZ}, which allows us to conclude
\begin{align}
    &K_{\mathbb{C}}^U(s,t;d+D,d_{\parallel}+D_{\parallel},0,0)\nonumber\\
    &=K_{\mathbb{C}}^U(s+1,t;d+D+1,d_{\parallel}+D_{\parallel},0,0)\nonumber\\
    &=K_{\mathbb{C}}^U(s+1,t+1;d+D+1,d_{\parallel}+D_{\parallel}+1,0,0).
\end{align}
This proof the relationship
\begin{align}
    &K_{\mathbb{C}}^U(s,t;d+D,d_{\parallel}+D_{\parallel},0,0)\nonumber\\
    &=K_{\mathbb{C}}^U(s-d+D,t-d_{\parallel}+D_{\parallel};0,0,0,0).
\end{align}
\subsection{Complex AZ with order-two anti-unitary symmetries}
As discussed in Sec.~\ref{sec:A and AIII+antyiunitary}, complex AZ classes with anti-unitary spatial symmetries will map the original class into a real AZ or AZ$^\dag$ class. The dimensional hierarchy is given by Eq.\eqref{eq:dimensional hierarchy complex AZ antiunitary}.
\subsection{Real AZ and AZ$^\dag$ with order-two symmetries}
We consider a Hamiltonian in class AI with symmetry $\prescript{c}{}{U}^+_+$. According to Table~\ref{Tab:Symmetry_type AZ}, this place the Hamiltonian in class $(s=0,t=0)$. The mapping 
\begin{equation}
    H_c(\textbf{k},\textbf{r},\theta)=\cos\theta \Tilde{H}_{nc}(\textbf{k},\textbf{r})\otimes\tau_z+\sin \theta 
         \textrm{i}\otimes\tau_0
\end{equation}
provided in Table~\ref{tab:dimensional hierarchy mappings} allows to map $H_{nc}$ to class BDI for a momentum-like $\theta$. If the $\theta$ is \emph{not} flipped under spatial symmetry i.e it is of the type $\textbf{k}_{\perp}$, the mapped Hamiltonian $H_c$ obeys $\prescript{c}{}{U}^+_{++}=\prescript{c}{}{\Tilde{U}}^+_+\otimes \tau_0$. This place $H_c$ in the class $(s=1,t=0)$. Recall that in Sec.~\ref{Sec:order-two spatial symmetry AZ and AZ dag} we showed that all anti-unitary spatial symmetries are equivalent to some of the unitary ones for real AZ classes (See Table~\ref{Tab:Symmetry_type AZ}). The process above then provide a homomorphsim $K_{\mathbb{R}}^{U/A}(s=0,t=0;d,d_{\parallel},D,D_{\parallel})\rightarrow K_{\mathbb{R}}^{U/A}(s=1,t=0;d+1,d_{\parallel},D,D_{\parallel})$. Following similar process for other spatial symmetries and AZ classes, we arrive at Table~\ref{tab:DH AZ nc to c} and~\ref{tab:DH AZ c to nc}. These two tables allow us to conclude the relationships
\begin{align}
    &K_{\mathbb{R}}^{U/A}(s,t;d,d_{\parallel},D,D_{\parallel})\nonumber\\
    &=K_{\mathbb{R}}^{U/A}(s+1,t;d+1,d_{\parallel},D,D_{\parallel})\nonumber\\
    &=K_{\mathbb{R}}^{U/A}(s+1,t+1;d+1,d_{\parallel}+1,D,D_{\parallel})\nonumber\\
    &=K_{\mathbb{R}}^{U/A}(s-1,t;d,d_{\parallel},D+1,D_{\parallel})\nonumber\\
    &=K_{\mathbb{R}}^{U/A}(s-1,t-1;d,d_{\parallel},D+1,D_{\parallel}+1)
    \label{eq:K group relation AZ}
\end{align}
which leads to
\begin{align}
    &K_{\mathbb{R}}^{U/A}(s,t;d,d_{\parallel},D,D_{\parallel})\nonumber\\
    &=K_{\mathbb{R}}^{U/A}(s-d+D,t-d_{\parallel}+D_{\parallel};0,0,0,0).
\end{align}
A similar analysis of AZ$^\dag$ classes with spatial symmetries results in Table~\ref{tab:DH AZ dag nc to c} and~\ref{tab:DH AZ dag c to nc}, which allows us to conclude the relationship
\begin{align}
    &K_{\mathbb{R}}^{\dag U/A}(s,t;d,d_{\parallel},D,D_{\parallel})\nonumber\\
    &=K_{\mathbb{R}}^{\dag U/A}(s+1,t;d+1,d_{\parallel},D,D_{\parallel})\nonumber\\
    &=K_{\mathbb{R}}^{\dag U/A}(s+1,t-1;d+1,d_{\parallel}+1,D,D_{\parallel})\nonumber\\
    &=K_{\mathbb{R}}^{\dag U/A}(s-1,t;d,d_{\parallel},D+1,D_{\parallel})\nonumber\\
    &=K_{\mathbb{R}}^{\dag U/A}(s-1,t+1;d,d_{\parallel},D+1,D_{\parallel}+1).
    \label{eq:K group relation AZ dag}
\end{align}
Notice that for AZ$^\dag$ class, the effect of $d_{\parallel}+1$ on $t$ is reversed [compare third and fifth row of Eqs.~\eqref{eq:K group relation AZ} and~\eqref{eq:K group relation AZ dag}]. Eq.~\eqref{eq:K group relation AZ dag} allows us to conclude
\begin{align}
    &K_{\mathbb{R}}^{\dag U/A}(s,t;d,d_{\parallel},D,D_{\parallel})\nonumber\\
    &=K_{\mathbb{R}}^{\dag U/A}(s-d+D,t+d_{\parallel}-D_{\parallel};0,0,0,0).
\end{align}

\section{Properties of Clifford Algebra}\label{Appendix:Properties of Clifford Algebra}
In this section, we list several useful properties that Clifford Algebra obeys~\cite{Morimoto_2013,Shiozaki_2014}.

The complex Clifford algebras obeys
\begin{align}
    Cl_1 &\simeq \mathbb{C}\oplus\mathbb{C}\\
    Cl_2 &\simeq \mathbb{C}(2)\\
    Cl_{n+2}&\simeq Cl_n \oplus \mathbb{C}(2).
\end{align}
Notably, $\mathbb{C}(2)$ that represents $2\times 2$ complex matrices does not affect the extension problem. This means the extension problem $Cl_p\otimes \mathbb{C}(2)\rightarrow Cl_p \otimes \mathbb{C}(2)$ is equivalent to the extension problem $Cl_p \rightarrow Cl_{p+1}$. 

The real Clifford algebras obeys
\begin{align}
    Cl_{0,1}&\simeq \mathbb{R}\oplus \mathbb{R}\\
    Cl_{0,2}&\simeq \mathbb{R}(2)\\
    Cl_{1,0}&\simeq \mathbb{C}\\
    Cl_{2,0}&\simeq \mathbb{H}\\
    Cl_{p+1,q+1}&\simeq Cl_{p,q}\otimes \mathbb{R}\\
    \label{eq:clifford algbra enlarge relation}
    Cl_{p,q}\otimes Cl_{0,1}&\simeq Cl_{p,q}\oplus Cl_{p,q}\\
    Cl_{p,q}\otimes Cl_{1,0}&\simeq Cl_{p+q}\\
    Cl_{p,q}\otimes Cl_{0,2}&\simeq Cl_{q,p+2}\\
    Cl_{p,q}\otimes Cl_{2,0}&\simeq Cl_{q+2,p},
\end{align}
where $\mathbb{H}$ denotes the set of quaternions, $\mathbb{R}(2)$ denotes the $2\times 2$ real matrices. Similar to the previous case, $\mathbb{R}(2)$ does not affect the classification. This means extension problem $Cl_{p,q}\otimes \mathbb{R}(2)\rightarrow Cl_{p,q+1}\otimes \mathbb{R}(2)$ is equivalent to the extension problem $Cl_{p,q}\rightarrow Cl_{p,q+1}$, and similar for $p$ to $p+1$.

\section{Classifying space of AZ and AZ$^\dag$ with spatial symmetries from Clifford algebra}\label{Appendix:classifying space}

\begin{table*}[!]
    \begin{center}
    \caption{Classification of $16$-fold AZ and AZ$^\dag$ classes under additional order-two unitary symmetries. Specific symmetries for each class is given in table~\ref{Tab:Symmetry_type AZ} for AZ classes and table~\ref{Tab:Symmetry_type AZ dag} for AZ$^\dag$ classes.}
    \label{tab:Clifford generators of AZ and AZ dag}
    \begin{tabular}[t]{ccccc}
\hline\hline
Class & Symmetry & Generators & Extensions & Classifying space \\ \hline
 \multirow{2}{*}{A}& $t_0$ & $\{\Tilde{H},\Sigma\}\otimes \{\Tilde{\prescript{c}{}{U}}\}$ & $Cl_1 \otimes Cl_1 \rightarrow Cl_2\otimes Cl_1$ & $\mathcal{C}_1\times \mathcal{C}_1$\\
 & $t_1$ & $\{\Tilde{H},\Sigma,\Sigma \Tilde{\prescript{g}{}{U}}\}$ & $Cl_2\rightarrow Cl_3$ & $\mathcal{C}_0$ \\ \hline
 
 \multirow{2}{*}{AIII}& $t_0$ & $\{\Tilde{H}, \Sigma,\Tilde{\Gamma}\}\otimes \{\Sigma \Tilde{\Gamma} \Tilde{\prescript{g}{}{U}}_-\}$ & $Cl_2\otimes Cl_1 \rightarrow Cl_3 \otimes Cl_1$ & $\mathcal{C}_0 \times \mathcal{C}_0$\\
 & $t_1$ & $\{\Tilde{H}, \Sigma,\Tilde{\Gamma},\Sigma \Tilde{\prescript{g}{}{U}}_+\}$ & $Cl_3 \rightarrow Cl_4$ & $\mathcal{C}_1$ \\ \hline
 
 \multirow{4}{*}{AI}& $t_0$ & $\{J \Tilde{H},J \Sigma,\Tilde{\mathcal{T}}_+, J\Tilde{\mathcal{T}}_+\}\otimes \{\Tilde{{\prescript{c}{}{U}}}^+_+\}$ & $Cl_{1,2}\otimes Cl_{0,1}\rightarrow Cl_{2,2}\otimes Cl_{0,1}$ & $\mathcal{R}_1\times \mathcal{R}_1$\\
 & $t_1$ & $\{J \Tilde{H},J \Sigma,\Tilde{\mathcal{T}}_+, J\Tilde{\mathcal{T}}_+, J \Sigma \Tilde{\prescript{g}{}{U}}^+_+\}$ & $Cl_{1,3}\rightarrow Cl_{2,3}$ & $\mathcal{R}_0$ \\
 & $t_2$ & $\{J \Tilde{H},J \Sigma,\Tilde{\mathcal{T}}_+, J\Tilde{\mathcal{T}}_+\}\otimes \{\Tilde{\prescript{c}{}{U}}^-_+\}$ & $Cl_{1,2}\otimes Cl_{1,0}\rightarrow Cl_{2,2}\otimes Cl_{1,0}$ & $\mathcal{C}_1$\\
 & $t_3$ & $\{J \Tilde{H},J \Sigma,\Tilde{\mathcal{T}}_+, J\Tilde{\mathcal{T}}_+, J \Sigma \Tilde{\prescript{g}{}{U}}^-_+\}$ & $Cl_{2,2}\rightarrow Cl_{3,2}$ & $\mathcal{R}_2$ \\ \hline

 \multirow{4}{*}{BDI} & $t_0$  & $\{\Tilde{H},\Tilde{\mathcal{C}}_-,J\Tilde{\mathcal{C}}_-,J\Tilde{\mathcal{C}}_- \Tilde{\mathcal{T}}_+,\Sigma\}\otimes\{\Tilde{\prescript{c}{}{U}}^+_{++}\}$ & $Cl_{1,3}\otimes Cl_{0,1}\rightarrow Cl_{1,4}\otimes Cl_{0,1}$ & $\mathcal{R}_2\times \mathcal{R}_2$ \\
 & $t_1$ & $\{\Tilde{H},\Tilde{\mathcal{C}}_-,J\Tilde{\mathcal{C}}_-,J\Tilde{\mathcal{C}}_- \Tilde{\mathcal{T}}_+,\Sigma,\Sigma\Tilde{\prescript{g}{}{U}}^+_{++}\}$ & $Cl_{2,3}\rightarrow Cl_{2,4}$ &  $\mathcal{R}_1$ \\
 &  $t_2$ & $\{\Tilde{H},\Tilde{\mathcal{C}}_-,J\Tilde{\mathcal{C}}_-,J\Tilde{\mathcal{C}}_- \Tilde{\mathcal{T}}_+,\Sigma\}\otimes\{\Tilde{\prescript{c}{}{U}}^-_{++}\}$ & $Cl_{1,3}\otimes Cl_{1,0}\rightarrow Cl_{1,4}\otimes Cl_{1,0}$ & $\mathcal{C}_0$\\
 & $t_3$ & $\{\Tilde{H},\Tilde{\mathcal{C}}_-,J\Tilde{\mathcal{C}}_-,J\Tilde{\mathcal{C}}_- \Tilde{\mathcal{T}}_+,\Sigma,\Sigma\Tilde{\prescript{g}{}{U}}^-_{++}\}$ &$Cl_{1,4}\rightarrow Cl_{1,5}$   & $\mathcal{R}_3$\\ \hline

 \multirow{4}{*}{D} & $t_0$ & $\{\Tilde{H},\Tilde{C}_-,J \Tilde{C}_-,\Sigma\}\otimes \{\Tilde{\prescript{c}{}{U}}^+_+\}$ &$Cl_{0,3}\otimes Cl_{0,1}\rightarrow Cl_{0,4}\otimes Cl_{0,1}$  & $\mathcal{R}_3\times \mathcal{R}_3$\\
 & $t_1$ & $\{\Tilde{H},\Tilde{C}_-,J \Tilde{C}_-,\Sigma,\Sigma \Tilde{\prescript{g}{}{U}}^+_+\}$ & $Cl_{1,3}\rightarrow Cl_{1,4}$ & $\mathcal{R}_2$\\
 & $t_2$ & $\{\Tilde{H},\Tilde{C}_-,J \Tilde{C}_-,\Sigma\}\otimes \{\Tilde{\prescript{c}{}{U}}^-_+\}$ & $Cl_{0,3}\otimes Cl_{1,0}\rightarrow Cl_{0,4}\otimes Cl_{0,4}\otimes Cl_{1,0}$ & $\mathcal{C}_1$\\
 & $t_3$ & $\{\Tilde{H},\Tilde{C}_-,J \Tilde{C}_-,\Sigma,\Sigma \Tilde{\prescript{g}{}{U}}^-_+\}$ & $Cl_{0,4}\rightarrow Cl_{0,5}$ & $\mathcal{R}_4$\\ \hline

  \multirow{4}{*}{DIII} & $t_0$  & $\{\Tilde{H},\Tilde{\mathcal{C}}_-,J\Tilde{\mathcal{C}}_-,J\Tilde{\mathcal{C}}_- \Tilde{\mathcal{T}}_+,\Sigma\}\otimes\{\Tilde{\prescript{c}{}{U}}^+_{++}\}$ & $Cl_{0,4}\otimes Cl_{0,1}\rightarrow Cl_{0,5}\otimes Cl_{0,1}$ & $\mathcal{R}_4\times \mathcal{R}_4$ \\
 & $t_1$ & $\{\Tilde{H},\Tilde{\mathcal{C}}_-,J\Tilde{\mathcal{C}}_-,J\Tilde{\mathcal{C}}_- \Tilde{\mathcal{T}}_+,\Sigma,\Sigma\Tilde{\prescript{g}{}{U}}^+_{++}\}$ & $Cl_{1,4}\rightarrow Cl_{1,5}$ &  $\mathcal{R}_3$ \\
 &  $t_2$ & $\{\Tilde{H},\Tilde{\mathcal{C}}_-,J\Tilde{\mathcal{C}}_-,J\Tilde{\mathcal{C}}_- \Tilde{\mathcal{T}}_+,\Sigma\}\otimes\{\Tilde{\prescript{c}{}{U}}^-_{++}\}$ & $Cl_{0,4}\otimes Cl_{1,0}\rightarrow Cl_{0,5}\otimes Cl_{1,0}$ & $\mathcal{C}_0$\\
 & $t_3$ & $\{\Tilde{H},\Tilde{\mathcal{C}}_-,J\Tilde{\mathcal{C}}_-,J\Tilde{\mathcal{C}}_- \Tilde{\mathcal{T}}_+,\Sigma,\Sigma\Tilde{\prescript{g}{}{U}}^-_{++}\}$ &$Cl_{0,5}\rightarrow Cl_{0,6}$   & $\mathcal{R}_5$\\ \hline

  \multirow{4}{*}{AII}& $t_0$ & $\{J \Tilde{H},J \Sigma,\Tilde{\mathcal{T}}_+, J\Tilde{\mathcal{T}}_+\}\otimes \{\Tilde{{\prescript{c}{}{U}}}^+_+\}$ & $Cl_{3,0}\otimes Cl_{0,1}\rightarrow Cl_{4,0}\otimes Cl_{0,1}$ & $\mathcal{R}_5\times \mathcal{R}_5$\\
 & $t_1$ & $\{J \Tilde{H},J \Sigma,\Tilde{\mathcal{T}}_+, J\Tilde{\mathcal{T}}_+, J \Sigma \Tilde{\prescript{g}{}{U}}^+_+\}$ & $Cl_{3,1}\rightarrow Cl_{4,1}$ & $\mathcal{R}_4$ \\
 & $t_2$ & $\{J \Tilde{H},J \Sigma,\Tilde{\mathcal{T}}_+, J\Tilde{\mathcal{T}}_+\}\otimes \{\Tilde{\prescript{c}{}{U}}^-_+\}$ & $Cl_{3,0}\otimes Cl_{1,0}\rightarrow Cl_{4,0}\otimes Cl_{1,0}$ & $\mathcal{C}_1$\\
 & $t_3$ & $\{J \Tilde{H},J \Sigma,\Tilde{\mathcal{T}}_+, J\Tilde{\mathcal{T}}_+, J \Sigma \Tilde{\prescript{g}{}{U}}^-_+\}$ & $Cl_{4,0}\rightarrow Cl_{5,0}$ & $\mathcal{R}_6$ \\ \hline

\multirow{4}{*}{CII} & $t_0$  & $\{\Tilde{H},\Tilde{\mathcal{C}}_-,J\Tilde{\mathcal{C}}_-,J\Tilde{\mathcal{C}}_- \Tilde{\mathcal{T}}_+,\Sigma\}\otimes\{\Tilde{\prescript{c}{}{U}}^+_{++}\}$ & $Cl_{3,1}\otimes Cl_{0,1}\rightarrow Cl_{3,2}\otimes Cl_{0,1}$ & $\mathcal{R}_6\times \mathcal{R}_6$ \\
 & $t_1$ & $\{\Tilde{H},\Tilde{\mathcal{C}}_-,J\Tilde{\mathcal{C}}_-,J\Tilde{\mathcal{C}}_- \Tilde{\mathcal{T}}_+,\Sigma,\Sigma\Tilde{\prescript{g}{}{U}}^+_{++}\}$ & $Cl_{4,1}\rightarrow Cl_{4,2}$ &  $\mathcal{R}_5$ \\
 &  $t_2$ & $\{\Tilde{H},\Tilde{\mathcal{C}}_-,J\Tilde{\mathcal{C}}_-,J\Tilde{\mathcal{C}}_- \Tilde{\mathcal{T}}_+,\Sigma\}\otimes\{\Tilde{\prescript{c}{}{U}}^-_{++}\}$ & $Cl_{3,1}\otimes Cl_{1,0}\rightarrow Cl_{3,2}\otimes Cl_{1,0}$ & $\mathcal{C}_0$\\
 & $t_3$ & $\{\Tilde{H},\Tilde{\mathcal{C}}_-,J\Tilde{\mathcal{C}}_-,J\Tilde{\mathcal{C}}_- \Tilde{\mathcal{T}}_+,\Sigma,\Sigma\Tilde{\prescript{g}{}{U}}^-_{++}\}$ &$Cl_{3,2}\rightarrow Cl_{3,3}$   & $\mathcal{R}_7$\\ \hline

\multirow{4}{*}{C} & $t_0$ & $\{\Tilde{H},\Tilde{C}_-,J \Tilde{C}_-,\Sigma\}\otimes \{\Tilde{\prescript{c}{}{U}}^+_+\}$ &$Cl_{2,1}\otimes Cl_{0,1}\rightarrow Cl_{2,2}\otimes Cl_{0,1}$  & $\mathcal{R}_7\times \mathcal{R}_7$\\
 & $t_1$ & $\{\Tilde{H},\Tilde{C}_-,J \Tilde{C}_-,\Sigma,\Sigma \Tilde{\prescript{g}{}{U}}^+_+\}$ & $Cl_{3,1}\rightarrow Cl_{3,2}$ & $\mathcal{R}_6$\\
 & $t_2$ & $\{\Tilde{H},\Tilde{C}_-,J \Tilde{C}_-,\Sigma\}\otimes \{\Tilde{\prescript{c}{}{U}}^-_+\}$ & $Cl_{2,1}\otimes Cl_{1,0}\rightarrow Cl_{2,2}\otimes Cl_{0,4}\otimes Cl_{1,0}$ & $\mathcal{C}_1$\\
 & $t_3$ & $\{\Tilde{H},\Tilde{C}_-,J \Tilde{C}_-,\Sigma,\Sigma \Tilde{\prescript{g}{}{U}}^-_+\}$ & $Cl_{2,2}\rightarrow Cl_{2,3}$ & $\mathcal{R}_0$\\ \hline

\multirow{4}{*}{CI} & $t_0$  & $\{\Tilde{H},\Tilde{\mathcal{C}}_-,J\Tilde{\mathcal{C}}_-,J\Tilde{\mathcal{C}}_- \Tilde{\mathcal{T}}_+,\Sigma\}\otimes\{\Tilde{\prescript{c}{}{U}}^+_{++}\}$ & $Cl_{2,2}\otimes Cl_{0,1}\rightarrow Cl_{2,3}\otimes Cl_{0,1}$ & $\mathcal{R}_0\times \mathcal{R}_0$ \\
 & $t_1$ & $\{\Tilde{H},\Tilde{\mathcal{C}}_-,J\Tilde{\mathcal{C}}_-,J\Tilde{\mathcal{C}}_- \Tilde{\mathcal{T}}_+,\Sigma,\Sigma\Tilde{\prescript{g}{}{U}}^+_{++}\}$ & $Cl_{3,2}\rightarrow Cl_{3,3}$ &  $\mathcal{R}_7$ \\
 &  $t_2$ & $\{\Tilde{H},\Tilde{\mathcal{C}}_-,J\Tilde{\mathcal{C}}_-,J\Tilde{\mathcal{C}}_- \Tilde{\mathcal{T}}_+,\Sigma\}\otimes\{\Tilde{\prescript{c}{}{U}}^-_{++}\}$ & $Cl_{2,2}\otimes Cl_{1,0}\rightarrow Cl_{2,3}\otimes Cl_{1,0}$ & $\mathcal{C}_0$\\
 & $t_3$ & $\{\Tilde{H},\Tilde{\mathcal{C}}_-,J\Tilde{\mathcal{C}}_-,J\Tilde{\mathcal{C}}_- \Tilde{\mathcal{T}}_+,\Sigma,\Sigma\Tilde{\prescript{g}{}{U}}^-_{++}\}$ &$Cl_{2,3}\rightarrow Cl_{2,4}$   & $\mathcal{R}_1$\\ \hline

 \multirow{4}{*}{AI$^\dag$}& $t_0$ & $\{J\Tilde{H},\Sigma,\Tilde{\mathcal{C}}_+,J\Tilde{\mathcal{C}}_+\}\otimes \{\Tilde{\prescript{c}{}{U}}^+_+\}$ & $Cl_{0,3}\otimes Cl_{0,1}\rightarrow Cl_{1,3} \otimes Cl_{0,1}$ & $\mathcal{R}_7 \times \mathcal{R}_7$\\
 & $t_1$ & $\{J\Tilde{H},\Sigma,\Tilde{\mathcal{C}}_+,J\Tilde{\mathcal{C}}_+,\Sigma\Tilde{\prescript{g}{}{U}}^+_+\}$ & $Cl_{1,3}\rightarrow Cl_{2,3}$ & $\mathcal{R}_0$\\ 
 & $t_2$ & $\{J\Tilde{H},\Sigma,\Tilde{\mathcal{C}}_+,J\Tilde{\mathcal{C}}_+\}\otimes \{\Tilde{\prescript{c}{}{U}}^-_+\}$ & $Cl_{0,3}\otimes Cl_{1,0}\rightarrow Cl_{1,3}\otimes Cl_{1,0}$ & $\mathcal{C}_1$ \\
 & $t_3$ & $\{J\Tilde{H},\Sigma,\Tilde{\mathcal{C}}_+,J\Tilde{\mathcal{C}}_+,\Sigma\Tilde{\prescript{g}{}{U}}^-_+\}$ & $Cl_{0,4}\rightarrow Cl_{1,4}$ & $\mathcal{R}_6$\\
 \hline

\multirow{4}{*}{BDI$^\dag$}& $t_0$  & $\{J\Tilde{H},\Tilde{\mathcal{C}}_+,J\Tilde{\mathcal{C}}_+,J\Tilde{\mathcal{C}}_+ \Tilde{\mathcal{T}}_-,\Sigma\}\otimes\{\Tilde{\prescript{c}{}{U}}^+_{++}\}$ & $Cl_{1,3}\otimes Cl_{0,1}\rightarrow Cl_{2,3}\otimes Cl_{0,1}$ & $\mathcal{R}_0\times \mathcal{R}_0$ \\
 & $t_1$ & $\{J\Tilde{H},\Tilde{\mathcal{C}}_+,J\Tilde{\mathcal{C}}_+,J\Tilde{\mathcal{C}}_+ \Tilde{\mathcal{T}}_-,\Sigma,\Sigma\Tilde{\prescript{g}{}{U}}^+_{++}\}$ & $Cl_{2,3}\rightarrow Cl_{3,3}$ &  $\mathcal{R}_1$ \\
 &  $t_2$ & $\{J\Tilde{H},\Tilde{\mathcal{C}}_+,J\Tilde{\mathcal{C}}_+,J\Tilde{\mathcal{C}}_+ \Tilde{\mathcal{T}}_-,\Sigma\}\otimes\{\Tilde{\prescript{c}{}{U}}^-_{++}\}$ & $Cl_{1,3}\otimes Cl_{1,0}\rightarrow Cl_{2,3}\otimes Cl_{1,0}$ & $\mathcal{C}_0$\\
 & $t_3$ & $\{J\Tilde{H},\Tilde{\mathcal{C}}_+,J\Tilde{\mathcal{C}}_+,J\Tilde{\mathcal{C}}_+ \Tilde{\mathcal{T}}_-,\Sigma,\Sigma\Tilde{\prescript{g}{}{U}}^-_{++}\}$ &$Cl_{1,4}\rightarrow Cl_{2,4}$   & $\mathcal{R}_7$\\ \hline

\multirow{4}{*}{D$^\dag$}& $t_0$&$\{\Tilde{H},J \Sigma,\Tilde{T}_-,J \Tilde{T}_-\}\otimes \{\prescript{c}{}{\Tilde{U}}^+_+\}$ & $Cl_{1,2}\otimes Cl_{0,1}\rightarrow Cl_{1,3}\otimes Cl_{0,1}$ & $\mathcal{R}_1 \times \mathcal{R}_1$\\
& $t_1$ & $\{\Tilde{H},J \Sigma,\Tilde{T}_-,J \Tilde{T}_-,J \Sigma \prescript{g}{}{\Tilde{U}}^+_+\}$ & $Cl_{1,3}\otimes Cl_{1,4}$ & $\mathcal{R}_2$\\
& $t_2$ & $\{\Tilde{H},J \Sigma,\Tilde{T}_-,J \Tilde{T}_-\}\otimes \{\prescript{c}{}{\Tilde{U}}^-_+\}$ & $Cl_{1,2}\otimes Cl_{1,0}\rightarrow Cl_{1,3}\otimes Cl_{1,0}$ & $\mathcal{C}_1$\\
& $t_3$ & $\{\Tilde{H},J \Sigma,\Tilde{T}_-,J \Tilde{T}_-,J \Sigma \prescript{g}{}{\Tilde{U}}^-_+\}$ & $Cl_{2,2}\otimes Cl_{2,3}$ & $\mathcal{R}_0$\\ \hline

 \multirow{4}{*}{DIII$^\dag$}& $t_0$  & $\{J\Tilde{H},\Tilde{\mathcal{C}}_+,J\Tilde{\mathcal{C}}_+,J\Tilde{\mathcal{C}}_+ \Tilde{\mathcal{T}}_-,\Sigma\}\otimes\{\Tilde{\prescript{c}{}{U}}^+_{++}\}$ & $Cl_{2,2}\otimes Cl_{0,1}\rightarrow Cl_{3,2}\otimes Cl_{0,1}$ & $\mathcal{R}_2\times \mathcal{R}_2$ \\
 & $t_1$ & $\{J\Tilde{H},\Tilde{\mathcal{C}}_+,J\Tilde{\mathcal{C}}_+,J\Tilde{\mathcal{C}}_+ \Tilde{\mathcal{T}}_-,\Sigma,\Sigma\Tilde{\prescript{g}{}{U}}^+_{++}\}$ & $Cl_{3,2}\rightarrow Cl_{4,2}$ &  $\mathcal{R}_3$ \\
 &  $t_2$ & $\{J\Tilde{H},\Tilde{\mathcal{C}}_+,J\Tilde{\mathcal{C}}_+,J\Tilde{\mathcal{C}}_+ \Tilde{\mathcal{T}}_-,\Sigma\}\otimes\{\Tilde{\prescript{c}{}{U}}^-_{++}\}$ & $Cl_{2,2}\otimes Cl_{1,0}\rightarrow Cl_{3,2}\otimes Cl_{1,0}$ & $\mathcal{C}_0$\\
 & $t_3$ & $\{J\Tilde{H},\Tilde{\mathcal{C}}_+,J\Tilde{\mathcal{C}}_+,J\Tilde{\mathcal{C}}_+ \Tilde{\mathcal{T}}_-,\Sigma,\Sigma\Tilde{\prescript{g}{}{U}}^-_{++}\}$ &$Cl_{2,3}\rightarrow Cl_{3,3}$   & $\mathcal{R}_1$\\ \hline

\multirow{4}{*}{AII$^\dag$}& $t_0$ & $\{J\Tilde{H},\Sigma,\Tilde{\mathcal{C}}_+,J\Tilde{\mathcal{C}}_+\}\otimes \{\Tilde{\prescript{c}{}{U}}^+_+\}$ & $Cl_{2,1}\otimes Cl_{0,1}\rightarrow Cl_{3,1} \otimes Cl_{0,1}$ & $\mathcal{R}_3 \times \mathcal{R}_3$\\
 & $t_1$ & $\{J\Tilde{H},\Sigma,\Tilde{\mathcal{C}}_+,J\Tilde{\mathcal{C}}_+,\Sigma\Tilde{\prescript{g}{}{U}}^+_+\}$ & $Cl_{3,1}\rightarrow Cl_{4,1}$ & $\mathcal{R}_4$\\ 
 & $t_2$ & $\{J\Tilde{H},\Sigma,\Tilde{\mathcal{C}}_+,J\Tilde{\mathcal{C}}_+\}\otimes \{\Tilde{\prescript{c}{}{U}}^-_+\}$ & $Cl_{2,1}\otimes Cl_{1,0}\rightarrow Cl_{3,1}\otimes Cl_{1,0}$ & $\mathcal{C}_1$ \\
 & $t_3$ & $\{J\Tilde{H},\Sigma,\Tilde{\mathcal{C}}_+,J\Tilde{\mathcal{C}}_+,\Sigma\Tilde{\prescript{g}{}{U}}^-_+\}$ & $Cl_{2,2}\rightarrow Cl_{3,2}$ & $\mathcal{R}_2$\\

    \end{tabular}
    \end{center}
\end{table*}

\begin{table*}
\begin{center}
\caption*{TABLE VI: (continue)}
\begin{tabular}{ccccc}\hline
\multirow{4}{*}{CII$^\dag$}& $t_0$  & $\{J\Tilde{H},\Tilde{\mathcal{C}}_+,J\Tilde{\mathcal{C}}_+,J\Tilde{\mathcal{C}}_+ \Tilde{\mathcal{T}}_-,\Sigma\}\otimes\{\Tilde{\prescript{c}{}{U}}^+_{++}\}$ & $Cl_{3,1}\otimes Cl_{0,1}\rightarrow Cl_{4,1}\otimes Cl_{0,1}$ & $\mathcal{R}_4\times \mathcal{R}_4$ \\
 & $t_1$ & $\{J\Tilde{H},\Tilde{\mathcal{C}}_+,J\Tilde{\mathcal{C}}_+,J\Tilde{\mathcal{C}}_+ \Tilde{\mathcal{T}}_-,\Sigma,\Sigma\Tilde{\prescript{g}{}{U}}^+_{++}\}$ & $Cl_{4,1}\rightarrow Cl_{5,1}$ &  $\mathcal{R}_5$ \\
 &  $t_2$ & $\{J\Tilde{H},\Tilde{\mathcal{C}}_+,J\Tilde{\mathcal{C}}_+,J\Tilde{\mathcal{C}}_+ \Tilde{\mathcal{T}}_-,\Sigma\}\otimes\{\Tilde{\prescript{c}{}{U}}^-_{++}\}$ & $Cl_{3,1}\otimes Cl_{1,0}\rightarrow Cl_{4,1}\otimes Cl_{1,0}$ & $\mathcal{C}_0$\\
 & $t_3$ & $\{J\Tilde{H},\Tilde{\mathcal{C}}_+,J\Tilde{\mathcal{C}}_+,J\Tilde{\mathcal{C}}_+ \Tilde{\mathcal{T}}_-,\Sigma,\Sigma\Tilde{\prescript{g}{}{U}}^-_{++}\}$ &$Cl_{3,2}\rightarrow Cl_{4,2}$   & $\mathcal{R}_3$\\ \hline

\multirow{4}{*}{C$^\dag$}& $t_0$&$\{\Tilde{H},J \Sigma,\Tilde{T}_-,J \Tilde{T}_-\}\otimes \{\prescript{c}{}{\Tilde{U}}^+_+\}$ & $Cl_{3,0}\otimes Cl_{0,1}\rightarrow Cl_{3,1}\otimes Cl_{0,1}$ & $\mathcal{R}_5 \times \mathcal{R}_5$\\
& $t_1$ & $\{\Tilde{H},J \Sigma,\Tilde{T}_-,J \Tilde{T}_-,J \Sigma \prescript{g}{}{\Tilde{U}}^+_+\}$ & $Cl_{3,1}\rightarrow Cl_{3,2}$ & $\mathcal{R}_6$\\
& $t_2$ & $\{\Tilde{H},J \Sigma,\Tilde{T}_-,J \Tilde{T}_-\}\otimes \{\prescript{c}{}{\Tilde{U}}^-_+\}$ & $Cl_{3,0}\otimes Cl_{1,0}\rightarrow Cl_{3,1}\otimes Cl_{1,0}$ & $\mathcal{C}_1$\\
& $t_3$ & $\{\Tilde{H},J \Sigma,\Tilde{T}_-,J \Tilde{T}_-,J \Sigma \prescript{g}{}{\Tilde{U}}^-_+\}$ & $Cl_{4,0}\otimes Cl_{4,1}$ & $\mathcal{R}_4$\\ \hline

 \multirow{4}{*}{CI$^\dag$}& $t_0$  & $\{J\Tilde{H},\Tilde{\mathcal{C}}_+,J\Tilde{\mathcal{C}}_+,J\Tilde{\mathcal{C}}_+ \Tilde{\mathcal{T}}_-,\Sigma\}\otimes\{\Tilde{\prescript{c}{}{U}}^+_{++}\}$ & $Cl_{0,4}\otimes Cl_{0,1}\rightarrow Cl_{1,4}\otimes Cl_{0,1}$ & $\mathcal{R}_6\times \mathcal{R}_6$ \\
 & $t_1$ & $\{J\Tilde{H},\Tilde{\mathcal{C}}_+,J\Tilde{\mathcal{C}}_+,J\Tilde{\mathcal{C}}_+ \Tilde{\mathcal{T}}_-,\Sigma,\Sigma\Tilde{\prescript{g}{}{U}}^+_{++}\}$ & $Cl_{1,4}\rightarrow Cl_{2,4}$ &  $\mathcal{R}_7$ \\
 &  $t_2$ & $\{J\Tilde{H},\Tilde{\mathcal{C}}_+,J\Tilde{\mathcal{C}}_+,J\Tilde{\mathcal{C}}_+ \Tilde{\mathcal{T}}_-,\Sigma\}\otimes\{\Tilde{\prescript{c}{}{U}}^-_{++}\}$ & $Cl_{0,4}\otimes Cl_{1,4}\rightarrow Cl_{4,1}\otimes Cl_{1,0}$ & $\mathcal{C}_0$\\
 & $t_3$ & $\{J\Tilde{H},\Tilde{\mathcal{C}}_+,J\Tilde{\mathcal{C}}_+,J\Tilde{\mathcal{C}}_+ \Tilde{\mathcal{T}}_-,\Sigma,\Sigma\Tilde{\prescript{g}{}{U}}^-_{++}\}$ &$Cl_{0,5}\rightarrow Cl_{1,5}$   & $\mathcal{R}_5$\\ \hline\hline
\end{tabular}
\end{center}
\end{table*}

In this section, we introduce the process of finding the classifying space of Hamiltonians in AZ and AZ$^\dag$ class with additional spatial symmetries. The process is identical to Ref.~\cite{Morimoto_2013,Shiozaki_2014}. Since we have showed in Sec.~\ref{Sec:order-two spatial symmetry AZ and AZ dag} that all antiunitary spatial symmetries can be mapped to unitary one, we only need to consider the addition of unitary symmetries to Hamiltonian. 
\subsection{Complex AZ classes}
First we consider complex AZ classes. The complex Clifford algebra $Cl_p$ is generated by a set of generators $Cl_p=\{e_1,e_2,\cdots,e_p\}$, where all of the generators anti-commute with each other $\{e_i,e_j\}=\delta_{i,j}$. Physically, these generators are symmetry operators of the Hamiltonian. The classification space can be found by the addition of extended Hamiltonian to the generators. Since the extended Hamiltonian obeys $\Tilde{H}^2=1$, we can put them into the generators without violating the properties of the set. The addition of Hamiltonian extends the generators by $Cl_p\rightarrow Cl_{p+1}$:
\begin{equation}
    \{e_1,e_2,\cdots,e_p\}\rightarrow \{\Tilde{H},e_1,e_2,\cdots,e_p\}.
\end{equation}
The mapping from $Cl_p \rightarrow Cl_{p+1}$ provides us the classification space $\mathcal{C}_p$. Furthermore, $\mathcal{C}_p$ obeys the so-called Bott periodicity $\mathcal{C}_{p+2}\simeq \mathcal{C}_p$. The Bott periodicity of complex Clifford algebra is the origin of two complex AZ classes. Hereafter, when referring to generator, we also include Hamiltonian. 

The generators for point-gapped Hamiltonian with \emph{only} internal symmetries~\eqref{eq:nH internal symmetries} are provided in Ref.~\cite{Zhou_2019}. Notice that Ref.~\cite{Zhou_2019} used a different notation systems for symmetry classes from this paper and Kohei~\cite{Kohei_classificaion_2019}. The relationship between these two different notation systems can be found in Ref.~\cite{Liu_2019_defects}. 

The addition of spatial symmetries would change the classification of original classes. To see this, let's consider a order-two spatial symmetry $L$ is added to the classification. There are two possible ways in which the addition of $L$ changes the classification. 

(1) Through combination of $L$ with other generators, the resulting generator $L'$ might \emph{anti-commute} with all other generators i.e. $\{L',e_i\}=0$ for all $i$ and $\{L',H\}=0$. In such a case, the generators would become $\{\Tilde{H},e_1,e_2,\cdots,e_p,L'\}$. Correspondingly, the classifying space would shift by $\mathcal{C}_p\rightarrow \mathcal{C}_{p+1}$. 

(2) It is also possible that $L'$ \emph{commutes} with all other generators i.e. $[L',e_i]=0$ for all $i$ and $[L',H]=0$. The generators then become $\{\Tilde{H},e_1,\cdots,e_p\}\otimes L'$. In this case, the Hilbert space splits into two eigenspaces of $L'$. Each eigenspace has the same Clifford algebra and hence the same classifying space. The classifying space then becomes $\mathcal{C}_p\times\mathcal{C}_p$. 

As an example, consider a point-gapped Hamiltonian in class A. The only symmetry $\Tilde{H}$ obeys in this class is the $\Sigma$ symmetry [Eq.~\eqref{eq:additional CS Sigma}]. Therefore, the Hamiltonian extends the generators as
\begin{equation}
    \{\Sigma\} \rightarrow \{\Tilde{H},\Sigma\}
\end{equation}
which has classification space $\mathcal{C}_1$. This is the classification of class A given in Table~\ref{tab: AZ}. Next, consider the addition of spatial symmetry $\prescript{g}{}{U}$ to the system. Recall that $\prescript{g}{}{\Tilde{U}}$ anti-commutes with $\Sigma$. Then the operator $\Sigma \prescript{g}{}{\Tilde{U}}$ anti-commutes with both $\Tilde{H}$ and $\Sigma$. Now, the generators are given by
\begin{equation}
     \{\Sigma,\Sigma \prescript{g}{}{\Tilde{U}}\} \rightarrow \{\Tilde{H},\Sigma,\Sigma \prescript{g}{}{\Tilde{U}}\}
\end{equation}
which has classifying space $\mathcal{C}_2\simeq \mathcal{C}_0$. 

As another example, we consider a point-gapped Hamiltonian in class A with the addition of $\prescript{c}{}{U}$. Recall that the extended operator $\prescript{c}{}{\Tilde{U}}$ that acts on extended Hamiltonian $\Tilde{H}$ commutes with both $\Tilde{H}$ and $\Sigma$. Therefore, the generators are now given by
\begin{equation}
    \{\Sigma\}\otimes \prescript{c}{}{\Tilde{U}}\rightarrow \{\Tilde{H},\Sigma\}\otimes\prescript{c}{}{\Tilde{U}},
\end{equation}
which has classifying space $\mathcal{C}_1\times\mathcal{C}_1$. 

Following a similar process, we summarize the classification of complex AZ classes under spatial symmetries in the first two rows of Table~\ref{tab:Clifford generators of AZ and AZ dag}. 
\subsection{Real AZ and AZ$^\dag$ classes}
Now consider real AZ and AZ$^\dag$ classes. Their classifying space are given by the real Clifford algebra $Cl_{p,q}=\{e_1,\cdots, e_p,e_{p+1},\cdots,e_1\}$, where  $\{e_i,e_j\}=\delta_{i,j}$, $e_i^2=-1$ for $i=1,\cdots,p$ and $e_i^2=1$ for $i=p+1,\cdots,p+q$. In this case, the Hamiltonian can extend the generators by
\begin{align}
    &Cl_{p,q}=\{e_1,\cdots,e_{p+q}\}\rightarrow Cl_{p+1,q}=\{\Tilde{H},e_1,\cdots,e_{p+q}\}\nonumber\\
    &\Tilde{H}=-1.
\end{align}
In this case, the classifying space is given by $\mathcal{R}_{p+2-q}$. Hamiltonian can also extend the generators by 
\begin{align}
    &Cl_{p,q}=\{e_1,\cdots,e_{p+q}\}\rightarrow Cl_{p,q+1}=\{\Tilde{H},e_1,\cdots,e_{p+q}\}\nonumber\\
    &\Tilde{H}=1.
\end{align}
The classifying space is given by $\mathcal{R}_{q-p}$. 

Now let's consider the addition of order-two spatial symmetry $L$. Similar to complex AZ classes, through the combination of $L$ and other generators as well as the imaginary unit $J$, the resulting generator $L'$ would commute or anti-commute with the rest of the generators. However, due to the presence of antiunitary symmetries, we need to consider the squared value of $L'$. Since $L'^2=\pm 1$, we have four possible situations. 

(1) $L'$ anti-commutes with all generators and $L'^2=-1$. In this case the generators would change by $p\rightarrow p+1$. The new classifying space would be $\mathcal{R}_{q-p-1}$ ($\mathcal{R}_{p+3-q}$) if $\Tilde{H}=-1$ ($1$)

(2) $L'$ anti-commutes with all generators and $L'^2=1$. In this case the generators would change by $q\rightarrow q+1$. The new classifying space would be $\mathcal{R}_{q+1-p}$ ($\mathcal{R}_{p+1-q}$) if $\Tilde{H}=-1$ ($1$)

(3) $L'$ commutes with all generators and $L'^2=-1$. In this case, $L'$ would introduce complex structure to real Clifford algebra. The set of generators without Hamiltonian is written as $Cl_{p,q}\otimes Cl_{1,0}$. Thus, the classifying space would be the same as complex AZ classes. In fact, we have the relationship $Cl_{p,q}\otimes Cl_{1,0}\simeq Cl_{p+q}$ (Appendix~\ref{Appendix:Properties of Clifford Algebra}). The classifying space is then given by $\mathcal{C}_{p+q}$. 

(4) $L'$ commutes with all generators and $L'^2=1$. In this case, the Hilbert space splits into two eigenspace of $L'$. The classifying space is given by $\mathcal{R}_{q-p}\times \mathcal{R}_{q-p}$ ($\mathcal{R}_{p+2-q}\times \mathcal{R}_{p+2-q}$) if $\Tilde{H}=-1$ ($1$). 

As an example, we consider the classification of class AI in addition to $\prescript{g}{}{U}^+_+$. The generators for point-gapped AI class is given by~\cite{Zhou_2019}
\begin{equation}
    Cl_{2,2}=\{J \Tilde{H},J\Sigma,\Tilde{\mathcal{T}}_+,J\Tilde{\mathcal{T}}_+\},
    \label{eq:generators AI}
\end{equation}
where $\mathcal{T}_+$ is the TRS operator and $\Sigma$ is the $\Sigma$ symmetry~\eqref{eq:additional CS Sigma}. Since $(J\Tilde{H})^2=-1$, the extension is $Cl_{1,2}\rightarrow Cl_{2,2}$ which gives classifying space $\mathcal{R}_1$. Now we consider the addition of spatial symmetry $\prescript{g}{}{U}^+_+$. Notice that $J\Sigma \prescript{g}{}{\Tilde{U}}^+_+$ anti-commutes with all the generators in Eq.~\eqref{eq:generators AI} and it squares to $1$ (Recall that $\Sigma$ and $\prescript{g}{}{\Tilde{U}}^+_+$ anti-commute). Therefore, the new set of generators is 
\begin{equation}
    Cl_{2,3}=\{J \Tilde{H},J\Sigma,\Tilde{\mathcal{T}}_+,J\Tilde{\mathcal{T}}_+,J \Sigma \prescript{g}{}{\Tilde{U}}^+_+\}.
\end{equation}
The extension is now given by $Cl_{1,3}\rightarrow Cl_{2,3}$, which gives classifying space $\mathcal{R}_0$.

Following these procedures, we arrive at Table~\ref{tab:Clifford generators of AZ and AZ dag} for real AZ and AZ$^\dag$ classes.

\bibliography{Biblio}

\end{document}